\documentclass[aps,floatfix,10pt,twocolumn,preprintnumbers,amsmath,amssymb,superscriptaddress,nofootinbib,longbibliography,prb]{revtex4-2}

\usepackage{graphicx}

\usepackage{bm}
\usepackage{dsfont}
\usepackage[usenames,dvipsnames]{xcolor}
\usepackage{pstricks}
\usepackage[tight]{subfigure}
\usepackage{verbatim}
\usepackage{units}
\usepackage{multirow}
\usepackage{enumitem}
\usepackage{mathrsfs}
\usepackage{leftidx}
\usepackage{xspace}
\usepackage{bbm}
\usepackage{mathtools}
\usepackage{amsfonts,amssymb,amsmath}
\usepackage{tabularx}
\usepackage[normalem]{ulem}
\usepackage{bbold}
\usepackage{capt-of}
\usepackage{hyperref}
\hypersetup{colorlinks=true,linktoc=all,linkcolor=blue,breaklinks=true,citecolor=blue,urlcolor=blue}

\usepackage[english]{babel}
\addto\captionsenglish{}

\AtBeginDocument{%
    \newwrite\bibnotes
    \def\bibnotesext{Notes.bib}
    \immediate\openout\bibnotes=\jobname\bibnotesext
    \immediate\write\bibnotes{@CONTROL{REVTEX42Control}}
    \immediate\write\bibnotes{@CONTROL{%
    apsrev42Control,author="08",editor="1",pages="1",title="0",year="1"}}
     \if@filesw
     \immediate\write\@auxout{\string\citation{apsrev42Control}}%
    \fi
}%

\renewcommand{\Re}{{\text{Re}}}
\renewcommand{\Im}{{\text{Im}}}

\newcommand{\one}{\mathds{1}}
\newcommand{\ket}[1]{|#1\rangle}
\newcommand{\bra}[1]{\langle#1|}

\newcommand{\braket}[2]{\langle #1 | #2 \rangle}

\DeclareMathOperator{\Tr}{Tr}

\newcommand{\Eq}[1]{Eq.~(\ref{#1})}
\newcommand{\Eqs}[1]{Eqs.~(\ref{#1})}
\newcommand{\BrackEq}[1]{[Eq.~(\ref{#1})]}
\newcommand{\eq}[1]{(\ref{#1})}
\newcommand{\Fig}[1]{Fig.~\ref{#1}}
\newcommand{\FigPart}[2]{Fig.~\ref{#1}(#2)}
\newcommand{\BrackFig}[1]{[Fig.~\ref{#1}]}
\newcommand{\BrackFigPart}[2]{[Fig.~\ref{#1}(#2)]}

\newcommand{\Sec}[1]{Sec.~\ref{#1}}
\newcommand{\BrackSec}[1]{$[$Sec.~\ref{#1}$]$}
\newcommand{\Refe}[1]{Ref.~\cite{#1}}
\newcommand{\Refs}[1]{Refs.~\cite{#1}}
\newcommand{\App}[1]{Appendix~\ref{#1}}
\newcommand{\fpop}{(-\one)^N}

\newcommand{\Htot}{H_{\text{tot}}}
\newcommand{\Htun}{H_{\text{tun}}}
\newcommand{\Hres}{H_{\text{res}}}

\newcommand{\lrsepa}{\quad , \quad}
\newcommand{\nbrack}[1]{\left(#1\right)}
\newcommand{\sqbrack}[1]{\left[#1\right]}
\newcommand{\cbrack}[1]{\left\{#1\right\}}

\newcommand{\Thetafn}[1]{\varTheta(#1)}

\renewcommand{\Re}{\mathrm{Re}}
\renewcommand{\Im}{\mathrm{Im}}
\newcommand{\cosfn}[1]{\cos{\left(#1\right)}}
\newcommand{\sinfn}[1]{\sin{\left(#1\right)}}

\newcommand{\equalby}[1]{\overset{#1}{=}}

\newcommand{\kB}{k_{\text{B}}}

\newcommand{\IN}{I_{N}}
\newcommand{\IE}{I_{E}}

\newcommand{\VdT}{V_{\text{d},2}}

\newcommand{\vV}{v_{\text{r}}}

\newcommand{\coulCon}{k_e}
\newcommand{\coulConTil}{\tilde{k}_e}
\newcommand{\Eb}{E_{\text{b}}}

\newcommand{\dEb}{\delta E_{\text{b}}}
\newcommand{\Ld}{L_{\text{d}}}
\newcommand{\Emit}{E_{\text{mit}}}
\newcommand{\Emitx}{E_{\text{mit},x}}
\newcommand{\Emitot}{E_{\text{mit},1}}
\newcommand{\Emitzo}{E_{\text{mit},2}}
\newcommand{\Etr}{E_{\text{tr}}}
\newcommand{\Vac}{\ket{\text{Vac}}}
\newcommand{\PsiDag}{\Psi^{\dagger}}
\newcommand{\PhiGS}{\text{GS}}
\newcommand{\PhiSeed}{\text{Seed}}
\newcommand{\HTP}{H_{\text{2P}}}
\newcommand{\HCoul}{H_{\text{Coul}}}
\newcommand{\Hpot}{H_{\text{pot}}}
\newcommand{\Hkin}{H_{\text{kin}}}
\newcommand{\meff}{m_{\text{eff}}}
\newcommand{\delx}{\delta x}
\newcommand{\Nl}{N^{l}}
\newcommand{\Hkinl}{H_{\text{kin}}^l}
\newcommand{\RlO}{R^{l}_1}
\newcommand{\RlT}{R^{l}_2}
\newcommand{\Nd}{N^{\text{d}}}

\newcommand{\Nf}{N^{\text{f}}}
\newcommand{\Hkinf}{H_{\text{kin}}^{\text{f}}}

\newcommand{\ECd}{E^{\text{d}}_{\text{C}}}
\newcommand{\totwo}{t_{1}}
\newcommand{\tzo}{t_{2}}
\newcommand{\nsepa}{\!\!\!\!\!\!}
\newcommand{\relTau}{X^{\text{RL}}}
\newcommand{\epso}{\epsilon_{0}}
\newcommand{\Nsw}{N_{\text{sw}}}
\newcommand{\Nt}{N_t}

\newcommand{\xdl}{x_\text{d,L}}
\newcommand{\xdc}{x_\text{d,C}}
\newcommand{\xdr}{x_\text{d,R}}
\newcommand{\xfl}{x_\text{f,L}}
\newcommand{\xfr}{x_\text{f,R}}
\newcommand{\xds}{x_{\text{d},s}}
\newcommand{\sdl}{\sigma_\text{d,L}}
\newcommand{\sdc}{\sigma_\text{d,C}}
\newcommand{\sdr}{\sigma_{\text{d,R}}}
\newcommand{\sds}{\sigma_{\text{d},s}}
\newcommand{\Vdl}{V_\text{d,L}}
\newcommand{\Vdc}{V_\text{d,C}}
\newcommand{\Vdr}{V_{\text{d,R}}}
\newcommand{\Vds}{V_{\text{d},s}}
\newcommand{\Ef}{E^{\text{f}}_{\text{kin}}}
\newcommand{\DEtrref}{\Delta E^{\text{ref}}_{\text{tr}}}
\newcommand{\Deps}{\Delta\epsilon}
\newcommand{\DEmit}{\Delta\Emit}
\newcommand{\epsb}{\epsilon_{\text{b}}}

\newcommand{\epsU}{\epsilon_{U}}
\newcommand{\depsU}{\delta\epsilon_{U}}
\newcommand{\vramp}{v_{\text{r},1}}
\newcommand{\NS}{N_{\text{S}}}
\newcommand{\NSt}{\tilde{N}_{\text{S}}}
\newcommand{\coup}{\tau}
\newcommand{\delt}{\delta t}
\newcommand{\NGS}{N_{\text{GS}}}
\newcommand{\NGSn}{N^{\text{norm}}_{\text{GS}}}
\newcommand{\Ntest}{N^{\text{test}}_{\text{GS}}}
\newcommand{\NCP}{N_{\text{CP}}}
\newcommand{\ECP}{E_{\text{CP}}}
\newcommand{\epsL}{\epsilon_{\text{L}}}
\newcommand{\epsR}{\epsilon_{\text{R}}}
\newcommand{\NL}{N_{\text{L}}}
\newcommand{\NR}{N_{\text{R}}}
\newcommand{\dL}{d_{\text{L}}}
\newcommand{\dR}{d_{\text{R}}}
\newcommand{\dDagL}{d^\dagger_{\text{L}}}
\newcommand{\dDagR}{d^\dagger_{\text{R}}}
\newcommand{\GamL}{\Gamma_{\text{L}}}
\newcommand{\GamR}{\Gamma_{\text{R}}}
\newcommand{\sumsub}[2]{\sum_{\substack{{#1}\\{#2}}}}

\begin{document}

\title{Spectroscopy of hot-electron pair emission from a driven quantum dot}

\author{Jens Schulenborg}
\affiliation{Department of Microtechnology and Nanoscience (MC2), Chalmers University of Technology, S-412 96 G\"oteborg, Sweden}
\affiliation{Center for Quantum Devices, Niels Bohr Institute, University of Copenhagen, 2100 Copenhagen, Denmark}

\author{Jonathan D. Fletcher}
\affiliation{National Physical Laboratory, Hampton Road, Teddington, Middlesex, TW11 0LW, UK}

\author{Masaya Kataoka}
\affiliation{National Physical Laboratory, Hampton Road, Teddington, Middlesex, TW11 0LW, UK}

\author{Janine Splettstoesser}
\affiliation{Department of Microtechnology and Nanoscience (MC2), Chalmers University of Technology, S-412 96 G\"oteborg, Sweden}

\date{\today}

\begin{abstract}
On-demand emission of individual electrons for the implementation of flying qubits and quantum electron-optics experiments requires precise knowledge and tunability of emission times and energies.
Crucially, 
for confined electron sources such as driven quantum dots, 
the effect of local Coulomb interaction on these emission properties needs to be understood, in particular if multiple particles are 
emitted close in time or near-simultaneously. This paper theoretically analyzes electron-pair emission from an ac driven quantum dot, detailing the competing effects of the electron-electron interaction, the time-dependent potential forming the quantum dot, and of the quantum-state properties, such as degeneracy, on the emission times and energies. We complement a numerical analysis of the coherent Schr\"odinger evolution of two particles in a driven potential with a master-equation description for strongly interacting electrons tunneling stochastically into a weakly coupled conductor. This captures a broad range of different influences on the emitted particles and thereby guides the development of single-electron sources with higher control over two-particle emission properties.
\end{abstract}

\maketitle

\section{Introduction}

The invention of on-demand single-electron sources~\cite{Feve2007May,Blumenthal2007May,Dubois2013Oct} has opened up for recent research on electron-based flying qubits~\cite{Edlbauer2022Dec} and more generally on quantum optics experiments based on the controlled emission of single electrons~\cite{Bocquillon2012May,Bocquillon2013Mar,Dubois2013Oct}. This endeavour relies not only on precisely \emph{timing} the electron emission, but also on controlling the \textit{energy} of the emitted particles. Indeed, recent experiments have manipulated electrons in an energy-selective manner~\cite{Fletcher2013Nov,Ubbelohde2015Jan} and time-resolved measurements~\cite{Fletcher2013Nov,Waldie2015Sep} and tomographic techniques~\cite{Fletcher2019Nov, Jullien2014Oct,Thibierge2016Feb,Bisognin2019Jul,Locane2019Sep} have revealed information about the emission characteristics of different sources.

\begin{figure}[!t]
    \centering
    \includegraphics[width=\linewidth]{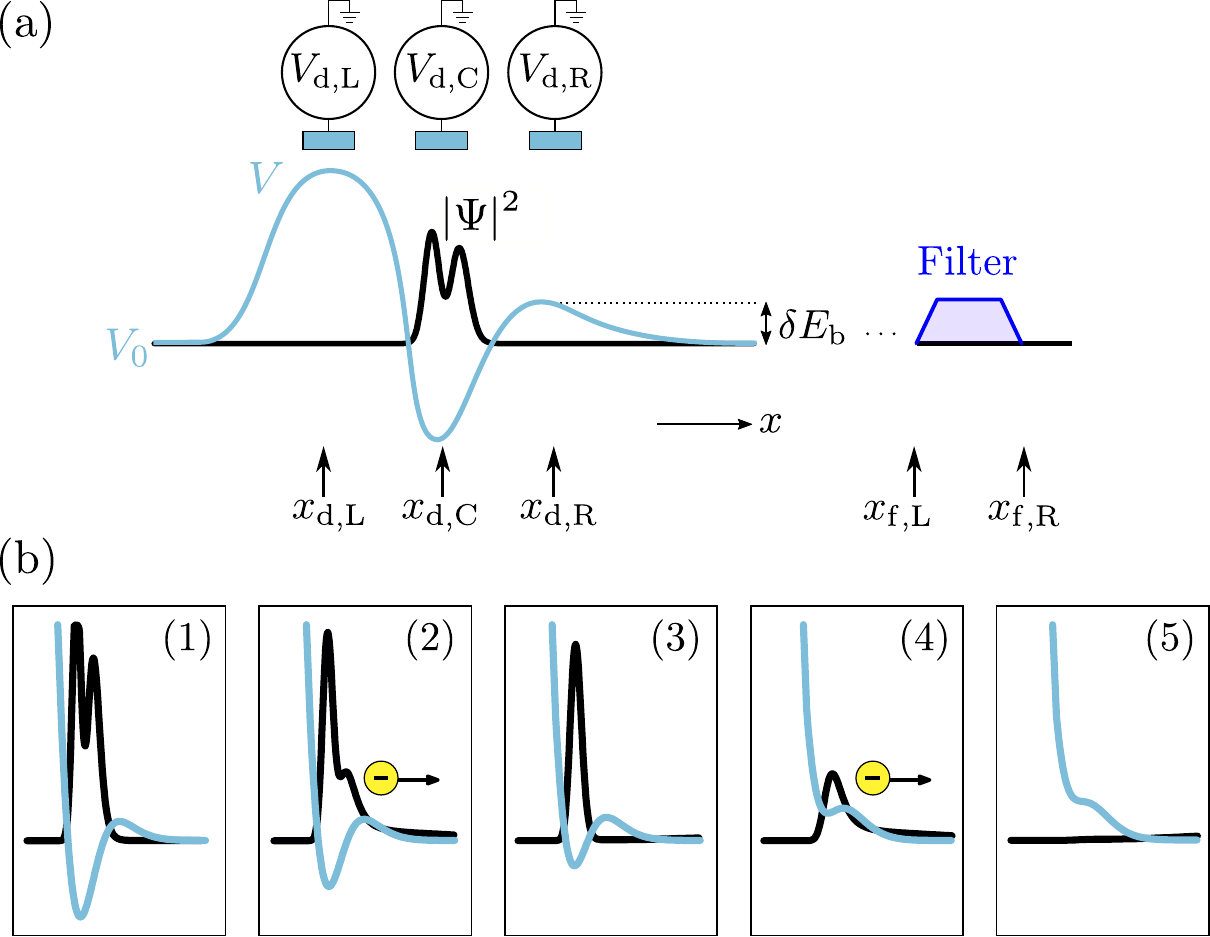}
    \caption{(a) Sketch of a tunable-barrier quantum-dot setup. Gates on top of a conductor time-dependently modulate the potential landscape [Eq.~(\ref{eq_potential_landscape})] to eject pairs of electrons. The positions $\xdl$, $\xdc$, and $\xdr$ indicate the boundaries and center of the quantum-dot region, $\xfl$ and $\xfr$ set the regions, where a filter would measure time and energy at which electrons propagate through it. (b) Sketch of the emission process: the potential is driven such that the first electron can tunnel out, until only one electron remains in the quantum dot. Further driving allows also the second electron to tunnel out until the potential well has basically disappeared.}
    \label{fig_setting_1d}
\end{figure}

Reliable two--qubit gates within the electron-optical flying qubit platform require further, crucial progress in manipulating two-electron quantum states~\cite{Bocquillon2013Mar,Dubois2013Oct,Ubbelohde2015Jan,Thibierge2016Feb,Fletcher2019Nov,Bisognin2019Jul,Fletcher2022Oct,WangSAW,UbbelohdeChargeDet}. The most important distinguishing feature of electronic devices in this context is the possibly strong Coulomb interaction between electrons. This can play a role during particle propagation along wave guides (implemented via quantum Hall edge states)~\cite{Freulon2015Apr,Sungguen2022}, or when particles are brought to collide in a tunable manner at quantum-point-contact based beam splitters~\cite{Ubbelohde2015Jan,Pavlovska2022Jan,Fletcher2022Oct,WangSAW,UbbelohdeChargeDet}.
In contrast, the study of the interplay with local Coulomb interaction, when electrons are emitted from  a periodically driven quantum dot~\cite{Blumenthal2007May,Feve2007May,Leicht2011Mar,Cho2022Dec,Pekola2013Oct},  has until now mostly been limited to single-level quantum dots with constant, energy-independent tunnel barriers~\cite{Schulenborg2016Feb,Vanherck2017Mar,Schulenborg2018Dec}. Previous analyses of the emission energy have emphasized the role of the time- and energy-dependence of the potential-well forming the quantum dot~\cite{Fletcher2019Nov,Locane2019Sep,Kashcheyevs2017Jun}. However, local Coulomb interaction is expected to strongly affect the emission times and energies 
of emitted electrons~\cite{Splettstoesser2010Apr,Filippone2011Oct,Kashuba2012Jun,Schulenborg2014May,Alomar2016Oct,Filippone2020Jul}, which in turn are important for propagation velocities and further processing of flying qubits.

The present paper thus theoretically 
analyses the emission times and energies of two-electron emission processes from a quantum dot into a conductor separated by a gate-tunable barrier~\cite{Kaestner2015Sep}, as sketched in Fig.~\ref{fig_setting_1d}(a). A particular focus of this study lies on the effect of local Coulomb interaction on such processes.

Typically, the emission into the conductor takes place far above the Fermi energy, and is hence referred to as hot-electron emission. The precise emission time and energy is controlled by a time-dependently modulated potential of the dot and its exit tunnel barrier, as sketched in Fig.~\ref{fig_setting_1d}(b).
Electrons escape the dot once any of the addition energies rises to a point above the conduction band bottom at which the exit barrier becomes sufficiently transparent. The emission process and the separation of the particles in time and energy~\cite{Ubbelohde2015Jan,Fletcher2013Nov}, as pictured in \FigPart{fig_setting_1d}{b}, is hence determined by a complex interplay between the modulated potential shape and the two-particle dot state dynamics due to quantum interference as well as Coulomb interaction.

To get hold of the various competing effects, we approach the problem with two complementary methods. First, we model the two-particle problem by numerically solving and analyzing the fully coherent two-particle Schr\"odinger dynamics in a 1 dimensional (1d) potential landscape with a dip between two barriers acting as the dot~\BrackFig{fig_setting_1d}. This captures how different addition energies for the two emitted electrons due to charging energy and single-particle level splitting lead to different emission times, since the time-dependent potential modulation causes these energies to reach the onset of finite exit-barrier transparency at subsequent times. The simulation, however, also exposes how this straightforward time-energy separation can be nontrivially modified by the possibility of Coulomb interaction transferring energy between the particles during the first emission event. Similar interaction-based energy transfer between particles scattered at a barrier has been identified in Ref.~\cite{Sungguen2022}.

Second, the emission times themselves can be enhanced or delayed differently for the two particles, thereby further altering the emission time-energy relations. Key underlying mechanisms include
single-particle state degeneracies leading to a faster first emission~\cite{Hofmann2016Nov,Beckel2014May,Splettstoesser2010Apr,Schulenborg2016Feb}, as well as time-dependent dot and exit barrier variations between the emission events shifting the required transition energy of the second particle relatively to the first. A suitable experiment to study these mechanisms is to measure the emission energies in sweeps of the dot potential ramp speed, as the latter changes the rate of energy increase relatively to the exit barrier escape rate. Simulating such sweeps using the full, interacting two-particle Schr\"odinger dynamics is challenging.  
Results may also depend on many experimentally inaccessible details of the potential landscape, and the numerical cost of repeating the full simulation for many different ramp speeds is rather high. We tackle these challenges by complementing the two-particle simulation in 1d with a time-dependent quantum master equation for an effective two-orbital quantum dot, capturing in a simple way the interplay between Coulomb interaction, spatial dependence of single-electron states with dot-internal dynamics, and energy-dependent dot-conductor couplings. Unlike the coherent 1d Schr\"odinger evolution, this 
only applies to the weak dot-conductor coupling regime, governed by electron emission processes due to stochastic tunneling. The crucial advantage is, however, that the much smaller state- and parameter space enables us to systematically classify and compare the competing effects impacting the emission times and energies as a function of the dot potential ramp speed. This approach in particular accounts for degeneracy-enhanced emission rates, and for classical ensemble-averaging which is typically performed in experiments as well.

We implement our investigation strategy by organizing the remainder of this paper as follows. The model and theoretical background of the coherent two-particle Schr\"odinger dynamics within the time-dependently modulated, 1d potential landscape is introduced in \Sec{sec_twoparticle_theory}; \Sec{sec_twoparticle_results} then discusses the emission times and energies obtained from the corresponding numerical simulations. The quantum master equation approach for the dynamics of the simplified quantum-dot model is set up in \Sec{sec_low_energy_theory}; the resulting effects in the energy-time spectroscopy of the emission process are detailed in \Sec{sec_low_energy_results}. We highlight the main insights and open questions emerging from the two complementary approaches of describing two-particle emission times and energies in the concluding section \ref{sec_conclusion}. This paper also contains an appendix with details on the numerical simulation of \Sec{sec_twoparticle_theory} and \Sec{sec_twoparticle_results} and on the quantum master equation approach. Throughout the manuscript, we set $\hbar = \kB = |e| = 1$.

\section{Two-particle simulation in 1d}\label{sec_twoparticle_theory}

Experimental realizations of periodically driven quantum-dot electron pumps typically rely on electrostatic confinement of two-dimensional electron gases via tunable gates, as sketched in Fig.~\ref{fig_setting_1d}(a). A full theory of two-electron emission from such dots would need to determine the time evolution of a many-body electron system with Coulomb interaction in an inhomogeneous and time-dependent potential. Some approximations hence need to be made to make the problem manageable.

In a first step, Secs.~\ref{sec_twoparticle_theory} and \ref{sec_twoparticle_results}, we make two approximations: First of all, we decide to treat a two-particle problem, which is motivated by the fact that the interplay between the emitted `hot' electrons and the Fermi sea in the contacts is rather weak. Furthermore, in order to keep the numerical simulation tractable, we model the system in one dimension.

\subsection{Hamiltonian with time-dependent potential for hot-electron emission}\label{sec_potential_landscape}

The potential landscape of the dot and its environment we consider is sketched in \Fig{fig_setting_1d}(a). With $V_0$ representing the conduction band bottom high above the Fermi sea, we envision the three gates defining the quantum dot to have a Gaussian-like effect on the potential landscape:
\begin{align}
 V(x \geq 0,t) &= V_0 - \Vdc(t) \exp\sqbrack{-\frac{(x - \xdc)^2}{2\sdc^2}} \label{eq_potential_landscape}\\
 &\phantom{=}+ \sum_{s=\text{L,R}}\Vds \exp\sqbrack{-\frac{(x - \xds)^2}{2\sds^2}}. \notag
\end{align}
The left Gaussian peak defines the left quantum dot `wall' of height $\Vdl$ at the reference coordinate $\xdl \equiv 0$ with width $\sdl$, and the right peak represents the exit barrier of height $\Vdr$ and width $\sdr$ at $\xdr$. 
The maximum of the resulting emission barrier with respect to the potential in the conductor $V_0$ at the initial time $t_0$ defines a characteristic energy scale of the potential and is indicated by $\delta E_\mathrm{b}$ in Fig.~\ref{fig_setting_1d}. The central, inverted Gaussian with constant width $\sdc$ but time-dependent depth $\Vdc(t)$ at $\xdc > \xdl = 0$ establishes a tunable potential well. At initial time $t_0 \equiv 0$, we choose $\Vdc(0) > 0$ such that a dot forms inside this potential dip,   
and confines two particles within the typical spatial dot range\footnote{Due to the slightly skewed potential, we choose 2 standard deviations to the left and 3 to the right.} $[0,\Ld = 5\sdc]$. The dip potential $\Vdc(t)$ is then raised until the particles are emitted into the flat channel right to the exit barrier, where we linearize $\Vdc(t) \approx \Vdc(0) - \vV t$ with corresponding ramp speed $\vV$. The latter approximation is appropriate if emission takes place during some small fraction of a periodic drive signal~\cite{Fletcher2019Nov} with an amplitude $A\sim V_0$, such that the only relevant transition energies $E$ lie close to the emission point, $|E-V_0|/A \ll 1$.

We set up the system Hamiltonian by discretizing the potential landscape $V(x \geq 0,t)$ \BrackEq{eq_potential_landscape} into $R+1$ points defining $R$ equidistant real space intervals $\delx = L/R$ on some large, but finite length $L \gg \Ld$. Importantly, we explicitly model only the dynamics of the two particles initially occupying the dot. This relies on the assumption that for hot-electron emission with sufficient energy splitting between empty conductance and completely filled valence band, the latter appears to be nearly chargeless to the emitted electrons, apart from some overall screening effect. Hence fixing the total particle number to $2$, we obtain a Hilbert space spanned by the $(R+1)(2R+1)$ orthonormal, anti-symmetric states $\ket{r\sigma,r'\sigma'} = \PsiDag_{r\sigma}\PsiDag_{r'\sigma'}\Vac = -\ket{r'\sigma',r\sigma}$ in discrete positions $x_r = r\delx$, $r\in\cbrack{0,\dotsc,R}$ with spin-z projection $\sigma = \uparrow,\downarrow$. The particles are created from the vacuum $\Vac$  by the fermionic field operators $\PsiDag_{r\sigma}$ obeying the usual anti-commutation rules $\cbrack{\Psi_{r\sigma},\Psi_{r'\sigma'}} = 0$, $\cbrack{\Psi_{r\sigma},\PsiDag_{r'\sigma'}} = \delta_{rr'}\delta_{\sigma\sigma'}$. 

The time-dependent two-particle Hamiltonian $\HTP(t) = \Hkin + \HCoul + \Hpot(t)$ includes the potential landscape
\begin{subequations}\label{eq_hamiltonian_2p_parts}
  \begin{align}
  \Hpot(t) &= \sum_{\sigma=\uparrow,\downarrow}\sum_{r = 0}^RV(x_r,t)N_{r\sigma}\label{eq_hamiltonian_2p_pot},
\end{align}
where $N_{r\sigma} = \PsiDag_{r\sigma}\Psi_{r\sigma}$ are the occupation numbers at positions $x_r$, and the kinetic energy 
\begin{align}
 \Hkin &= \sum_{\sigma=\uparrow,\downarrow}\sum_{r = 0}^{R}2\lambda N_{r\sigma}\label{eq_hamiltonian_2p_kin}\\
 &-\sum_{\sigma=\uparrow,\downarrow}\sqbrack{\sum_{r = 0}^{R-1}\lambda\PsiDag_{(r+1)\sigma}\Psi_{r\sigma} + \sum_{r = 0}^{R-1}\lambda\PsiDag_{r\sigma}\Psi_{(r+1)\sigma}}\notag\ . 
\end{align}
The coupling $\lambda = 1/(2\meff\delx^2)$ with effective electron mass $\meff$ arises when discretizing the kinetic term $\PsiDag_{r\sigma}\frac{-\partial_x^2}{2\meff}\Psi_{r\sigma}$ as the centered difference, $\partial_x^2\Psi_{r\sigma} \approx \sqbrack{\nbrack{\Psi_{(r+1)\sigma} - \Psi_{r\sigma}} - \nbrack{\Psi_{r\sigma} - \Psi_{(r-1)\sigma}}}/\delx^2$.  The Coulomb interaction potential
 \begin{align}
 \HCoul &= \sum_{\sigma,\sigma'=\uparrow,\downarrow}\sum_{r' > r}^{R}\frac{\coulConTil}{\delx|r' - r|}N_{r'\sigma'} N_{r\sigma} \label{eq_hamiltonian_2p_coul}     
 \end{align} 
 \end{subequations}
contains the effective Coulomb constant $\coulConTil = \kappa\coulCon$, where $\coulCon$ is the bare value and $\kappa$ accounts for material properties screening. 
Note that for a spin-up and spin-down electron in the same position $x_r$, we regularize the Coulomb potential heuristically by setting the denominator in \Eq{eq_hamiltonian_2p_coul} to $\delx/2$. This is appropriate if the discretization step $\delx$ is small enough for the two electrons to reside in a state with a typical separation larger than $\delx$ prior to emission. Only in situations irrelevant to our analysis may our $\delx$ choice still be insufficient, such as collisions of counter-propagating particles. 
While we here put forward one ---experimentally relevant--- choice for the confinement and interaction potentials, modifications of this could be envisioned in the future. One extension would be to implement a Thomas-Fermi model for the effect of a background electronic density on the screening of interactions or via image-charge screening~\cite{Skinner2010Oct}, as used for example in Ref.~\cite{Fletcher2022Oct}. Furthermore, one could optimize the confinement potential for the emission, similarly to how it has recently been shown for the process of loading the dot~\cite{Akmentinsh2023Jan}.

\subsection{Unitary two-particle evolution and spectroscopy}\label{sec_two_particle}

We analyze the emission process, which is induced by ramping up $\Vdc(t)$ until the two electrons have enough energy to pass through the exit barrier at $\xdr$ into the environment channel, at a potential energy $V_0$. We start by setting the initial state $\ket{\Phi(t = 0)}$ to the ground state $\ket{\PhiGS}$ of $\HTP(t = 0)$, for which the potential dip $\VdT(0) \gg \ECd$ is deep enough for the dot to be stably occupied by both electrons, given a typical dot charging energy $\ECd = \coulConTil/\Ld$. 
See, e.g., Ref.~\cite{Kashcheyevs2010May,Fricke2013Mar,Wenz2019May} for details of the loading process.

The ground state and its subsequent evolution
\begin{equation}
 \partial_t\ket{\Phi(t)} = -i\HTP(t)\ket{\Phi(t)} \lrsepa \ket{\Phi(0)} = \ket{\PhiGS}\label{eq_schrodinger}
\end{equation}
due to the gradually lifted potential dip $\VdT(t) \rightarrow 0$ are obtained numerically using the Hamiltonian given in \Eq{eq_hamiltonian_2p_parts}.  
We have developed a GPU implementation of a direct, norm-preserving leap-frog solver for a large, but bounded system, see Appendix~\ref{sec_2p_numerics}. This offers non-prohibitive run times even for boundaries far enough away that unphysical reflections do not affect the particle emission. In particular, it enables us to use the kernel polynomial method~\cite{Weisse2006Mar} to compute the full emission energy distribution as a function of time.

With the time-dependent two-particle state $\ket{\Phi(t)}$ at hand, we track the emission of the two electrons from the dot with the expectation values $\langle O\rangle(t) = \bra{\Phi(t)}O\ket{\Phi(t)}$ of various observables $O$. Namely, next to the Coulomb energy $\HCoul$ and the spatial charge density $|\Psi_1(r,t)|^2 = \sum_{\sigma=\uparrow,\downarrow}\bra{\Phi(t)}N_{r\sigma}\ket{\Phi(t)}/\delx$, we extract the charge  and kinetic energy in the dot $(l = \text{d})$ as well as in a region away from the dot representing the energy filter $(l = \text{f})$, see Fig.~\ref{fig_setting_1d}(b). Concretely, we obtain the charge $\Nl = \sum_{\sigma=\uparrow,\downarrow}\sum_{r = \RlO}^{\RlT}N_{r\sigma}$ 
and the kinetic energy $\Hkinl$ by modifying Eq.~(\ref{eq_hamiltonian_2p_kin}) by replacing in all sums the lower limit $0$ by $\RlO$ and $R$ in the upper limit by $\RlT$; 
 the indices $\RlT > \RlO$ delimit the spatial ranges $[\RlO\delx,\RlT\delx]$ on which these regions are defined. The time-dependent charge and kinetic energy in the filter region as function of time are the main quantities of interest in the analysis of Sec.~\ref{sec_twoparticle_results}. Their properties are further supported by the 
 time- and energy resolved emission distribution 
extracted from the spectral density in the filter,
\begin{equation}
 \phi_{2\text{P}}(t,E) = \bra{\Phi(t)}\delta(E - \Hkinf)\ket{\Phi(t)}\label{eq_dos}.
\end{equation}
Technically, this distribution is analogous to spectral decompositions in Green's function approaches, see e.g. Ref.~\cite{Bjornson2016}. From a practical point of view, it connects to the experimentally broadened distribution functions of Refs.~\cite{Fletcher2013Nov,Fletcher2019Nov}, see also the averaged emission distribution~\eqref{eq_dist_approx} of Sec.~\ref{sec_low_energy_theory}. However, in contrast to the stochastic effects discussed later in Secs.~\ref{sec_low_energy_theory} and \ref{sec_low_energy_results}, Eq.~(\ref{eq_dos}) only emerges from quantum effects and two-particle interaction.
Note that this quantity should not be confused with quasi probabilities, such as for example the Wigner function.

\begin{figure*}[t]
    \centering
    \includegraphics[width=\linewidth]{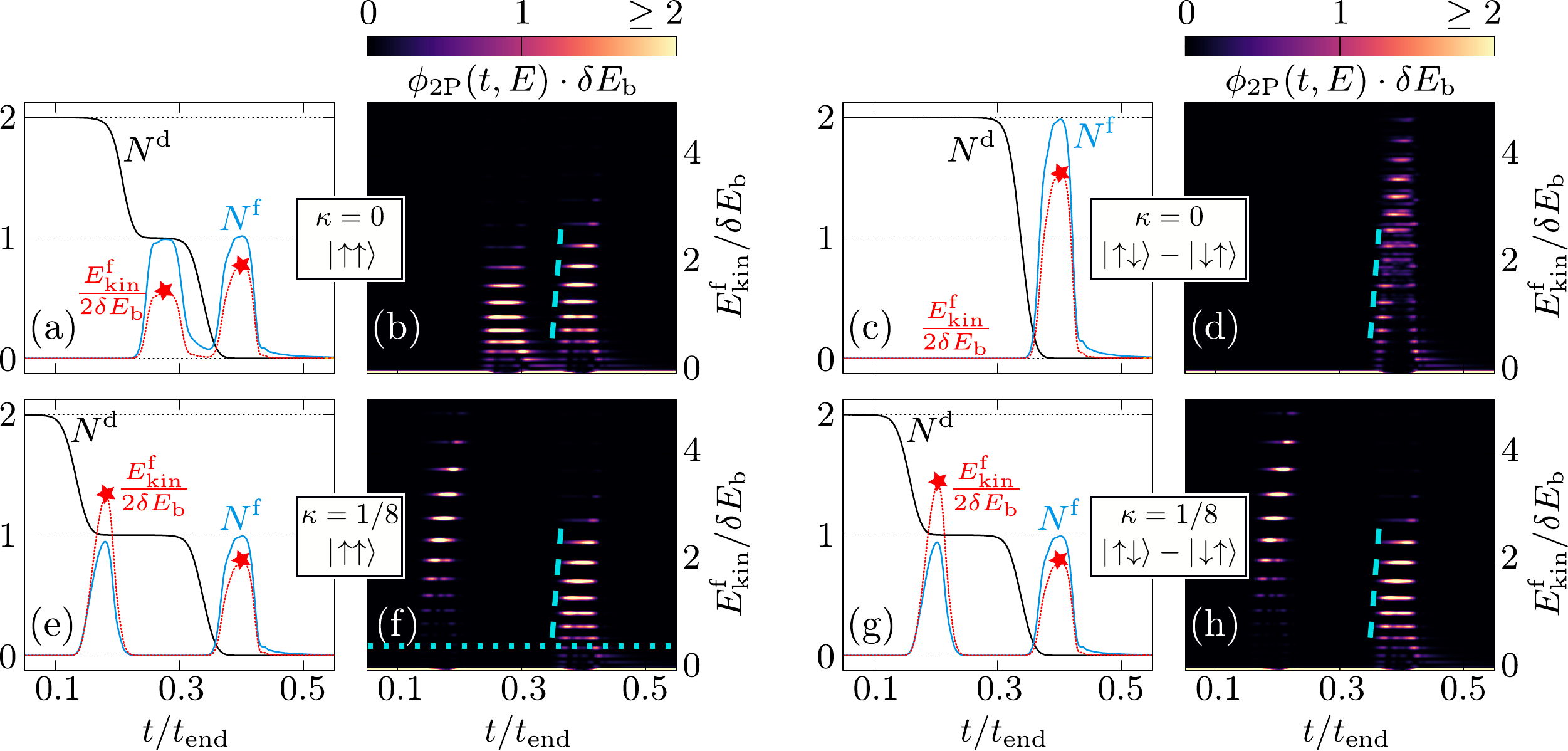}
    \caption{
    Energy- and time dependent emission spectroscopy. Upper row: Vanishing Coulomb interaction, $\kappa=0$, for equal spins, $\uparrow,\uparrow$, in panels (a) and (b) as well as for degenerate spin states $\uparrow, \downarrow$ in panels (e) and (f). Bottom row: Finite Coulomb interaction $\kappa=0.125$, for equal spin states $\uparrow, \uparrow$ in (c) and (d) and for  degenerate spin states $\uparrow, \downarrow$ in (g) and (h). We show the number of electrons in the dot $\Nd$, and the average kinetic energy and particle density in the filter region, $E_\text{kin}^\text{f}$ and $\Nf$ as function of time (left columns, a,c,e,g); the right columns (b,d,f,h) show the time-resolved energy-density, $\phi_{2\text{P}}(t,E)$, where the discrete contributions stem from the discrete energies of the finite filter region. The typical energy scale $\dEb$ is the numerically determined difference between the initial ($t = 0$) exit barrier height and the conductance band potential bottom $V_0$ \BrackFigPart{fig_setting_1d}{a}. 
 The red stars mark the time-energy peaks of the average $E^{\text{f}}_{\text{kin}}$, the turquoise dashed and dotted lines, respectively, indicate the ramp speed $\vV$ and the lowest level with sizable density once the wave package is fully located in the filter region. We set $t_{\text{end}} = 307.2\,\text{ps}$, $\meff = 0.016m_e, L = 12000\,\text{nm}$, $\xdl = 0$, $\xdc = 115\,\text{nm}$, $\xdr = 255\,\text{nm}$, $\Ld = 300\,\text{nm}$, $x_{\text{f,L}} = 1000\,\text{nm}$, $x_{\text{f,R}} = 2350\,\text{nm}$, and a potential landscape $V(x,t)$ \BrackEq{eq_potential_landscape} with $\vV \approx 0.39\,\text{meV/ps}$, $\sdl = 90\,\text{nm}$, $\sdc = 60\,\text{nm}$, $\sdr = 50\,\text{nm}$, $\Vdl = 2.2V_0$, $\Vdc(0) = 0.8V_0$, $\Vdr = 0.006V_0$, $\dEb \approx 0.0057V_0$, where $V_0 = 150\,\text{meV}$. With these parameters, the initial single-particle energy splitting of electrons with equal spin in the lowest energy states in the potential well ((a),(b),(e),(f)) is $4.69 \text{meV}\approx 5.5\dEb$ and the initial charging energy in the interacting cases ((e),(f),(g),(h)) is $\langle H_\text{Coul}\rangle(t=0)=8.35 \text{meV}\approx 9.83\dEb$. \App{sec_simulation_overview} provides all parameters related to the numerical implementation.
    \label{fig_2p}}
\end{figure*}

\section{Two-particle dynamics - Results}\label{sec_twoparticle_results}

We now discuss the emission dynamics from the two-particle simulation described in Sec.~\ref{sec_twoparticle_theory},  analyzing the time-dependent expectation values of the charge $\Nd$ in the dot, of the charge $\Nf$ and kinetic energy $\Ef$ in the filter region, as well as of the energy-and time-resolved filter region spectral density \BrackEq{eq_dos}.
We compare the cases of vanishing $(\kappa=0)$ and finite $(\kappa\neq0)$ Coulomb repulsion strength, both for opposite spins $\uparrow,\downarrow$ initially occupying dot orbitals of equal single-particle energy, and for two $\uparrow$ spins initially residing in different, energy-split dot orbitals due to Pauli exclusion. 

Videos of the time-dependent particle density $|\Psi|^2$ along the 1d potential landscape for the different scenarios can be found in the supplementary material~\cite{SchulenborgZen2023Apr}.

In the following, we focus on the situation where particles are emitted separately in time, such as found in recent experiments~\cite{Ubbelohde2015Jan,Fletcher2013Nov}. This regime furthermore enables to clearly attribute which of the effects described in the following stem from Coulomb interaction, and which are merely related to energy splittings or to time-dependent potential variations. To achieve comparable situations independently of the absence or presence of Coulomb interaction, we choose for the simulation in Fig.~\ref{fig_2p}, a smaller effective mass $\meff$ than the literature values known, e.g., for InAs or GaAs. However, the interaction-related effects identified below remain equally important for larger effective masses $\meff$, see \App{sec_2p_eff_mass}.

\subsection{Effects of single-particle level splitting}\label{sec_twoparticle_Pauli}

We start by analyzing the case of fully screened Coulomb interaction $(\kappa = 0)$ between two spin-$\uparrow$ particles with different orbital energies in the dot before emission, see \FigPart{fig_2p}{a,b}. Due to this energy splitting, the time-dependent $V(x,t)$ modulation causes the dot to emit the first particle clearly separated from the second, as shown by the successive dot particle number reduction from two over one to zero (black line in panel (a)). The charge and energy expectation values $\Nf,\Ef$ (blue and red line) accordingly indicate one particle after another to arrive in the filter region, with a time difference to emission set by the propagation velocity of the coherent wave packet~\cite{SchulenborgZen2023Apr}. Notably, the second emitted particle, once fully inside the filter region $(\Nf \rightarrow 1)$, has an overall higher energy than the first one. We attribute this to the fact that by lifting the Gaussian potential dip in \Eq{eq_potential_landscape} with the prefactor $\Vdc(t) = \Vdc(0) - \vV t$, the potential barrier around $\xdr$ is also slightly raised, as demonstrated in the wave propagation animations~\cite{SchulenborgZen2023Apr}; the second particle emitted at a later time therefore needs to be lifted to a higher energy before it can escape~\cite{Fletcher2013Nov}.

While the left panel visualizes the main results to be presented here, this trend is also confirmed by the behavior of the spectral density. The spectral density $\phi_{2\text{P}}(t,E)$ in \FigPart{fig_2p}{b} is the time-dependent representation of the traveling wave packets in terms of the discrete energy eigenstates of the filter region, i.e., of a one-dimensional 
problem with finite length $x_\text{f,R}-x_\text{f,L}= 1350\text{nm}$; it hence reveals how the particle energies are distributed as a function of time, where the peak at $E = 0$ means that no particle is present in the filter region. The plot exhibits two time intervals with non-zero weight at finite energies, corresponding to the two emitted particles just as in \FigPart{fig_2p}{a}. The discrete values with nonvanishing density stem from the finite size of the 
filter region; their broadening stems from the truncation scheme that we employ here.
Interestingly, we observe a steep yet finite $E-t$-slope of the spectral density, comparable to the potential ramp speed $\vV$ (see the turquoise dashed lines). This reflects the experimentally observed effect of the potential drive increasing the energy of a wave packet during its emission, a so-called energy-time chirp~\cite{Fletcher2019Nov}.

Since the high-energy wave function components also propagate faster through the filter, the spectral density at lower energies remains finite for the longest time duration. In fact, these slowly traveling components of the wave function are also the main reason for the non-vanishing $\Nf$ in between the two peaks in (a), and for an $\Nf$ even slightly larger than $1$ in the second peak. Furthermore note that during periods in which the wave packets are only partly localized in the filter region $(\Nf < 1)$, there is also partial overlap with low-energy filter states that do not exhibit any significant spectral density once the particle is fully inside the filter region. This is highlighted by the horizontal, turquoise-dotted line in \FigPart{fig_2p}{f}, and creates the visual appearance of voids in the spectral density at low energy.

In \FigPart{fig_2p}{c,d}, we contrast the above case of two equal spins subject to Pauli exclusion against the degenerate situation with two opposite spins initially occupying the same dot orbitals. In the absence of Coulomb interaction, $\kappa=0$, this implies equal addition energies for the two particles, and thus simultaneous emission. The filter-region charge and kinetic energy accordingly exhibit only one peak, where the energy is approximately twice as large compared to the second peak of the spin-split case \BrackFigPart{fig_2p}{a,b}. The spectral density in panel~(d) furthermore features much more closely spaced lines once both particles are localized well within the filter region. 
The reason for this is that the spectral density is composed of all possible \emph{pairs} of (not equally spaced) single-particle energies stemming from the contributions of the two emitted particles.

\subsection{Energy transfer via Coulomb interaction}\label{sec_twoparticle_Ufin}

Having covered the non-interacting limit, we proceed with \FigPart{fig_2p}{e-h} showing the corresponding dynamics in the presence of a finite Coulomb interaction between the two particles, $\kappa\neq0$.
Focussing on the relative emission energy of the first and second particle, we see that the first emitted particle carries significantly more energy than the second. This is the opposite outcome to the $\kappa = 0$ case \BrackFigPart{fig_2p}{a-d}, and largely independent of the particle spin orientations, which generally play a less prominent role, c.f., \FigPart{fig_2p}{e-h}.

The weak sensitivity to the particles' spins is intuitively clear since the Coulomb repulsion keeps the particles in the dot apart, such that fermionic anti-bunching becomes less of a factor. 
The main feature we highlight here is, however, the fact that 
the first emitted particle now always has \emph{more} energy than the second. 
We attribute this to the combination of our system being 1d, and a Coulomb-interaction mediated energy exchange between the particles at the first emission event. 
Namely, if both electrons are inside a one-dimensional dot in which they cannot move about each other, they are bound to repel each other into orbital configurations higher in energy both in response to Coulomb or exchange interaction. 
However, once the first electron starts to leave, the two-particle Coulomb interaction allows the remaining electron inside the dot to relax to a lower energy by transferring the energy difference as additional kinetic energy to the escaping electron. This is the key difference between a mere single-particle level splitting related to Pauli exclusion, and the transition energy difference introduced by Coulomb repulsion. In fact, a close comparison between panels (e,f) and (g,h) reveals that the $\Ef$ difference between first and second electron is even slightly higher for opposite spins without Pauli exclusion. This suggests an even more efficient energy exchange, possibly due to the fact that the particles can approach each other even more without fermionic anti-bunching.

These features described above are confirmed by the properties of the spectral density, plotted in panels (f) and (h). The higher-energy contributions of the first emitted particle compared to the second are clearly visible, whereas the second emission strongly resembles the non-interacting, spin-split case, shown in panel (b). The higher kinetic energy of the first emitted particle goes along with a higher propagation velocity, which can be seen in the smaller time-window in which the high-energy states are occupied in the filter region. 

Another Coulomb-repulsion related effect on the first emitted particle is simply the static, $1/r$-dependence of the potential originating from the second particle residing in the dot. This features most prominently when $V(x,t)$ is chosen to be a sharp potential well with a flat bottom shifted up in time. In this case, the first emitted wave packet remains sharper over time and travels faster than the second, since the first particle effectively runs down the $1/r$ potential while the second diffuses out into a flat potential landscape, see box-potential animations in~\Refe{SchulenborgZen2023Apr}. This also highlights the more general fact that the potential landscape needs to fall off into the environment in order to see well confined wave packets\footnote{In the tunneling regime captured by the master equation in \Sec{sec_low_energy_theory} and \Sec{sec_low_energy_results}, this difference is insignificant because tunneling is due to stochastic wave function projections.}.

Additional smaller features occur in the spectral function of particles in the filter region, which are expected to derive from the precise realization of the potential landscape and the resulting complex coherent two-particle wave-packet dynamics. Attributing these features to specific physical mechanisms is hindered by the large available parameter space and the numerical cost of sweeping this space. This includes variations of the ramp speed $\vV$ as an experimentally feasible method to expose nontrivial deviations from the above identified time-energy chirp. In the following, we therefore switch to the complementary quantum master equation description of an effective, two-orbital quantum-dot, in which these effects can be studied more systematically.

\section{Master equation for effective dot model}\label{sec_low_energy_theory}

In addition to highlighting the importance of dot-internal Coulomb scattering, the numerical analysis in \Sec{sec_twoparticle_results} shows more generally that the emission times and energies are dependent on a complex interplay between interaction, interference and the spatial structure of the potential landscape. However, isolating these features within the full two-particle model is tricky and time-consuming due to the many different parameter inter-dependencies, and the analysis would be tied to a 1d setting.  We therefore approximate the emission dynamics with a simpler, fermionic two-orbital dot with local Coulomb interaction and a tunnel-coupled reservoir. Given sufficient spatial confinement for two-particle emission, this model is appropriate for zero- to three dimensional quantum dots. It furthermore allows for a full many-body master-equation study that can to a significant extent be carried out analytically. This analysis yields the average emission energies at a given emission time but does not provide information about the subsequent particle propagation, which is instead accessible via the complementary numerical approach of Secs.~\ref{sec_twoparticle_theory} and \ref{sec_twoparticle_results} and the related movies in Ref.~\cite{SchulenborgZen2023Apr}.

\subsection{Quantum dot coupled to reservoir}\label{sec_pump}

\begin{figure}[t]
    \centering
    \includegraphics[width=\linewidth]{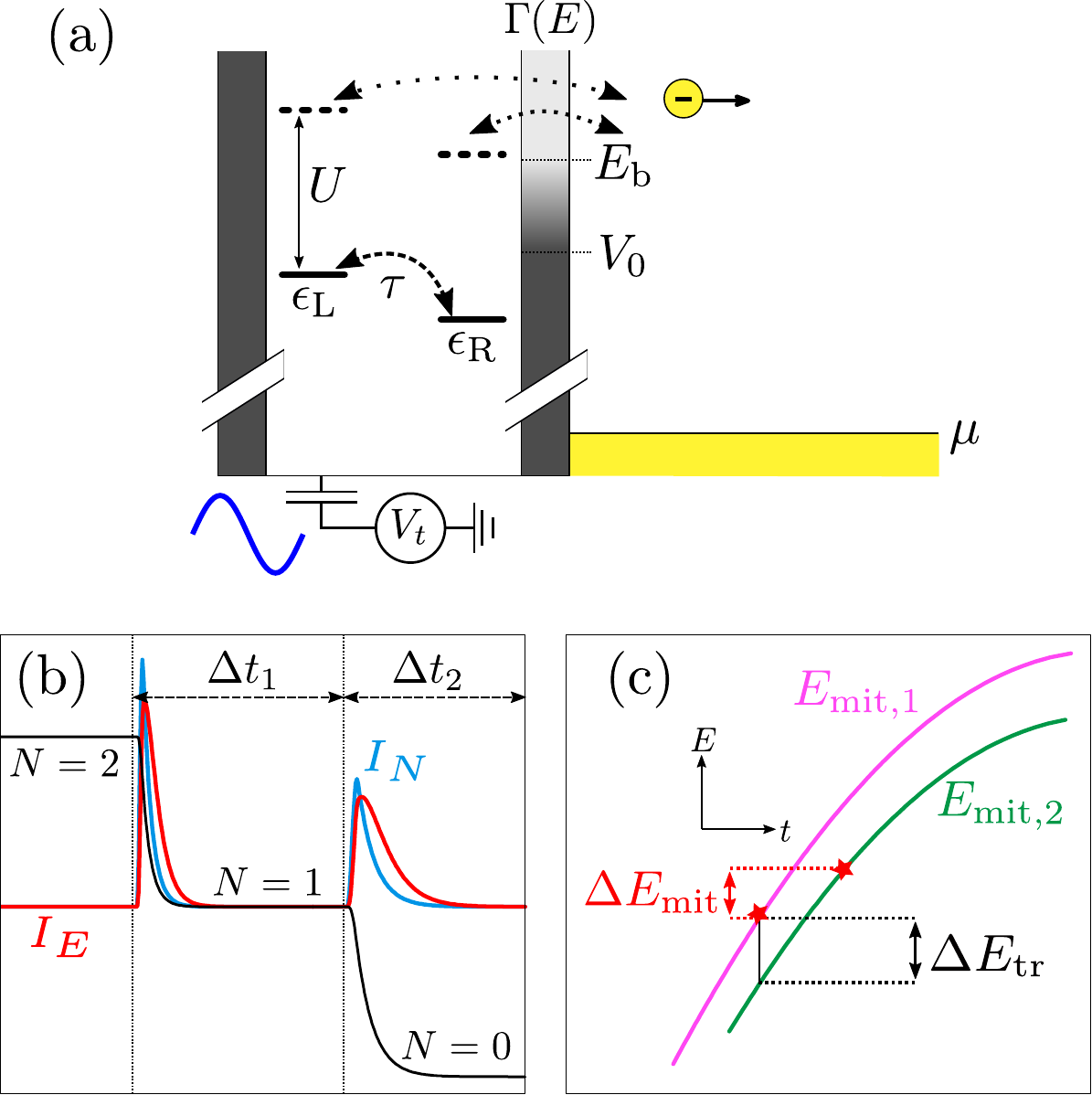}
    \caption{(a)~Sketch of the discrete single-particle energy levels $\epsilon_{\text{L/R}}$, their coupling $\coup$, as well as two-particle interaction $U$ in the quantum dot. Electrons can be emitted from the dot far above the Fermi level $\mu$ due to the time-dependent driving $V_t$ and energy-dependent coupling $\Gamma(E)$ to the reservoir. (b)~Example for the dot occupation number and the emitted charge and energy currents as function of time due to the modulated potential. (c)~Example for emission energies of the two electrons as function of emission time. Stars indicate energy-time pairs of one driving protocol on a line extrapolated by the time-dependent addition energies, see Sec.~\ref{sec_Etr} for definitions of $\Delta\Emit$ and $\Delta\Etr$.}
    \label{fig_setting_0d}
\end{figure}

Our model contains two fermionic single-particle states $\ell = \text{R,L}$ with local Coulomb charging energy in a time-dependently driven potential coupled to a reservoir \BrackFigPart{fig_setting_0d}{a}. In the corresponding Hamiltonian $\Htot = H(t) + \Hres + \Htun(t)$, the dot dynamics are given by
\begin{equation}
 H(t) = \epsL(t)\NL + \epsR(t) \NR  + \coup(\dDagR \dL + \dDagL \dR)  + U(t) \NL \NR.\label{eq_hamiltonian}
\end{equation}
 This includes the driven onsite-energy $\epsR(t) = \epsilon(t) = \epso - A\cosfn{\Omega t}$ with offset $\epso$ and amplitude $A$, a single-particle level splitting $\Delta\epsilon = \epsL(t) - \epsR(t)$, an internal transition amplitude $\coup \geq 0$, and the spatially independent, but possibly time-dependent onsite interaction strength $U(t) > 0$. While the latter cannot describe dot-internal Coulomb scattering as seen for the extended 1d model \BrackSec{sec_two_particle}, 
 the time-dependence does reflect how possible dot size variations during the driving cycle can affect the Coulomb energy, as detailed in \Sec{sec_onsite_interaction}. The occupation number operators $N = \NR + \NL$, $N_\ell = d_\ell^\dagger d_\ell$ are expressed in terms of the dot creation $(d^\dagger_\ell)$ and annihilation $(d_\ell)$ operators for single-particle state $\ell = \text{R,L}$. Note that despite the suggestive notation, we do not yet specify the physical nature of these states: $\ell$ may, at this point, not only refer to two localized levels $\ell=\text{R,L}$, but also to a spin projection $(\ell = \uparrow,\downarrow)$ in a spinful setup, or 
 to two orbitals $(\ell = \text{s,p})$ in a strongly spin-polarized system.

The effectively non-interacting Hamiltonian $\Hres = \sum_{k,\nu}\epsilon_{k\nu}c_{k \nu}^{\dagger} c_{k\nu}$ of the electronic reservoir describes a quasi-continuum of fermionic modes with wave number $k$ and all discrete quantum numbers $\nu$ necessary to characterize these modes. The respective creation and annihilation operators are denoted by $c^\dagger_{k\nu}, c_{k\nu}$. The charge conserving coupling between these modes and the dot state is governed by
\begin{equation}
 \Htun = \sum_{k,\nu}\coup_{k\nu}(t)c_{k\nu}^\dagger \nbrack{\dR + \relTau_{k\nu}\dL} + \mathrm{H.c.}\label{eq_hamiltonian_tunnel}
\end{equation}
The coupling amplitudes $\coup_{k\nu}(t)$ account for time-dependently varying couplings due to the driven potential landscape; the relative amplitude $\relTau_{k\nu}$ models that depending on the setup, one dot mode $\ell$ may couple more strongly to the environment mode $k\nu$ than the other dot mode. Averaged over all environment modes as carefully prescribed in \Sec{sec_low_energy_dynamics}, the coupling amplitudes determine the typical rates
\begin{equation}
 \Gamma_\ell(E,t) = 2\pi\sum_{k,\nu}\delta(E - \epsilon_{k\nu})\left|\nbrack{\delta_{\ell\text{R}} + \relTau_{k\nu}\delta_{\ell \text{L}}}\coup_{k\nu}(t)\right|^2\label{eq_tunneling}.
\end{equation}
for tunneling in or out of dot mode $\ell = \text{R,L}$ with Kronecker deltas $\delta_{\ell\text{R}},\delta_{\ell\text{L}}$. Importantly, 
the energy-dependent barrier transparency $\Gamma_\ell(E)$, beyond  
the common wideband limit~\cite{Jauho1994Aug}, should capture that the potential $V(x,t)$ \BrackEq{eq_potential_landscape} induces emission once $\Vdc(t)$ is small enough for a dot transition energy $E$ to exceed the potential energy outside the dot. We achieve this with a barrier height $V_0 > \mu$ above which $\Gamma_\ell(E)$ increases smoothly from an uncoupled dot with $\Gamma_\ell(\mu \leq E \leq V_0) = 0$ to a weakly coupled dot with $\Gamma_\ell(E \geq \Eb) = \Gamma_\ell$:
\begin{equation}
 \Gamma_\ell(E \geq \mu,t) = \Gamma_\ell S\left(E,V_0,\Eb\right).
 \label{eq_barrier_transparency}
\end{equation}
The modified sigmoid 
\begin{equation}
    S(E,x,y) = \begin{cases} 0 & E \leq x \\ \sqbrack{1+\exp\left(\frac{y-x}{E-x}+\frac{y-x}{E-y}\right)}^{-1} & x < E < y \\ 1 &  E\geq y \end{cases}\label{eq_sigmoid}
\end{equation} 
implements the ramp-up from exactly $0$ to $\Gamma_\ell$ using an analytic function in the finite interval\footnote{unlike a regular sigmoid, which is only exponentially close to its limits within any finite interval.} $[V_0,\Eb]$. The explicit time-dependence of $\Gamma_\ell(E,t)$ can arise from a possibly time-dependent exit barrier height, $V_0 \rightarrow V_0(t) = V_0(\epsilon(t))$ and $E_\mathrm{b} \rightarrow E_\mathrm{b}(t) = E_\mathrm{b}(\epsilon(t))$. 
We address this further below, see Sec.~\ref{sec_deltaEb}, to analyze the effect of dot potential modulations on the tunnel barrier, which we have already identified for the unitary two-particle evolution above as the cause of the different energy peak heights in \FigPart{fig_2p}{a}.

\subsection{Dot state dynamics}
\label{sec_low_energy_dynamics}

We describe the time-dependent dot state with a master equation for the reduced density matrix $\rho(t)$, i.e., with the reservoir modes traced out as detailed, e.g., in \cite{Breuer2007Jan,Koenig1999}. 
Indeed, the emission of electrons far above the Fermi energy of the conductors is equivalent to a situation with infinitely large bias, for which a weak-coupling master equation approach is applicable~\cite{Gurvitz1996Jun,Gurvitz1997Dec,Oguri2013Oct}.
The broadening in the currents calculated in this section hence stem from the stochastic nature of the tunneling.

With the above in mind, we use the approach of Ref.~\cite{Nathan2020Sep} to derive a Lindblad master equation for the dot tunnel-coupled to the bath. Care needs to be taken since the tunneling \eq{eq_hamiltonian_tunnel} couples a single environment mode $k\nu$ to both dot states $\ell = \mathrm{L,R}$ for $\relTau_{k\nu} \neq 0$. This introduces up to four independent decay channels per particle- and hole process
corresponding to the two possible orthogonal hole-like operators $\alpha \dR + \beta \dL$ and the analogous particle-like fields. As we have not yet tied $\ell$ to any specific single-particle basis, and since the shape of $H(t)$ is invariant under any unitary transform (after properly regauging the fields to ensure $\coup \geq 0$ and by removing any constant energy shift) of the single-particle basis \BrackEq{eq_hamiltonian}, we demand that each of the maximally $4$ orthogonal channels in $\Htun$ are proportional to only one of the four dot operators $d_\ell,d^\dagger_\ell$. The corresponding master equation for the density operator $\rho(t)$ then reads
\begin{equation}
 \partial_t\rho = -i[H + \Lambda,\rho] + \sum_{\eta\ell}\sqbrack{L_{\eta\ell}\rho L^\dagger_{\eta\ell} - \frac{1}{2}\cbrack{L^\dagger_{\eta\ell}L_{\eta\ell},\rho}}\label{eq_master}
\end{equation}
with Lamb shift $\Lambda$ \BrackEq{eq_lamb} and Lindblad operators
\begin{equation}
 L_{\eta\ell} = \sum_{i,j}\sqrt{\Gamma_{\ell}(\eta E_{ij},t)f^{\eta}(\eta E_{ij})}\bra{i}d_{\eta\ell}\ket{j}\times \ket{i}\bra{j}.\label{eq_lindblad}
\end{equation}
The states $\ket{i},\ket{j}$ are the four instantaneous many-body energy eigenstates of the dot Hamiltonian $H(t)$, and $E_{ij} = E_i - E_j$ the corresponding energy differences, where our notation suppresses their time-dependence for better readability. The operator $d_{\eta\ell} = \delta_{\eta +}d^\dagger_\ell + \delta_{\eta -}d_\ell$ combines the corresponding dot creation and annihilation operators, and $f^\eta(x) = \sqbrack{\mathrm{exp}\left(\eta\frac{x - \mu}{T}\right) + 1}^{-1}$ is the Fermi function $f$ for $\eta=+$, respectively $1-f$ for $\eta=-$. The master equation \eq{eq_master} relies on the instantaneous-time approximation, in which the Lindblad operators \eq{eq_lindblad} only depend on the parameters at the current emission time $t$~\cite{Nathan2020Sep,Splettstoesser2006Aug,Reckermann2010Jun}. 
We expect that this leads to reliable results for driving that is limited by  
$A\Omega \ll (E_b-V_0)^2$~\cite{Riwar2014,Oguri2013Oct}. For fast driving of tunable-barrier dots, see also Ref.~\cite{Kashcheyevs2012Nov,Ubbelohde2015Jan,Kashcheyevs2017Jun}.

As discussed above, we represent the hot electron setting by taking the $T\rightarrow 0$ limit of the Fermi function, i.e., $f_{\eta}(x) \rightarrow \Thetafn{\eta\sqbrack{\mu - x}}$ with Heaviside function $\Thetafn{x}$. Moreover, in the lowest-order coupling approximation assumed here, the environment-induced Lamb shift $\Lambda$ only modifies the unitary dynamics in the single-particle sector of the local Hamiltonian $H$. The splitting $\Deps$ and coupling $\tau$ are hence generally shifted for coherent L-R rotations, but the jump rates \eq{eq_lindblad} are only affected by the \emph{bare} energies.
Also, our discussion focuses 
on a tunnel coupling $\Gamma/U = 0.001$ much smaller than the range $|\Eb - V_0| \sim 0.05U$ on which the barrier changes its opacity \BrackEq{eq_barrier_transparency}, so that $|\Lambda| \sim \Gamma \ll |\Eb - V_0|$. Thus, while we do account for the Lamb shift $\Lambda$ in solving \Eq{eq_master}, it is in fact irrelevant for the emission dynamics in all cases considered in this paper, i.e., both for $\Deps = \tau = 0$ and for $|\Delta\epsilon|,|\tau| \gg \Gamma$. 
More detailed derivations of the master equation 
are given in \App{sec_master_detail}.

\subsection{Time-resolved emission energy}

The observables of interest for our master-equation based analysis are the ensemble-averaged time-dependent charge current and energy current carried by particle transfer into the environment. Local charge conservation $[H(t),N] = 0$ and charge conserving tunneling implies that the particle current into the environment equals the particle current out of the dot. Furthermore, in the weak coupling limit, no energy is stored in the barriers~\cite{SchulenborgLic}, $\langle\Htun\rangle = 0$, so that the energy flow into the reservoir is the flow out of the dot minus the energy exchange with the work source. We hence write
\begin{equation}
 \IN(t) = -\Tr[N\partial_t{\rho}] \lrsepa \IE(t) = -\Tr[H(W^{I,E}\times\rho)]. \label{eq_currents}
\end{equation}
In the regime studied here, the energy flow is concomitant with net particle transfer; hence the kernel $W^{I,E}$ involves only those state transitions of the master equation \eq{eq_master} with changing dot occupation, as detailed in Appendix~\ref{sec_current_kernel}.

We start the dynamics \eq{eq_master} from double occupation, \mbox{$\bra{\text{d}}\rho(t = 0)\ket{\text{d}} = 1$} with \mbox{$\ket{\text{d}} = \dDagR\dDagL\ket{0}$} created from the dot vacuum $\ket{0}$, and an initial dot potential \mbox{$\epsilon(t = 0) = \epso - A$}, sufficiently below the energy where the exit barrier becomes opaque to keep the electrons stable in the dot, \mbox{$2(\epso - A)+U\ll V_0$}. 
The driven transition energies $E_{ij}$ eventually reach the opaque-barrier interval $\Gamma_\ell(E > V_0) > 0$ causing a particle to be emitted.
The resulting particle current $\IN(t)$ is a measure for the emission probability at time $t$, and the ratio \mbox{$\Emit(t) = \IE(t)/\IN(t)$} is associated to the energy emitted per particle.
We combine the above quantities to define the time- and energy-resolved emission distribution
\begin{equation}
 \phi(t,E) = \IN(t)\delta(E - \Emit(t))\lrsepa \Emit(t) = \frac{\IE(t)}{\IN(t)}.\label{eq_dist_approx}
\end{equation}
Unlike experimental data in, e.g.,~\cite{Fletcher2013Nov,Fletcher2019Nov,Fletcher2022Oct} or the two-particle spectral density \eq{eq_dos}, this distribution is sharp in $E$ because the energy $\Emit(t)$ per emitted particle is already defined in terms of both classical and quantum averaging. 
The observed energy spread in experiments~\cite{Waldie2015Sep,Fletcher2019Nov,Fletcher2022Oct} is considered to stem to a large extent from the noise and timing jitter of pump and detector drive signals, obscuring the quantum uncertainty. The reconstructed Wigner distribution \cite{Fletcher2019Nov} is likely of mixed states, in other words, a classical ensemble of different emitted states, rather than of a pure state.  
Our simplified, low-energy model in \Eq{eq_dist_approx} captures how an uncertainty in emission time results in a classical uncertainty in emission energy; quantum mechanical energy smearing is hence not at the focus of the present paper, but could instead potentially
be obtained from the full two-particle approach 
previously described in Sec.~\ref{sec_twoparticle_theory}.

We consider a central question to be how the two emitted particles differ in the distribution $\phi(t,E)$ ---and in particular in their average emission times and energies--- as a function of the system- and driving-parameters, with special focus on the potential ramp speed $\partial_t\epsilon(t)$. Assuming two particles released within two separated time intervals, as sketched in \FigPart{fig_setting_0d}{b}, with the first transition from double to single occupied dot and the second from single occupied to empty dot indicated by $x \in\cbrack{1,2}$, we obtain these averages by integrating over $\phi(t,E)$:
\begin{align}
 t_x &= \int_{\Delta t_x}\nsepa dt\int\!\!dE\, \sqbrack{t\times\phi(t,E)} = \int_{\Delta t_x}\nsepa dt\,\sqbrack{t\times\IN(t)} \notag\\
 \Emitx &= \int_{\Delta t_x}\nsepa dt\int\!\! dE\, \sqbrack{E\times \phi(t,E)} = \int_{\Delta t_x}\nsepa dt\, \IE(t)\label{eq_averages}.
\end{align}
Sweeps of the offset potential $\epso$, within a range fulfilling $V_0 \leq \epso + U + A$, modify the potential ramp speed $\partial_t\epsilon(t)$ in the time in which the addition energies are in the opaque-barrier interval. This modifies the emission energies and emission times $t_{x}$ (as well as the ramp speed at the time of emission). We sketch in \FigPart{fig_setting_0d}{c} a parametric time-energy plot of $(t_x,\Emitx - \epso)$ for the two-particle emissions, indicated by stars for one specific choice of $\epso$. When sweeping $\epso$, we expect the emission energies to lie on the indicated lines, extrapolating with the cosine-shape of the addition energies, time-dependent via $\epsilon(t)$. 

To connect the obtained trajectories to available experimental data~\cite{Fletcher2013Nov,Ubbelohde2015Jan,Jonunpublished}, we compare the two emission curves in two different ways. First, we fix the offset $\epso$, meaning we compare the emission times and energies of the first and second particle for \emph{the same driving protocol}:
\begin{equation}
\Delta\Emit(\epso) = \Emitzo(\epso) - \Emitot(\epso) \label{eq_emission_energy_diff}.
\end{equation}
Compare to the horizontal distance betwen the stars shown as an example in \FigPart{fig_setting_0d}{c}. If a particle with energy $E$ was instantaneously emitted as soon as $E > V_0$ with constant $V_0$, we would not expect any difference between first and second particle, $\Delta\Emit \rightarrow 0$. We, however, show in \Sec{sec_low_energy_results} that $\Delta\Emit \neq 0$ arises due to the finite emission time $\sim1/\Gamma_\ell$ competing with the driving parameters $\Omega,\epso$ determining the $\epsilon(t)$-ramp speed at emission, and with dot-internal transitions $\sim1/\tau$.

Second, we compare \emph{different driving protocols with different} $\epso,\epso' > \epso$, black vertical line in \FigPart{fig_setting_0d}{c}, to obtain the \emph{apparent transition energy difference} between first and second particle:
\begin{equation}
\Delta\Etr(t) = \Emit(\totwo[\epso] = t) - \Emit(\tzo[\epso'] = t)\label{eq_apparent_transition_diff}.
\end{equation}
For well separated emission events, we intuitively expect $\Delta\Etr$ to be given by the difference between the largest possible double-to-single transition energy and the largest available single-to-zero transition energy after the first emission event. Based on the dot Hamiltonian \Eq{eq_hamiltonian}, this difference is given by the sum of interaction strength and splitting between the two single-particle states, $\Delta\Etr \approx \DEtrref = U + 2\sqrt{|\tau|^2 + \Deps^2/4}$, see \Eq{eq_energies}. While we indeed find this to be mostly the case, there are, however, deviations from $\Delta\Etr \approx \DEtrref$ that we further explore in \Sec{sec_Etr}.

\section{Time-resolved emission spectroscopy --- results}\label{sec_low_energy_results}

To discuss the time-resolved emission spectroscopy obtained from the master equation, we first analyze the difference in emission energy of the two particles $\Delta\Emit$ and in particular its dependence on the ramp speed $\vramp = A\Omega\sin(\Omega t_1)$, as shown in \Fig{fig_en_diffs_time_independent}. 
To stay in line with previous experiments~\cite{Fletcher2013Nov,Fletcher2022Oct}, we tune $\vramp$ by sweeping the offset $\epsilon_0$ determining at which phase, and thus at which slope of the cosine-shaped $\epsilon(t)$ the particles are emitted. This entails two features visible in all results shown in the following subsections: first, all graphs begin at a finite $\vramp$, determined by the minimum offset requirement $\epso + A \geq V_0$ for the driving cycle with amplitude $A$ to emit both particles from the dot. Second, the $\DEmit$-lines bend downwards when approaching this minimum ramp speed from above. This stems from the stronger curvature of the cosine-shaped driving potential around the turning point $\epsilon(t) \approx \epsilon_0 + A \approx V_0$; it causes the two ramp speeds to differ significantly, $\vramp\gg v_{\text{r},2}=  A\Omega\sin(\Omega t_2)$, and thus the first particle to be emitted at a higher energy, $\Delta\Emit < 0$.

In the following we provide a detailed analysis of how emission times and energies are impacted by local Coulomb interaction effects as compared to effetcs due to the potential landscape creating the dot (here visible as energy- and dot-state-dependent couplings). We address the impact that quantum-state degeneracies have enabled by Coulomb interaction, Sec.~\ref{sec_onsite_interaction}, the effect of coupling asymmetry and internal dynamics (as they could realistically arise also from local Coulomb repulsion, see Sec.~\ref{sec_twoparticle_results}), Sec.~\ref{sec_asymmetry}, compare to the effect that results from driving-dependent tunnel couplings, Sec.~\ref{sec_deltaEb}, and show the effect of a time-varying local Coulomb interaction, Sec.~\ref{sec_Etr}.

\subsection{Interplay between interaction and degeneracy}\label{sec_onsite_interaction}

\begin{figure}[t]
    \centering
    \includegraphics[width=\linewidth]{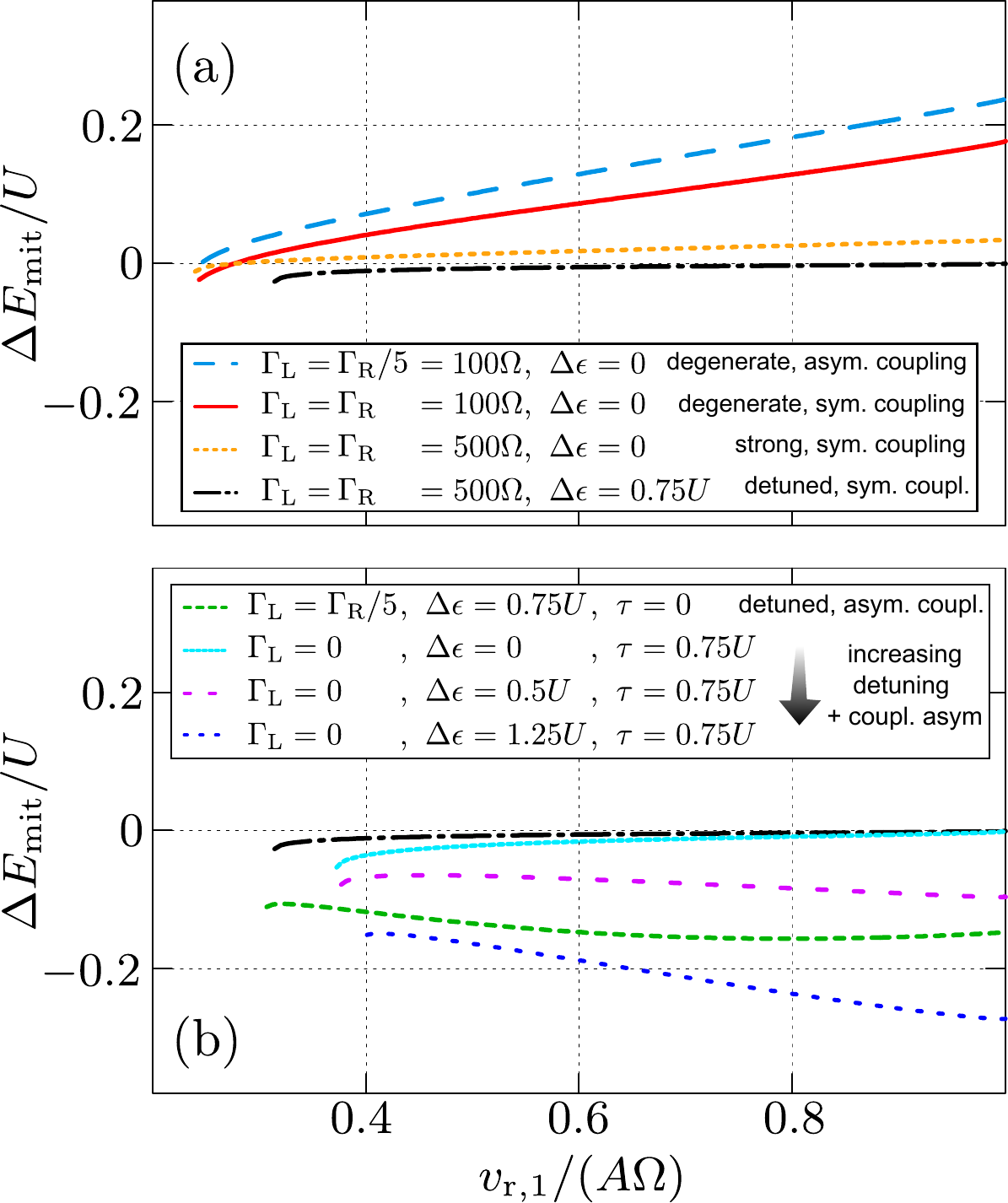}
    \caption{
    Emission energy differences $\Delta\Emit$ \BrackEq{eq_emission_energy_diff} as function of the ramp speed $\vramp$. (a) shows the case of zero level-coupling $\tau=0$. In (b) we set for all lines $\GamR=500\Omega$. The black dashed-dotted 
    lines in panels (a,b) are identical for reference.
    The time-independent parameters in both panels are $U = 1000\Gamma$, $A = 34U$, $V_0 = 35U$, $\Eb = 35.05U$. \App{sec_master_detail} states all parameters relevant for the numerical implementation.}
    \label{fig_en_diffs_time_independent}
\end{figure}

We start by studying the influence of degenerate orbitals and of the coupling asymmetry, see results presented in \FigPart{fig_en_diffs_time_independent}{a}. We, therefore, first set $\Delta\epsilon=\tau=0$ in the Hamiltonian \eq{eq_hamiltonian}. 
In this case of degenerate single-particle levels, as is commonly the case for spin-degeneracies for example, the first emitted electron can be either the one occupying level R or level L, leading to an increased rate $\GamR+\GamL$ for the emission of the first particle~\cite{Schulenborg2016Feb}. Since the second particle is emitted more slowly than the first, it is also emitted at a higher energy due to the continuous lifting of the energy level by the driving potential. 
This effect of the level degeneracy can be seen in Fig.~\ref{fig_en_diffs_time_independent}(a), where $\Delta\Emit$ is shown for 3 different situations  with zero detuning, $\Delta\epsilon=0$. The red and the orange-dashed lines show this degeneracy-induced effect, demonstrating a $\Delta\Emit$ that increases with increasing ramp speed $\vramp$. This effect is stronger for the red line, where the equal rates $\GamR=\GamL$ are smaller than for the orange line. Therefore the time that passes until the emission of the second particle is larger such that the level has been shifted to a higher energy. 
The blue-dashed line shows that this degeneracy effect indeed yields similar results to having one of the levels being coupled more strongly than the other. 
In contrast, as soon as the dot-orbital degeneracy of two equally coupled levels is lifted, $\Delta\epsilon\neq0$, the degeneracy-induced delay of the second emission vanishes, as shown by the black dashed-dotted line that is close to zero and almost independent of the driving speed.

This effect due to the degeneracy of single-particle energy levels can only be observed in the presence of strong Coulomb interaction, separating the emission of the two particles from the dot.
Note that the impact of degeneracy on decay rates~\cite{Bonet2002Jan} of a quantum dot has been observed in various experimental settings~\cite{Hofmann2016Nov,Hartman2018Nov,Kleeorin2019Dec} and further schemes have been theoretically proposed to read out such differences in decay rates~\cite{Schulenborg2014May,Riwar2016Jun}.

\subsection{Coupling asymmetry and internal dynamics}\label{sec_asymmetry}

One possible factor governing the emission spectrum in the 1D two-particle dynamics discussed in \Sec{sec_twoparticle_results} is the dot-internal repulsion due to Pauli exclusion and Coulomb interaction, pushing one particle further away from the tunnel barrier than the other. The resulting asymmetric coupling to the environment can be reflected within the master equation approach by setting $\GamL \neq \GamR$. Calculations which illustrate the result of an (effective) coupling asymmetry are shown in Fig.~\ref{fig_en_diffs_time_independent}(b).

We have previously seen in \FigPart{fig_en_diffs_time_independent}{a} that $\Delta\Emit$ grows with increasing $\vramp$ for degenerate levels $(\Deps = 0)$ or if the higher-lying state was at least equally if not stronger coupled to the environment. If we instead assume a much more weakly coupled high-lying state ($\Deps > 0, \GamL\ll\GamR$), the slower tunneling rate in the first emission event leads to a significantly higher emission energy than for the second electron emitted from $\epsR(t)$ with rate $\GamR$. As long as the two emission events remain clearly separated in time, this leads to exactly the opposite situation with a negative $\DEmit < 0$ becoming more negative with growing $\vramp$, as shown by the green-dashed line in Fig.~\ref{fig_en_diffs_time_independent}(b). Note however, that if $\vramp$ is increased even further, approaching or exceeding $\GamL$, the lower addition energy $\epsR+U$ may cross $V_0$ and induce emission from state L before the higher-lying particle from state R could leave the dot. This can result in almost simultaneous emission and a $\Delta\Emit$ approaching positive values again, as indicated by the positively curving green-dashed line in \FigPart{fig_en_diffs_time_independent}{b} for $\vramp/(A\Omega) \rightarrow 1$.

A relevant situation where such an asymmetric tunnel coupling, and hence a negative $\DEmit(\vramp)$-slope can be realized is when two localized orbitals can be occupied in the dot, and tunneling to the environment from one of them  can occur only through coherent coupling to the other. Concretely, we realize this by setting $\GamL=0$ and turning on a finite $\tau > 0$ in the Hamiltonian \eq{eq_hamiltonian}. The ratio $|\Deps/\tau|$ then determines to what degree the single-particle eigenstates of the dot are localized in orbitals L and R, interpolating between perfect (anti-)bonding states $(|\Deps/\tau| \rightarrow 0)$ and perfectly localized states $(|\Deps/\tau| \rightarrow \infty)$. With state $2$ decoupled from the bath $(\GamL)$, a splitting $\Deps = 0$ results in equally coupled bonding- and anti-bonding states, whereas a positive splitting $\Deps > 0$ yields a higher-lying state with weaker effective environment coupling, see Appendix~\ref{sec_master_detail}. Figure~\ref{fig_en_diffs_time_independent}(b) accordingly shows a very weak $\vramp$-dependence of $\DEmit$ for $\Deps = 0$ (turquoise solid line) due to the nearly symmetric environment coupling, and a clearly negative $\DEmit(\vramp)$-slope due to the strong coupling-asymmetry for finite splitting $\Deps \sim \tau > 0$  (violet-dashed and blue-dotted line). In particular, the effect gets larger with larger splittings $\Delta\epsilon$, localizing the single-particle eigenstates more strongly into orbital L and R.

Finally, it is interesting to compare the above described effect leading to higher emission energy of the first emitted particle $(\DEmit < 0)$ in the weak tunneling regime to the mechanism causing $\DEmit < 0$ in the coherent two-particle simulation \BrackFig{fig_2p}. In the latter case, the first emitted particle attains additional energy because energy is transferred from the particle remaining in the dot due to a dot-internal rearrangement process. This is physically distinct from the above described effect of asymmetric coupling, which can exist even in the absence of Coulomb interaction. 

\subsection{Impact of time-modulated barrier potential}\label{sec_deltaEb}

\begin{figure}[t]
    \centering
    \includegraphics[width=\linewidth]{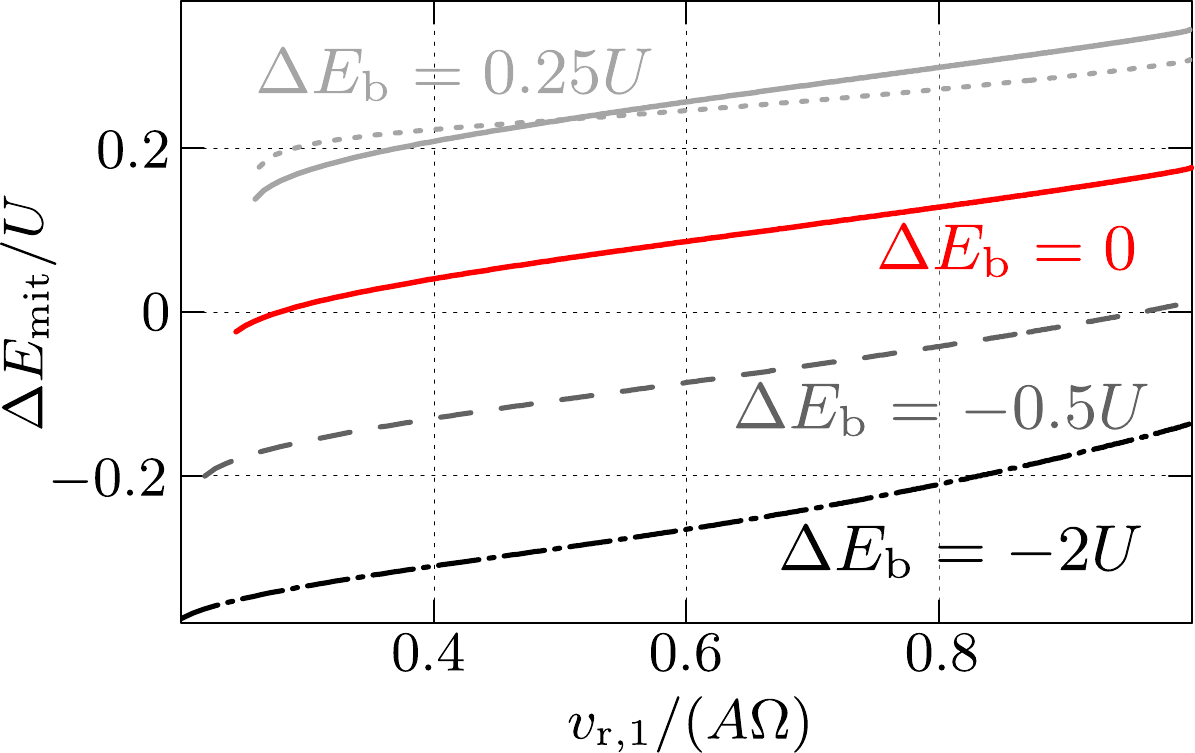}
    \caption{Emission energy differences $\Delta\Emit$ \BrackEq{eq_emission_energy_diff} as function of the ramp speed $\vramp$ for different potential-dependent barrier height shifts $\Delta\Eb$ \BrackEq{eq_potential_dependent_coupling}. We fix $\GamR = \GamL = 100\Omega, \epsilon_V = V_0 - U = 34U$, $\tau = \Deps = 0$. For the light grey dashed curve $(\Delta\Eb = 0.25U)$, we set $\epsb-\epsilon_V = 0.25U$, whereas all other graphs are taken at $\epsb-\epsilon_V = 2U$. The remaining parameters equal those stated in the caption of \Fig{fig_en_diffs_time_independent}.}
    \label{fig_en_diffs_barrier_potential_dependent}
\end{figure}

Until here, we have kept the coupling $\Gamma_\ell$ constant in time, while only the energy levels were modulated. However, a major cause of finite $\DEmit$ between the two emitted particles, which we have already identified from the coherent two-particle simulation, Sec.~\ref{sec_twoparticle_results}, can be the effect of the dot potential modulation on the tunnel barrier. We can systematically analyze this within the master equation framework by letting the bare coupling strengths $\Gamma_\ell(E,t)$ explicitly depend on the time-dependent potential $\epsilon(t)$. Our approach is to model a situation where the energy window in which the coupling is modulated from zero to $\Gamma_\ell$, is shifted upwards during the driving by a total amount $\Delta\Eb$. We therefore add to $V_0$ and $\Eb$ in \Eq{eq_barrier_transparency} the expression
\begin{equation}
\Delta\Eb S(\epsilon(t),\epsilon_V,\epsb),\label{eq_potential_dependent_coupling}
\end{equation}
interpolating with the smooth sigmoid-like function \eq{eq_sigmoid} between an initial $(V_0,\Eb)$ and final $(V_0+\Delta\Eb,\Eb + \Delta\Eb)$ within the interval $[\epsilon_V,\epsb]$ traversed by $\epsilon(t)$. 

The influence of the varying exit barrier \eq{eq_potential_dependent_coupling} on the ramp-speed dependent emission-energy difference $\DEmit$ is illustrated in \Fig{fig_en_diffs_barrier_potential_dependent}. 
We find that if the total barrier-height shift $\Delta\Eb$ takes place well between the two emission events on the scale of the typical tunneling time,  the barrier shift expectedly leads to a shift of $\DEmit(\vramp)$ in equal direction, $\text{sgn}\left(\Delta\Eb\right) = \text{sgn}\left(\DEmit\right)$, as illustrated by the light grey solid, 
red and dark grey dashed lines in \Fig{fig_en_diffs_barrier_potential_dependent}. 
The precise magnitude $|\DEmit|$, however, depends on how fast the barrier height is ramped compared to both the emission time $~1/\Gamma_\ell$ and the dot potential ramp time $~(E_\mathrm{b}-V_0)/\vramp$ itself. This becomes even more relevant if the barrier height is ramped instead within a dot potential interval $[\epsilon_V,\epsb]$ close to, or even containing a particle emission event. In this case, comparable time and energy scales $\dot{\epsilon}(t) \sim \vramp \sim \frac{\Delta\Eb}{\epsilon_V-\epsb}\vramp$ can also change the slope of $\DEmit(\vramp)$ compared to the case of constant $V_0,\Eb$. For example, the light grey dashed line in \Fig{fig_en_diffs_barrier_potential_dependent} for $\Delta\Eb = \epsilon_V-\epsb = 0.25U$ and $\epsilon_V = V_0 - U$ corresponds to a case in which the onset of the opaque region of the barrier $V_0$ is essentially raised together with the addition energy $\epsilon(t) + U$ of the first particle once the latter has reached $V_0$. The imminent particle emission is thereby continuously deferred for $\epsilon_V < \epsilon(t) < \epsb$, and $\vramp$ affects $\DEmit$ more strongly nonlinearly, via the combination of $\epsilon(t)$ itself and $V_0(\epsilon(t)),\Eb(\epsilon(t))$. Apart from the opposite sign of barrier height shift, a similar nonlinearity is also seen in the dashed-dotted black line in \Fig{fig_en_diffs_barrier_potential_dependent} for $\Delta\Eb = -2U$ and $\epsilon_V-\epsb = 2U$.

\subsection{Apparent transition energy changes due to time-dependent charging energy}\label{sec_Etr}

\begin{figure}[t]
    \centering
    \includegraphics[width=\linewidth]{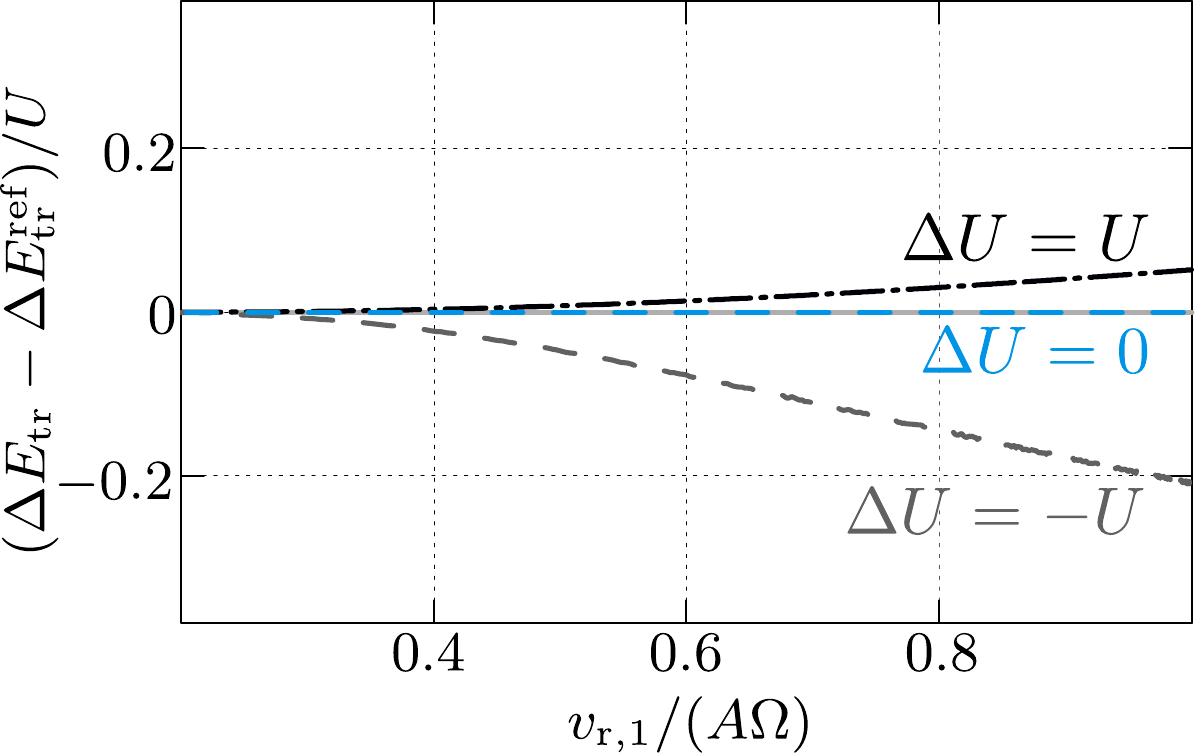}
    \caption{Apparent transition energy difference $\Delta\Etr$ \BrackEq{eq_apparent_transition_diff} as function of the ramp speed $\vramp$ for a potential-dependent interaction strength $U(\epsilon(t))$ as given in \Eq{eq_potential_dependent_interaction}. We fix $\GamR = \GamL = 500\Omega, \epsU = \Eb - U + 0.01U = 34.01U$, $\depsU = 0.29U$. For the light grey solid line underneath the blue dashed line, we set $\Deps = 0.5U, \tau = 0.75U$, whereas $\Deps = \tau = 0$ in all other graphs. The remaining parameters equal those stated in the caption of \Fig{fig_en_diffs_time_independent}. }
\label{fig_etr}
\end{figure}

We finish our analysis by studying how the apparent equal-time transition energy difference $\Delta\Etr$ between first and second emitted particle \BrackEq{eq_apparent_transition_diff} is affected by the dot parameters. The effects analyzed up to here are found to not significantly affect $\Delta\Etr$. This changes when the time-dependent dot potential $\epsilon(t)$ also influences the charging energy $U$, via the changing spatial confinement in the dot's potential landscape. A similar effect could in practice also arise via a modulated single-particle splitting $\Deps$, but we here only consider $U$ for simplicity and concreteness. We model this in analogy to \Eq{eq_potential_dependent_coupling} with the smooth transition
\begin{equation}
U(\epsilon) = U + \Delta U S(\epsilon(t),\epsU,\epsU + \depsU),\label{eq_potential_dependent_interaction}
\end{equation}
where $S(\epsilon,\epsU,\epsU+\depsU)$ \eq{eq_sigmoid} establishes a smooth transition between the initial $(U)$ and final $(U + \Delta U)$ interaction strength within the potential interval $[\epsU,\epsU + \depsU]$. 

In \Fig{fig_etr}, we compare $\Delta\Etr(\vramp)$ with constant $U$ to cases in which the interaction strength either disappears $(\Delta U = -U)$ or doubles when $\epsilon$ is ramped up through $[\epsU,\epsU + \depsU]$. First note that the expected reference energy difference $\Delta E^{\text{ref}}_{\text{tr}} = U + 2\sqrt{|\tau|^2 + \Deps^2/4}$ includes both the interaction strength $U$ for $\epsilon < \epsU$ and the single-particle orbital splitting due to $\Deps,\tau$. The light-grey solid and blue dashed lines in \Fig{fig_etr} show results for constant $U$, 
where we choose $\Delta\epsilon=\tau=0$ (blue) and $\Delta\epsilon=0.5U$, $\tau=0.75U$. These lines clearly show that
the apparent transition energy difference closely approaches this expected difference $\Delta\Etr \rightarrow \Delta E^{\text{ref}}_{\text{tr}}$.

Instead, a finite $\Delta U \neq 0$, quantifying the time-dependent shift of the interaction energy, can affect $\Delta\Etr$ if the interaction strength changes prior to or during the first emission event: A $U$-shift prior to emission only leads to a constant, $\vramp$-independent shift to $\Delta\Etr \rightarrow \Delta E^{\text{ref}}_{\text{tr}} + \Delta U$, as intuitively expected. However, if the interaction strength changes closely to the likely emission time on the scale of the typical emission rate, the effect depends on how fast $\epsilon$, and hence $U(\epsilon)$ are shifted relatively to this emission rate. The resulting $\vramp$-dependence in this case ---with $\epsU + U \approx V_0$ purposely coinciding with the addition energy of the first particle reaching $V_0$--- is illustrated by the dark-grey dashed/black dashed-dotted line in \Fig{fig_etr}. Note that we consider not only decreasing $(\Delta U = - U)$, but also increasing interaction strength $(\Delta U = +U)$: while $\Delta U < 0$ reflects the physically more intuitive scenario in which a rising $\epsilon$ reduces the spatial confinement, one could also imagine a special potential shape in which this confinement tightens at least within a limited $\epsilon$ interval, meaning $\Delta U > 0$. As apparent from the concrete example $\Delta U = +U$ in \Fig{fig_etr}, $\Delta\Etr - \Delta E^{\text{ref}}_{\text{tr}}$ is then accordingly shifted in positive direction, albeit with smaller magnitude than the negative bending for $\Delta U = -U$. This different magnitude occurs because a shift $\Delta U < 0$ close to the emission time lowers the addition energy and hence further defers the emission time, thereby providing more time during which the emission energy also decreases.

\section{Conclusion and outlook}\label{sec_conclusion}

This manuscript has provided a detailed analysis of the energy- and time-dependent emission of a pair of hot electrons from a driven quantum-dot potential. To capture a broad range of physical mechanisms impacting the emission process, we have employed two complementary models and ensuing approaches: a numerical two-particle simulation in a time-modulated 1d potential including Coulomb interaction, and an effective master equation description of a two-orbital quantum dot with time-dependent energy levels. Both methods have generally revealed the impact of the driven, dot-internal dynamics and of the possibly time-dependent dot exit-barrier height on the emission spectrum. A specific insight of the coherent two-particle simulation is how dot-internal energy transfer via Coulomb interaction can influence the emission process ---especially in an effectively one-dimensional geometry. By contrast, the master equation analysis has demonstrated how the emission energies are impacted by dot-level modulation speed, dot-degeneracy effects typical of zero- or higher-than-one dimensional systems, and by orbital dependent dot-environment coupling asymmetries.

With the theory results obtained here, a systematic comparison with experiments is enabled. This is thanks to the largely isolated discussion of the above stated physical mechanisms that are furthermore to a good extent separately accessible/tunable in experiment. 

For very fast dot level modulation, we expect the master equation approach used here to break down and we anticipate that nonadiabatic effects can become important. The treatment of these effects is however beyond the scope of the current manuscript and left open to future studies.

\acknowledgements

We thank Max Geier for very helpful discussions. This work has received funding from the European Union's H2020 research and innovation program under grant agreement No.~862683.
Funding from the Knut and Alice Wallenberg Foundation through the Academy Fellow program (J.Sp. and J.Sc.), from the Danish National Research Foundation (J.Sc.), and the Danish Council for Independent Research Natural Sciences (J.Sc.) is also gratefully acknowledged.

\section*{Appendix}\label{sec_appendix}
\appendix

\section{Details of two-particle dynamics simulation}\label{sec_2p_numerics}

This appendix details how we perform the two-particle simulation outlined in \Sec{sec_twoparticle_theory} on a GPU. Generally, we need to overcome two numerical challenges. First, the two-particle interaction complicates the common approximation of unbounded systems by finite systems with an absorbing, non-Hermitian boundary Hamiltonian~\cite{Antoine2017Aug,Selsto2010Mar}. As pointed out in \Sec{sec_twoparticle_theory}, our solution to this is to resort to a reflecting boundary in a large system, such that unphysical boundary reflections happen far enough away from the dot to not interfere with the emission process. Second, the time-dependent potential significantly increases the numerical cost of the conventional Lanczos method~\cite{Park1986Nov,Kosloff1988Apr} of solving Schrödingers equation for time-independent Hamiltonians, since the Krylov subspace truncation must eventually be repeated after a few potential update. We thus instead choose a direct leap-frog~\cite{Visscher1991Nov} solver as discussed below.

More precisely, we start by describing how we map out and index the Hilbert space \BrackSec{sec_triangle_grid} and Hamiltonian \BrackSec{sec_hamiltonian_matrix}. We then detail the leap-frog time evolution scheme as well as the calculation of expectation values \BrackSec{sec_time_evolution}. This appendix finishes with an overview of the simulation steps \BrackSec{sec_simulation_overview}. The full code implemented in C++ and in the Nvidia CUDA progamming language is provided in \Refe{SchulenborgZen2023Apr}.

\begin{figure}[t!!]
    \centering
    \includegraphics[width=\linewidth]{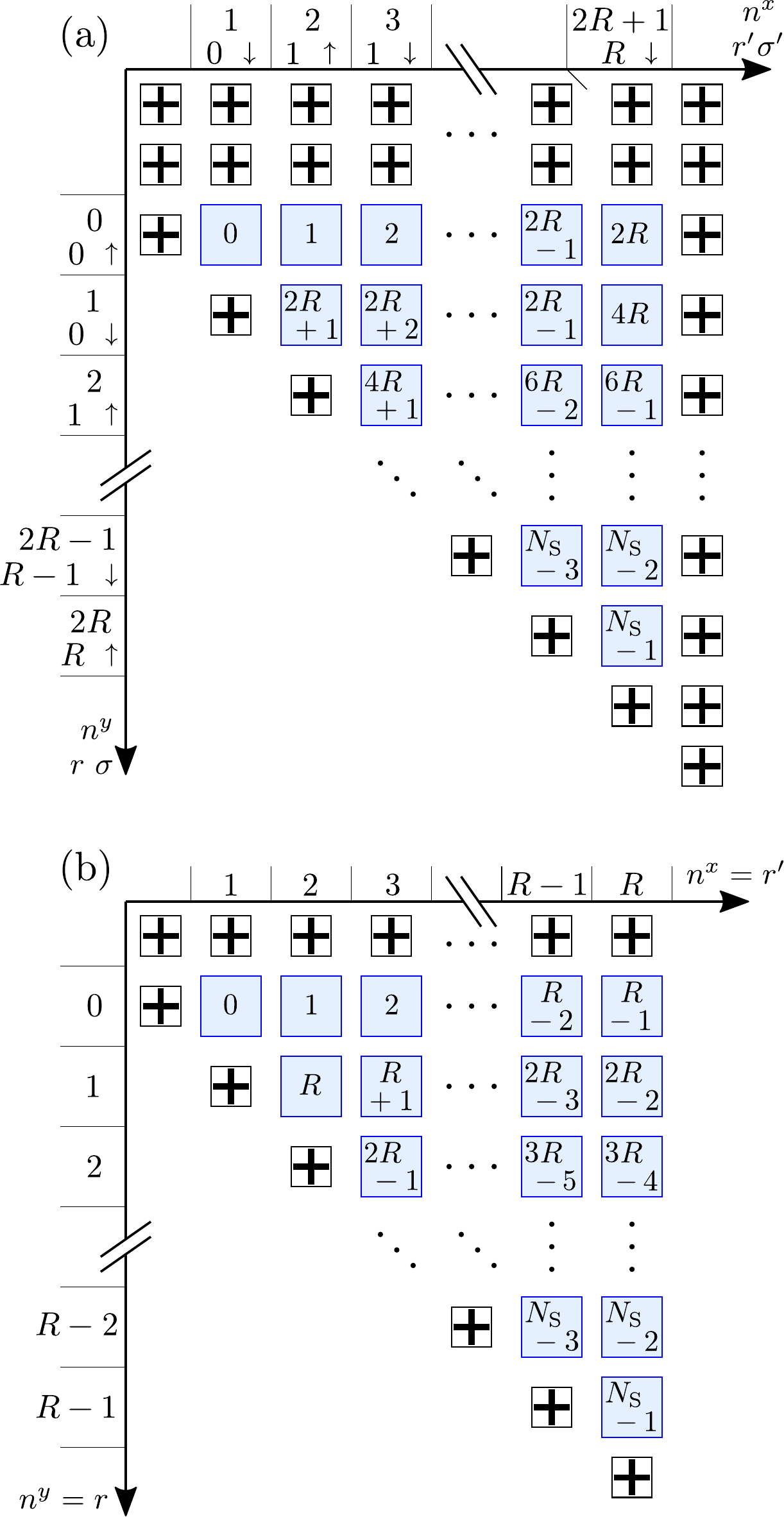}
    \caption{Two-dimensional triangle-grid mapping of the two-particle state $\ket{\Phi}$ with(a)/without(b) spin degree of freedom $\sigma$ in a potential landscape with $R + 1$ sites $r$. The-blue shaded boxes with number $i = 0,\dotsc,\NS - 1$ indicate the corresponding projection $\braket{i}{\Phi}$ onto the $i$-basis state, given by $\ket{i} = \ket{r_i\sigma_i,r'_i\sigma_i}$ for the spinful, and by $\ket{r_i,r'_i}$ for the spinless case. The smaller black-crossed boxes represent empty, zero-valued grid points acting as padding elements in memory. The horizontal and vertical axes display the grid coordinates \eq{eq_grid_coords} corresponding to the state numbers $i$.}
\label{fig_grid}
\end{figure}

\subsection{State representation on triangular 2d grid}
\label{sec_triangle_grid}
We start by considering the Hilbert space basis of occupation number states $\ket{r\sigma,r'\sigma'} = \PsiDag_{r\sigma}\PsiDag_{r'\sigma'}\Vac = -\ket{r'\sigma',r\sigma}$ in discrete positions $x_r = r\delx$, $r\in\cbrack{0,\dotsc,R}$ with spin-z projection $\sigma = \uparrow,\downarrow$ and fixed total occupation number $N = 2$. The full Hilbert space of two particles in 1d can be arranged in a 2d grid of projections of the full state $\ket{\Phi}$ onto these $\NS$ linearly independent basis states $\ket{i} := \ket{r_i\sigma_i,r'_i\sigma'_i}$. However, due to the anti-symmetry of these states under exchange of spin and position, this does not form a full square, but rather a triangular arrangement. The grid used in our simulations is illustrated in \Fig{fig_grid}: The blue-shaded boxes in this grid number the $i = 0,\dotsc,\NS - 1$ projections onto the basis states $\ket{i}$. The second particle in these states with quantum number pair $r'\sigma'$ is tied to the discrete horizontal grid coordinate $n^x = 0,1,\dotsc$, where $(r'\sigma')$-index pairs with larger $r'$ are located farther to the right, and spin $\sigma' = \downarrow$ is always to the right of $\sigma' = \uparrow$ for equal $r'$. The vertical coordinate $n^y = 0,1,\dotsc$ likewise indicates the single-particle state of the first particle with quantum numbers $r\sigma$, where larger $r$ are further below and spin $\downarrow$ is below $\uparrow$ at equal $r$. Without loss of generality, we arrange the triangular grid to ensure $r \leq r'$, i.e., in upper-right format. The grid points with $r\sigma = r'\sigma'$ are forbidden by Pauli's principle, but are nevertheless considered part of the triangle: together with two additional rows (one per spin-projection) at the top, and one additional column to the right of the triangle (smaller black-crossed boxes in \Fig{fig_grid}), they are regarded as constant, zero-weight grid points. Our numerical implementation allocates these points in memory merely to act as padding elements for sparse matrix-vector multiplications, see \App{sec_hamiltonian_matrix}. 

Considering only the state-representing grid points, the $R + 1$ possible $r$-values and two spin-projections yield, by application of the Gauss sum rule, one grid point/vector components for each of the $\NS = (2R + 1)(2R + 1 + 1)/2 = (R + 1)(2R + 1)$ linearly independent basis states. We enumerate these non-empty points from $i = 0$ to $i = \NS - 1$ by starting from the top non-empty grid row, advancing from left to right, and then by jumping to the left of the next row, skipping all empty grid points. The top-left non-empty grid point thereby corresponds to $i = 0$, and the bottom right to $i = \NS - 1$. Our GPU-based numerical approach associates GPU threads directly to the index $i$, but not to the quantum numbers $r_i\sigma_i,r'_i\sigma'_i$ which can be inferred from the grid-location $(n^y_i,n^x_i)$. Each thread hence first determines $n^y_i,n^x_i$ from $i$. Carefully contemplating the above defined grid point arrangement and again making use of the Gauss sum rule, we derive and use
\begin{align}
n^y_i &= \left\lfloor 2(R + 1) - \frac{1}{2} - \sqrt{2(\NS - i) + \frac{1}{4}}\right\rfloor\notag\\
n^x_i &= n^y_i + \frac{(2R + 1 - n^y_i)(2R + 2 - n^y_i)}{2} + 1 + i - \NS\notag\\
r_i &= \left\lfloor n^y_i/2\right\rfloor \lrsepa r'_i = \left\lfloor n^x_i/2\right\rfloor\notag\\
\sigma_i &= n^y_i \,\,\text{mod}\,\, 2 \lrsepa \sigma'_i = n^x_i \,\,\text{mod}\,\, 2
\label{eq_grid_coords},
\end{align}
where $\sigma = 0/1$ is synonymous to  $\sigma = \uparrow/\downarrow$. The memory address of the state vector component at grid coordinate $(n^i_y,n^i_x)$ is given by the basis state index $i$ ---shifted by all preceding empty padding elements--- to ensure sufficient data locality, see full code~\cite{SchulenborgZen2023Apr} for details. 

Note that while the square root in \Eq{eq_grid_coords} is in principle computationally demanding, our implementation nevertheless calculates it on-the-fly in every GPU thread without any severe performance impact. This works because our simulation mostly relies on sparse matrix-vector multiplication, whose performance is in any case primarily memory-bandwidth limited. To efficiently use the available bandwidth, we compute and store the vector components only in single precision (32 bit) instead of double precision (64 bit), as our comparisons between the two have not revealed any appreciable inaccuracies for the former. Moreover, due to the absence of any Zeeman field and spin-orbit coupling, the equal-spin case effectively reduces to a spinless system. Neglecting the quantum number $\sigma$ altogether for this case, we obtain a triangle grid with approximately 4 times fewer points compared to the spinful case, $\NS \rightarrow \NSt =  \frac{R(R+1)}{2}$, see \FigPart{fig_grid}{b}. Moreover, we then only add a single row of top-padding empty points, and no padding column to the right. Grid and particle coordinate also become directly related, i.e., we set $r_i = n^y_i$ and $r'_i = n^x_i$ in \Eq{eq_grid_coords} without dividing by the spin factor $2$.

\subsection{Implementing the sparse Hamiltonian matrix}
\label{sec_hamiltonian_matrix}
The absent Zeeman and spin-orbit coupling terms in our model mean that any Hamiltonian appearing in our simulation can always be gauged to have a fully real matrix representation with respect to the above defined occupation number basis. Moreover, the only non-diagonal elements within this matrix appear for the kinetic energy $\Hkin$ \BrackEq{eq_hamiltonian_2p_kin}, yielding a sparse matrix containing non-zero elements only in the main diagonal and in maximally 4 sub-diagonals, i.e., 2 subdiagonals per particle/grid dimension. Exploiting this simple structure, we implement the sparse matrix-vector multiplication of any Hamiltonian with a state as a CUDA kernel acting on an array that stores the grid-ordered vector components according to the previous \Sec{sec_triangle_grid}. An actual calculation is of course only performed for output-array elements corresponding to non-empty grid points associated with a basis state index $i$, skipping the computation for constantly zero-valued padding elements. These padding elements are nevertheless important for any input array, as explained below.

The main-diagonal elements of the Hamiltonian matrix, i.e., those elements relating grid points $i$ to themselves, consist of three different contributions.  One stems from the onsite term $4\lambda$ of $\Hkin$ \BrackEq{eq_hamiltonian_2p_kin} for two particles, which is independent of time $t$ and state $i$. The second contribution comes from the Coulomb potential \eq{eq_hamiltonian_2p_coul}, which only requires a single division by $\delx|r_i - r'_i|$ per index $i$, and is hence always computed on-the-fly. This is in contrast to third contribution, given by the diagonal elements of $\Hpot(t)$. The latter namely equal the sum of two single-particle potentials $V(\delx r_i,t) + V(\delx r'_i,t)$ \BrackEq{eq_potential_landscape}, meaning that while several computationally expensive exponential functions are required to evaluate $V(x,t)$, the same value for $V$ enters many different main-diagonal elements of $\Hpot(t)$. The potential $V(x,t)$ is thus pre-computed only once for each of the $R + 1$ different spatial coordinates $x = r\delx$, and stored in the GPU memory. The subsequent calculation of the main-diagonal matrix elements of $\Hpot(t)$ and of the full two-particle Hamiltonian $\HTP(t)$ then simply reads these values.

The only contribution to the sub-diagonals of the Hamiltonian matrix come from the bilinear coupling terms $\sim\lambda$ in $\Hkin$ \BrackEq{eq_hamiltonian_2p_kin}. As such, they relate state vector components associated with different indices $i' \neq i$, both in horizontal $(n^x)$ and vertical $(n^y)$ grid direction, see full code at~\cite{SchulenborgZen2023Apr}. Care needs to be taken in determining the correct fermionic exchange sign in the spinful case, and in correctly truncating the coupling at any edge of the non-empty grid. Points on these edges namely coincide with a particle at one of the ends of the 1d potential landscape, where the coupling beyond the edge no longer contributes to $\Hkin$. The above mentioned zero-value padding elements are added precisely to make any branching logic for these edge cases in the application of $\Hkin$ unnecessary. This helps not only with code readability, but also with avoiding divergent GPU threads.

\subsection{Time evolution of state, expectation values, and spectral density}
\label{sec_time_evolution}
Having defined the state space and set up the Hamiltonian matrix, we can now describe how we perform the Schr\"odinger time evolution \eq{eq_schrodinger} and obtain the expectation values for the charge and energy in the dot and in the filter region.

For a time-independent Hamiltonian, the Schr\"odinger equation \eq{eq_schrodinger} can be solved efficiently by truncating the evolution to the relevant Krylov subspace, such as in the often employed time-dependent Lanczos approach~\cite{Park1986Nov,Kosloff1988Apr}. However, the modulated potential $V(x,t)$ entering our time-dependent two-particle Hamiltonian $\HTP(t)$ \BrackEq{eq_hamiltonian_2p_parts} reduces the efficiency and increases the complexity of this method, because the truncation must be repeated regularly, with possibly $V(x,t)$-dependent truncation cutoffs.
Our numerical implementation avoids these complications by instead solving the Schr\"odinger equation \eq{eq_schrodinger} directly, demanding a spatial$(\delx)$- and temporal$(\delt)$ resolution to reliably capture the desired maximum kinetic energy $E_{\text{max}}$, i.e., $E_{\text{max}} \leq 2\frac{\pi^2}{2\meff\delx^2}$ (lowest energy of two opposite-spin electrons in a box of size $\delx$) and $E_{\text{max}} \leq 2\pi/\delt$ (using $E = \omega = 2\pi/\Delta t$). Given kinetic energies $E \sim \text{meV}$ in a potential landscape of typical size $L \sim 10^4\text{nm}$, we choose a leap-frog solver suitable for the quite large number of basis states $\NS \sim 10^6$ required by these criteria. This method saves memory bandwidth by allocating only few temporary resources per time step, and in particular preserves the state norm without any explicit renormalization~\cite{Visscher1991Nov}: As the latter would be computationally expensive due to the necessary overflow protection for the given $\NS$, this represents a crucial advantage.
To check whether our discretization steps are actually small enough for our results to converge, we also run a few higher-resolution test-runs for comparison.

The leap-frog scheme begins by initializing the system in the ground state $\ket{\PhiGS}$ of $\HTP(0)$, i.e., with respect to the initial potential configuration $V(x,t=0)$. We achieve this by first seeding a two-particle state $\ket{\PhiSeed}$ that has only non-zero vector components inside the dot region, $0 < r\delx,r'\delx < \Ld$, and then by evolving this state along the imaginary time axis to project out any excited state:
\begin{align}
\sqbrack{\ket{\Phi(\tilde{t} + \delta\tilde{t})}}_i &= \sum_{j = 0}^{\NS - 1}\sqbrack{e^{-\HTP(0)\delta\tilde{t}}}_{ij}\sqbrack{\ket{\Phi}(\tilde{t})}_j \notag\\
&\approx \sum_{j = 0}^{\NS - 1}\sqbrack{\one - \HTP(0)\delta\tilde{t}}_{ij}\sqbrack{\ket{\Phi(\tilde{t})}}_j \notag\\
\ket{\Phi(\tilde{t} = 0)} &= \ket{\PhiSeed},\label{eq_gen_gs}
\end{align}
for which we have implemented a time-independent CUDA kernel to apply $\one - \HTP(0)\delta\tilde{t}$ to the state vector. Altogether performing $\NGS$ imaginary-time steps \eq{eq_gen_gs} to obtain the ground state $\ket{\PhiGS}$, these steps are interspersed with explicit state renormalizations $\ket{\Phi(\tilde{t})} \leftarrow \ket{\Phi(\tilde{t})}/\sqrt{\braket{\Phi(\tilde{t})}{\Phi(\tilde{t})}}$ after each set of $\NGSn$ subsequent evolution steps \eq{eq_gen_gs}. We then define the ground state as
\begin{equation}
\ket{\PhiGS} := \ket{\Phi(\tilde{t} = \NGS\delta\tilde{t})}.\label{eq_gs}
\end{equation}
Each renormalization by the inverse square-root norm $1/\sqrt{\braket{\Phi(\tilde{t})}{\Phi(\tilde{t})}}$ is computed on the GPU using the cuBLAS Level-1 API from the standard CUDA library~\cite{SchulenborgZen2023Apr}.

To prepare the actual real-time leap-frog evolution, we first divide the state into real and imaginary part. Due to measurably better performance, the latter are not stored in a block format, but instead in an interleaved fashion. This means that we do not distribute real and imaginary parts into separate grids, but rather split each state component $\sqbrack{\ket{\Phi(t = 0)}}_i$ and each empty point in the grid shown in \Fig{fig_grid} into a pair of successively stored single-precision floating point numbers, the first corresponding to the real part and the second to the imaginary part; the empty padding elements are represented by two constant zero-values. The crucial aspect of the leap-frog method is now to introduce a half-time step between real and imaginary part. This means that the initial-state components $\sqbrack{\ket{\Phi(t = 0)}}_i$ read
\begin{gather}
\Re\cbrack{\sqbrack{\ket{\Phi(t = -\delt)}}_i} = \sqbrack{\ket{\PhiGS}}_i\notag\\
\Im\cbrack{\sqbrack{\ket{\Phi(t = -\delt)}}_i} = \Im\cbrack{\sqbrack{\frac{\ket{\delt/2}}{\sqrt{\braket{\delt/2}{\delt/2}}}}_i}\notag\\
\Re\cbrack{\sqbrack{\ket{\Phi(t = 0)}}_i} = \Re\cbrack{\sqbrack{\frac{\ket{\delt}}{\sqrt{\braket{\delt}{\delt}}}}_i}\notag\\
\Im\cbrack{\sqbrack{\ket{\Phi(t = 0)}}_i} = \Im\cbrack{\sqbrack{\frac{\ket{3\delt/2}}{\sqrt{\braket{3\delt/2}{3\delt/2}}}}_i}\notag\\
\ket{\delt/2} = \nbrack{\one - i\HTP(0)\frac{\delt}{2}}\ket{\PhiGS}\notag\\
\ket{\delt} = \nbrack{\one - i\HTP(0)\frac{\delt}{2}}\frac{\ket{\delt/2}}{\sqrt{\braket{\delt/2}{\delt/2}}}\notag\\
\ket{3\delt/2} = \nbrack{\one - i\HTP(0)\frac{\delt}{2}}\frac{\ket{\delt}}{\sqrt{\braket{\delt}{\delt}}} \label{eq_gen_in_state},
\end{gather}
where we have implemented a custom CUDA kernel to perform the sparse matrix-vector multiplication $\nbrack{\one - i\HTP(0)\frac{\delt}{2}}\ket{x}$, and exploited that $\ket{\PhiGS}$ is already normalized and has purely real components. Using the likewise purely real matrix representation of $\HTP(t)$, the leap-frog time evolution then proceeds according to~\cite{Visscher1991Nov}
\begin{subequations}
\begin{align}
\Re\cbrack{\sqbrack{\ket{\Phi(t + \delt)}}_i} &= \sum_{j = 0}^{\NS - 1}\Re\cbrack{\sqbrack{\one - i\HTP(t)\delt }_{ij}\sqbrack{\ket{\Phi(t)}}_j}\notag\\
&= \Re\cbrack{\sqbrack{\ket{\Phi(t)}}_i}\\
&\phantom{=}+ \sum_{j = 0}^{\NS - 1}\sqbrack{\HTP(t)\delt}_{ij}\Im\cbrack{\sqbrack{\ket{\Phi(t)}}_j}\notag
\end{align}
\begin{align}
\Im\cbrack{\sqbrack{\ket{\Phi(t + \delt)}}_i} &= \sum_{j = 0}^{\NS - 1}\Im\cbrack{\sqbrack{\one - i\HTP(t)\delt }_{ij}\sqbrack{\ket{\Phi(t)}}_j}\notag\\
&= \Im\cbrack{\sqbrack{\ket{\Phi(t)}}_i}\\
&\phantom{=}- \sum_{j = 0}^{\NS - 1}\sqbrack{\HTP(t)\delt}_{ij}\Re\cbrack{\sqbrack{\ket{\Phi(t)}}_j}\notag,
\end{align}
\label{eq_evolution}
\end{subequations}
where the CUDA kernel for the sparse matrix-vector product $\sqbrack{\HTP(t)\delt}\ket{\Phi}$ reads the single-particle potential $V(x,t)$ \BrackEq{eq_potential_landscape} pre-generated for time $t$ from GPU memory prior to evaluating the time step \eq{eq_evolution}.

As already highlighted above, the key and name-giving feature of the leap-frog scheme is the initialization \eq{eq_gen_in_state} with half-time separation $\delt/2$ between real and imaginary part. This stabilizes the state norm during the time evolution \eq{eq_evolution} ---and hence renders explicit renormalizations unnecessary--- if that norm and the expectation value are slightly redefined~\cite{Visscher1991Nov}:
\begin{subequations}
\begin{align}
\langle O\rangle(t) &= \frac{\bra{\Phi(t)}\sqbrack{O\ket{\Phi(t)}}}{\braket{\Phi(t)}{\Phi(t)}}\label{eq_expectation_value}\\
\braket{\Phi(t)}{\Phi(t)} &:= \sum_{i = 0}^{\NS - 1}\sqbrack{\Re\cbrack{\sqbrack{\ket{\Phi(t)}}_i}}^2\label{eq_norm}\\
&\phantom{:=} + \sum_{i = 0}^{\NS - 1}\Im\cbrack{\sqbrack{\ket{\Phi(t)}}_i}\Im\cbrack{\sqbrack{\ket{\Phi(t-\delt)}}_i}\notag\\
\bra{\Phi(t)}\sqbrack{O\ket{\Phi(t)}} &:= \sum_{i = 0}^{\NS - 1}\Re\cbrack{\sqbrack{O\ket{\Phi(t)}}_i}\Re\cbrack{\sqbrack{\ket{\Phi(t)}}_i}\notag\\
&\phantom{} + \sum_{i = 0}^{\NS - 1}\Im\cbrack{\sqbrack{O\ket{\Phi(t)}}_i}\Im\cbrack{\sqbrack{\ket{\Phi(t-\delt)}}_i}.\label{eq_braket}
\end{align}
\label{eq_overlaps}
\end{subequations}
Note that the interleaved memory storage of real and imaginary parts enables us to compute the norm \eq{eq_norm} on the GPU with a single call of the cuBLAS Level-1 scalar-product routine, thereby preventing numerical overflow while maintaining a high runtime performance.

A particularly important expectation value for our simulation is the time-dependent spectral density $\phi_{2\text{P}}(t,E)$ \BrackEq{eq_dos}. We evaluate this density by expanding the $\Hkinf$-dependence of $\delta(E - \Hkinf)$ in terms of a finite number $\NCP$ of Chebyshev polynomials of first kind~\cite{Weisse2006Mar}:
\begin{align}
&\phi_{2\text{P}}(t,E) = \langle\delta(E - \Hkinf)\rangle(t)\label{eq_dos_cheby}\\
&\approx \sum_{n = 0}^{\NCP-1}\frac{K_n(E/\ECP)}{\ECP\braket{\Phi(t)}{\Phi(t)}}\bra{\Phi(t)}\sqbrack{p_n\nbrack{\frac{\Hkinf}{\ECP}}\ket{\Phi(t)}}\notag
\end{align}
with energy scale $\ECP$ chosen such that all eigenvalues of $\Hkinf/\ECP$ lie within the open interval $(-1,1)$, with the recursively defined Chebyshev vectors
\begin{align}
p_0(x)\ket{\Phi(t)} &= \ket{\Phi(t)}\label{eq_chebyshev_vectors}\\
p_1(x)\ket{\Phi(t)} &= x\ket{\Phi(t)}\notag\\
p_{n \geq 2}(x)\ket{\Phi(t)} &= 2x\sqbrack{p_{n-1}(x)\ket{\Phi(t)}} - \sqbrack{p_{n-2}(x)\ket{\Phi(t)}},\notag
\end{align}
and with the Chebyshev expansion coefficients
\begin{align}
K_n(x) &= \frac{(2 - \delta_{n0})\cosfn{\text{acos}(x)n}}{\pi\sqrt{1 - x^2}}J_n\label{eq_chebyshev_coeffs}\\
J_n &= \frac{1}{(\NCP + 1)}\sinfn{\frac{n\pi}{\NCP + 1}}\text{cot}\nbrack{\frac{\pi}{\NCP + 1}}\notag\\
&\phantom{=}+ \nbrack{1 - \frac{n}{\NCP + 1}}\cosfn{\frac{n\pi}{\NCP + 1}}.\label{eq_jackson_kernel}
\end{align}
The Jackson kernel $J_n$ suppresses unphysical oscillations in the Chebyshev polynomial expansion of $\delta(E - x)$ due to the Dirac $\delta$ being sharp in $x$.
Note again that the norm and vector overlaps in \Eq{eq_dos_cheby} are defined by  \Eq{eq_overlaps}, which, due to the employed leap-frog scheme, slightly differs from the usual definition.

To compute the sum over $n$ in \Eq{eq_dos_cheby} for a fixed time $t$ on the GPU, we first obtain all $\bra{\Phi(t)}\sqbrack{p_n\nbrack{\frac{\Hkinf}{\ECP}}\ket{\Phi(t)}}$ and buffer them in a single vector of size $\NCP$; all coefficients $K_n(E/\ECP)$ for all considered ratios $E/\ECP$ are stored into a matrix. Performing the sum in \Eq{eq_dos_cheby} thereby reduces to a single call of the cuBLAS dense matrix-vector product of the coefficient matrix with the Chebyshev polynomial vector. Note, also, that we do not call a scalar-product routine for each individual polynomial $\bra{\Phi(t)}\sqbrack{p_n\nbrack{\frac{\Hkinf}{\ECP}}\ket{\Phi(t)}}$ entering \Eq{eq_dos_cheby}. We instead employ a custom CUDA kernel for the matrix-vector product $\Hkinf\ket{\Phi(t)}$ to recursively compute a subset of all $\NCP$ required Chebyshev vectors (limited by the GPU memory capacity) according to \Eq{eq_chebyshev_vectors}, and first buffer these vectors as columns of a large matrix. Then, we calculate the given subset of Chebyshev polynomials with a single cuBLAS dense vector-matrix multiplication of $\bra{\Phi(t)}$ ($\bra{\Phi(t - \delt)}$ for the imaginary part \BrackEq{eq_braket}) with this matrix. We thereby benefit from cuBLAS-specific optimizations of the dense vector-matrix product, which sizably reduces the time to evaluate \Eq{eq_dos_cheby} compared to calculating all scalar products individually ($\sim 10 - 20\%$, depending on vector size).

\subsection{Overview of simulation steps}
\label{sec_simulation_overview}
\begin{figure}[t!!]
    \centering
    \includegraphics[width=\linewidth]{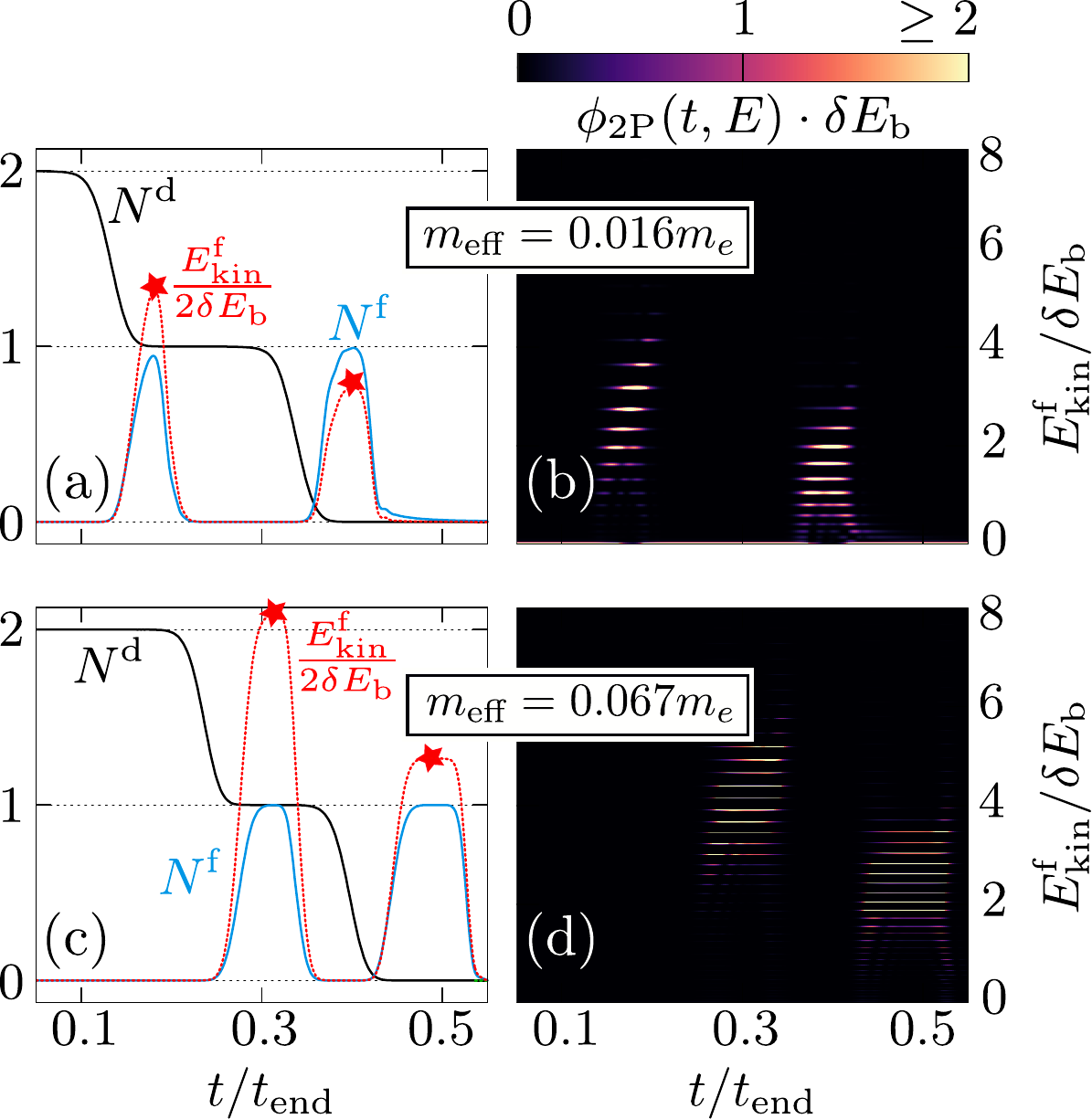}
    \caption{Energy-and time dependent emission spectroscopy for effective mass $\meff = 0.016m_e$  as in main text (a,b) compared to the corresponding quantities for $\meff = 0.067m_e$ as in GaAs (c,d). We assume two equal $\uparrow$-spins, $\kappa = 0.125$, and all other parameters as stated in the caption of \Fig{fig_2p} and at the end of \App{sec_simulation_overview}.}
\label{fig_eff_mass}
\end{figure}

Having detailed the individual aspects of the simulation, we conclude \App{sec_2p_numerics} with an overview of the steps performed, and the system parameters used to obtain the data presented in \Fig{fig_2p} and \Fig{fig_eff_mass}.

\begin{enumerate}
    \item Generate the seed state $\ket{\PhiSeed}$ \BrackEq{eq_gen_gs}, pre-calculate all Chebyshev coefficients $K_n(E/\ECP)$ \BrackEq{eq_chebyshev_coeffs} for all $n = 0,1,\dotsc,\NCP - 1$ and for a discrete set of energy ratios $E/\ECP$, and store these $K_n(E/\ECP)$ into a matrix.
    \item Compute the initial potential landscape $V(x,0)$ \BrackEq{eq_potential_landscape} and initialize the system state into the ground state $\ket{\PhiGS}$ of the initial Hamiltonian $\HTP(0)$ \BrackEq{eq_hamiltonian_2p_parts} with the scheme described in and around \Eqs{eq_gen_gs}-\eq{eq_gs}.
    \item Starting from $\ket{\PhiGS}$, prepare the initial state of the leap-frog time evolution scheme according to \Eq{eq_gen_in_state}.
    \item Perform a number $\Ntest$ of time evolution steps \eq{eq_evolution} with a constant Hamiltonian $\HTP(0)$ to let the system state settle to the actual initial simulation state $\ket{\Phi(0)}$, and confirm that $\HTP(0)$ does not further evolve this state $\ket{\Phi(0)}$ in time apart from a global phase.
    \item Calculate the initial expectation values for the observables of interest with respect to $\ket{\Phi(0)}$ using \Eqs{eq_overlaps}-\eq{eq_jackson_kernel}.
    \item Finally, run the simulation for a desired time $t_{\text{end}}$ by repeating the evolution step \eq{eq_evolution} with a continuously updated, time-dependent potential $V(x,t)$ \BrackEq{eq_potential_landscape} entering the two-particle Hamiltonian $\HTP(t)$ \BrackEq{eq_hamiltonian_2p_parts}. Use \Eqs{eq_overlaps}-\eq{eq_jackson_kernel} to calculate observable expectation values in regular time intervals.
\end{enumerate}

For all simulations shown in \Fig{fig_2p} and \Fig{fig_eff_mass}, the system is discretized in $R + 1 = 2^{11} = 2048$ sites. The total simulation time $t_{\text{end}} = 307.2\,\text{ps}$ is divided into numerical time steps of length $\delta t = 2.5\times 10^{-5}\,\text{ps}$, and we compute all observables for $2^9 = 512$ equidistant time points within $[0,t_{\text{end}}]$, each using $\NCP = 2^{17} = 131072$ Chebyshev polynomials to generate $\phi_{2\text{P}}$. We use $\NGS = 5\times 2^{16} = 327680$ imaginary time steps $\delta\tilde{t} = 1.25\times 10^{-5}\text{ps}$ to generate the ground state $\ket{\PhiGS}$ of $\HTP(0)$, where we renormalize every $\NGSn = 2^3 = 8$ step, and we have checked for convergence in a test run with larger $\NGS$ at lower $\NGSn, \delta\tilde{t}$.
Prior to the actual time evolution, we use $\Ntest = 2^{15} = 32768$ time steps $\delta t$ to initialize the state $\ket{\Phi(0)}$ from $\ket{\PhiGS}$.

\section{Two-particle dynamics for larger effective mass}\label{sec_2p_eff_mass}

This short appendix compares the two-particle simulation from \Sec{sec_twoparticle_results} for two spin-$\uparrow$ electrons with finite interaction $\kappa = 1/8$ and effective mass $\meff = 0.016$ \BrackFigPart{fig_2p}{e,f} to simulations with larger effective mass $\meff = 0.067m_e$ typical of GaAs, but otherwise equal parameters. 
In particular, we confirm in \Fig{fig_eff_mass} that the first emitted particle still receives additional kinetic energy from the particle relaxing inside the dot \BrackFigPart{fig_eff_mass}{a,c}. 
By contrast, differences between the two cases are the smaller energy splitting \BrackFigPart{fig_eff_mass}{b,d}, due to $E^{\text{f}}_{\text{kin}} \sim 1/\meff$, and the generally higher kinetic energy for the larger $\meff = 0.067$. 
The latter stems from a combination of the smaller single-particle dot energy splitting and the exit-barrier potential rising with higher dot potential. Namely, compared to the case $\meff = 0.016m_e$, the smaller dot addition energies cause emission at smaller dip depth $\Vdc$, and hence at later times $t$.
At these later times, the emission is even further deferred and, altogether pushed to higher emission energies by the increasing tunnel barrier height \BrackFigPart{fig_eff_mass}{a,c}, see also videos in \cite{SchulenborgZen2023Apr}. 
Moreover, we observe an initially sharper and, due to the larger $\meff$, also more slowly traveling wave package, as evident from a comparison of $N^{\text{f}}$, see turquoise lines in \FigPart{fig_eff_mass}{a,c}. 

\section{Master equation evolution}\label{sec_master_detail}
This appendix details some analytical and numerical aspects of our solution to the master equation \eq{eq_master}.
The Lindblad operators $L_{\eta\ell}$ \BrackEq{eq_lindblad} and the Lamb shift
\begin{align}
&\Lambda = \sumsub{\eta\ell}{ii'j}F_{\eta\ell}(E_{ii'},E_{i'j})\bra{i}d^\dagger_{\eta\ell}\ket{i'}\bra{i'}d_{\eta\ell}\ket{j}\times\ket{i}\bra{j}\label{eq_lamb}\\
 &F_{\eta\ell}(E,E') = -\mathcal{P}\int_{-\infty}^\infty d\omega\left[\frac{\sqrt{\Gamma_{\ell}(\eta\omega - \eta E)f^\eta(\eta\omega - \eta E)}}{2\pi\omega}\right.\notag\\
 &\phantom{F_{\eta\ell}(E,E') = -\mathcal{P}\int_{-\infty}^\infty}\times\left.\sqrt{\Gamma_{\ell}(\eta\omega + \eta E')f^{\eta}(\eta\omega + \eta E')}\right].\notag
\end{align}
as derived from \Refs{Nathan2020Sep,Schulenborg2023Mar} (all symbols including creation/annihilation operators $d_{\eta\ell}$ defined below \Eq{eq_lindblad}) depend on the energies $E_i$ and eigenstates $\ket{i}$ of the dot Hamiltonian \eq{eq_hamiltonian}. We hence start with these quantities:
\begin{gather}
E_{i = 0} = 0 \lrsepa E_{i = \text{d}} = 2\epsilon  + \Deps + U\notag\\
E_{i = \chi = \pm} = \epsilon + \frac{\Deps}{2} + \chi\sqrt{|\tau|^2 + \frac{\Deps^2}{4}}\label{eq_energies}\\
\ket{i = 0} = \ket{0} \,\,,\,\, \ket{i = \text{d}} = \ket{\text{d}}\notag\\
\ket{i = \chi = \pm} = \sum_{\ell = \text{L,R}}C^{\ell}_\chi \ket{\ell}\label{eq_states}, 
\end{gather}
with
\begin{equation}
C^{\text{L}}_\chi = \frac{\text{sgn}(\tau)\sqbrack{E_{\chi} - \epsilon}}{\sqrt{|\tau|^2 + (E_\chi - \epsilon)^2}} \lrsepa C^{\text{R}}_\chi = \frac{|\tau|}{\sqrt{|\tau|^2 + (E_\chi - \epsilon)^2}}\label{eq_coeffs}
\end{equation}
interpolating between $\ket{i}$ being either the "L"
and "R"
states $\ket{\text{L}}$ and $\ket{\text{R}}$, or the bonding/antibonding states $\ket{\text{L}} \pm \ket{\text{R}}$, depending on $\Deps,\tau$.

Next, we use that each Lindblad operator flips the dot fermion parity $\fpop = (-\one)^{\NL + \NR}$ while the Hamiltonian $H$ and the Hermitian Lamb shift $\Lambda = \Lambda^\dagger$ conserve this parity, meaning $\fpop L_{\eta\ell} = - L_{\eta\ell}\fpop$ and $\fpop H = H \fpop$, $\fpop\Lambda = \Lambda\fpop$. Together with the absence of any pairing terms $\sim \dDagL\dDagR,\dL\dR$ in the Hamiltonian and any higher-order $\Gamma$ terms, this means that the reduced density operator $\rho$ can only have $6$ non-zero elements\footnote{This presupposes that any element other than the $6$ mentioned are initialized to 0. If not, those elements would, however, still evolve independently from the $6$ relevant ones.} during the entire time evolution \eq{eq_master}. In the occupation number basis $\nbrack{\ket{0},\ket{\ell = \text{L}} = \dDagL\ket{0},\ket{\ell = \text{R}} = \dDagR\ket{0},\ket{\text{d}} = \dDagL\dDagR\ket{0}}$, these elements are
\begin{align}
\mathbf{P} &= \nbrack{P_0, P_{\text{L}}, P_{\text{R}}, P_{\text{d}}, P_{\text{LR}}, P_{\text{RL}}}^T \label{prob_vector}\\
&= \nbrack{\bra{0}\rho\ket{0}, \bra{\text{L}}\rho\ket{\text{L}}, \bra{\text{R}}\rho\ket{\text{R}}, \bra{\text{d}}\rho\ket{\text{d}}, \bra{\text{L}}\rho\ket{\text{R}}, \bra{\text{R}}\rho\ket{\text{L}}}^T\notag
\end{align}
This vector evolves according to \BrackEq{eq_master}
\begin{equation}
\partial_t \mathbf{P} = \mathbf{W}\cdot\mathbf{P}\label{eq_master_vector},
\end{equation}
where the elements $W_{y,z} = \sqbrack{\mathbf{W}}_{y,z}$ of the transition matrix $\mathbf{W}$ are obtained from an expansion of the master equation \eq{eq_master} with the help of \Eqs{eq_energies}-\eq{eq_coeffs}:
\begin{gather}
W_{0,d} = W_{d,0} = W_{\text{L},\text{R}} = W_{\text{R},\text{L}} = W_{\text{RL},\text{LR}} = W_{\text{LR},\text{RL}} = 0\notag\\
W_{0,\ell = \text{L/R}} = \sum_{\ell'=\text{L,R}}\sqbrack{\gamma^{\ell'}_{0\ell}}^2 \lrsepa W_{\ell = \text{L/R},0} = \sum_{\ell'=\text{L,R}}\sqbrack{\gamma^{\ell'}_{\ell 0}}^2\notag\\
W_{\text{d},\ell = \text{L/R}} = \sum_{\ell'=\text{L,R}}\sqbrack{\gamma^{\ell'}_{\text{d}\ell}}^2 \lrsepa W_{\ell = \text{L/R},\text{d}} = \sum_{\ell'=\text{L,R}}\sqbrack{\gamma^{\ell'}_{\ell \text{d}}}^2\notag\\
W_{0,0} = -\sum_{\ell = \text{L,R}}W_{\ell,0} \lrsepa W_{\text{d},\text{d}} = -\sum_{\ell = \text{L,R}}W_{\ell,\text{d}}\notag\\
W_{\ell,\ell} = -W_{0,\ell} - W_{\text{d},\ell}\notag\\
W_{(0/\text{d}),\text{LR}} = W_{(0/\text{d}),\text{RL}} = \sum_{\ell'=\text{L,R}}\gamma^{\ell'}_{(0/\text{d})\text{L}}\gamma^{\ell'}_{(0/\text{d})\text{R}}\notag\\
W_{\text{LR},(0/\text{d})} = W_{\text{RL},(0/\text{d})} = \sum_{\ell'=\text{L,R}}\gamma^{\ell'}_{\text{L}(0/\text{d})}\gamma^{\ell'}_{\text{R}(0/\text{d})}\notag\\
W_{\ell,\ell\bar{\ell}} = W_{\ell\bar{\ell},\ell} = (W_{\ell,\bar{\ell}\ell})^* = i\tilde{\tau} - \sum_{\ell'=\text{L,R}}\frac{\gamma^{\ell'}_{0\ell}\gamma^{\ell'}_{0\bar{\ell}} + \gamma^{\ell'}_{\text{d}\ell}\gamma^{\ell'}_{\text{d}\bar{\ell}}}{2}\notag\\
W_{\text{LR},\text{LR}} = (W_{\text{RL},\text{RL}})^* = -i\Delta\tilde{\epsilon} -\frac{\Gamma_{\text{de}}}{2}\label{eq_rates},
\end{gather}
defining $\bar{\ell} = \text{R}$ if $\ell = \text{L}$ and $\bar{\ell} = \text{L}$ if $\ell = \text{R}$. 
These equations \eq{eq_rates} introduce the shifted single-particle energies
\begin{align}
\Delta\tilde{\epsilon} &= \Delta\epsilon + \bra{\text{L}}\Lambda\ket{\text{L}} - \bra{\text{R}}\Lambda\ket{\text{R}}\notag\\
\tilde{\tau} &= \tau + \sqbrack{\bra{\text{L}}\Lambda\ket{\text{R}} + \bra{\text{R}}\Lambda\ket{\text{L}}}/2\label{eq_lamb_shifted_energies}
\end{align}
as well as the following (square-root) rates:
\begin{align}
\gamma^{\ell'}_{0\ell} &= \sum_{\chi = \pm}\sqrt{\Gamma^-_{\ell'}(E_{\chi0})}C^{\ell}_{\chi}C^{\ell'}_{\chi}\label{eq_rates_functions_step}\\
\gamma^{\ell'}_{\ell0} &= \sum_{\chi = \pm}\sqrt{\Gamma^+_{\ell'}(E_{\chi0})}C^{\ell}_{\chi}C^{\ell'}_{\chi}\notag\\
\gamma^{\ell'}_{\text{d}\ell} &= \sum_{\chi = \pm}\sqrt{\Gamma^+_{\ell'}(E_{\text{d}\chi})}C^{\ell}_{\chi}C^{\bar{\ell}'}_{\chi}\notag\\
\gamma^{\ell'}_{\ell\text{d}} &= \sum_{\chi = \pm}\sqrt{\Gamma^-_{\ell'}(E_{\text{d}\chi})}C^{\ell}_{\chi}C^{\bar{\ell}'}_{\chi}\notag
\end{align}
and
\begin{align}
\Gamma_{\text{de}} &= \sum_{\chi = \pm}\sqbrack{\Gamma^+_{\text{L}}(E_{\text{d}\chi}) + \Gamma^-_{\text{R}}(E_{\chi0})}(C^{\text{R}}_{\chi})^2 \notag\\
&\phantom{=}+\sum_{\chi = \pm}\sqbrack{\Gamma^+_{\text{R}}(E_{\text{d}\chi}) + \Gamma^-_{\text{L}}(E_{\chi0})}(C^{\text{L}}_{\chi})^2\label{eq_rates_functions},
\end{align}
where we use
\begin{equation}
\Gamma^\pm_{\ell'=\text{L,R}}(E) = \Gamma_{\ell'}(E)f^\pm(E).
\end{equation}
As in the main text, we denote energy differences as $E_{ij} = E_i - E_j$.

In the following, for completeness, we describe how to evaluate the Lamb shift. Note, however, that for the parameter regimes considered here, with $\Gamma\ll U,\delta E_\text{b}$, the Lamb shift has no visible impact on the presented results and could therefore equally be neglected.

Equation~\eqref{eq_lamb_shifted_energies} relies on the symmetry of the Lamb shift $\Lambda$ both in the eigenbasis and L-R basis: $\bra{i}\Lambda\ket{j} = \bra{j}\Lambda\ket{i}$, $\bra{\ell}\Lambda\ket{\ell'} = \bra{\ell'}\Lambda\ket{\ell}$. The latter follows from $\Lambda^\dagger = \Lambda$ and the fact that the eigenstates $\ket{i}$ admit the real representation \eq{eq_states} in the local dot basis. Using the identity $\one = \ket{0}\bra{0} + \sum_{\chi=\pm}\ket{\chi}\bra{\chi} + \ket{\text{d}}\bra{\text{d}}$ and the definition \Eq{eq_lamb}, we can further expand \Eq{eq_lamb_shifted_energies} to
\begin{align}
\Delta\tilde{\epsilon} &= \Delta\epsilon + \sqbrack{\Lambda_{++} - \Lambda_{--}}\sqbrack{|C^{\text{L}}_+|^2 - |C^{\text{R}}_+|^2}\notag\\
&\phantom{=}+ 2\Lambda_{+-}\sqbrack{C^{\text{L}}_+C^{\text{L}}_- - C^{\text{R}}_+C^{\text{R}}_-}\label{eq_lamb_shifted_energies_expansion}\\
\tilde{\tau} &= \tau + \sqbrack{\Lambda_{++} - \Lambda_{--}}C^{\text{L}}_+C^{\text{R}}_+ + \Lambda_{+-}\sqbrack{C^{\text{L}}_+C^{\text{R}}_- + C^{\text{L}}_-C^{\text{R}}_+}\notag.
\end{align}
The matrix elements
\begin{align}
\Lambda_{\chi\chi} &= \bra{\chi}\Lambda\ket{\chi}\notag\\
&= \sqbrack{F_{-\text{L}}(E_{\chi0},E_{0\chi}) + F_{+\text{R}}(E_{\chi\text{d}},E_{\text{d}\chi})}|C^{\text{L}}_\chi|^2\notag\\
&\phantom{=} + \sqbrack{F_{-\text{R}}(E_{\chi0},E_{0\chi}) + F_{+\text{L}}(E_{\chi\text{d}},E_{\text{d}\chi})}|C^{\text{R}}_\chi|^2\notag\\
\Lambda_{+-} &= \Lambda_{-+} = \bra{+}\Lambda\ket{-}\label{eq_lamb_shift_matrix_elements}\\
&= \sum_{\ell=\text{L,R}}\sqbrack{F_{-\ell}(E_{+ 0},E_{0-}) - F_{+\ell}(E_{+ \text{d}},E_{\text{d}-})}C^{\ell}_+C^{\ell}_-\notag
\end{align}
contain structure factors $F_{\eta\ell}$ given in \Eq{eq_lamb}. To calculate these, we assume the here relevant zero-temperature limit and model the barrier transparency as in \Eqs{eq_barrier_transparency},\eq{eq_sigmoid} for energies $E > \mu$; to enable the dot to be recharged for $E < \mu$, we also model the barrier as reopening for decreasing energies below $\mu$, choosing for convenience the same relative energy range $\delta\Eb = |\Eb - V_0|$ on which the barrier becomes transparent:
\begin{equation}
 \Gamma_\ell(E \leq \mu,t) = \Gamma_\ell\sqbrack{1 - S\left(E,\mu-\delta\Eb,\mu\right)}.
 \label{eq_barrier_transparency_reopening}
\end{equation}
Physically, the details of this barrier reopening is irrelevant for the emission process, apart from the mere fact that it brings the dot back to its original doubly occupied state at the end of the $\epsilon(t)$ driving cycle. However, \Eq{eq_barrier_transparency_reopening} formally enters $F_{+\ell}$ just as \Eq{eq_barrier_transparency} enters $F_{-\ell}$:
\begin{align}
F_{+\ell}(E_{\chi\text{d}},E_{\text{d}\chi}) &\equalby{T\rightarrow 0} -\frac{1}{2\pi}\mathcal{P}\int_{-\infty}^{\mu} d\omega\frac{\Gamma_{\ell}(\omega)}{\omega-E_{\text{d}\chi}}\notag\\
&= -\frac{1}{2\pi}\mathcal{P}\int_{\mu-\delta\Eb}^{\mu} d\omega\frac{\Gamma_{\ell}(\omega)}{\omega-E_{\text{d}\chi}}\notag\\
&\phantom{=} -\lim_{|D|\rightarrow\infty}\frac{\Gamma_\ell}{2\pi}\ln\left|\frac{\mu - \delta\Eb - E_{\text{d}\chi}}{|D|}\right| \label{eq_lamb_hot_electron_plus}\\
F_{-\ell}(E_{\chi0},E_{0\chi}) &\equalby{T\rightarrow 0} \frac{1}{2\pi}\mathcal{P}\int^{\infty}_{\mu} d\omega\frac{\Gamma_{\ell}(\omega)}{\omega - E_{\chi}}\notag\\
 &= \frac{1}{2\pi}\mathcal{P}\int^{\Eb}_{V_0} d\omega\frac{\Gamma_{\ell}(\omega)}{\omega - E_{\chi}}\notag\\
 &\phantom{=} -\lim_{|D|\rightarrow\infty}\frac{\Gamma_\ell}{2\pi}\ln\left|\frac{\Eb - E_\chi}{|D|}\right|\label{eq_lamb_hot_electron_minus}\\
 F_{+\ell}(E_{+\text{d}},E_{\text{d}-}) &\equalby{T\rightarrow 0} \mathcal{P}\int_{-\infty}^{\mu - E_{\text{d}-}} d\omega\frac{\prod_{\chi=\pm}\sqrt{\Gamma_{\ell}(\omega + E_{\text{d}\chi})}}{-2\pi\omega}\notag\\
 &= -\mathcal{P}\int_{\mu - \delta E_{\text{b}}}^{\mu} d\omega\frac{\sqrt{\Gamma_{\ell}(\omega)}\sqrt{\Gamma_{\ell}(\omega - E_{+-})}}{2\pi(\omega - E_{\text{d}-})}\notag\\
 &\phantom{=} -\lim_{|D|\rightarrow\infty}\frac{\Gamma_\ell}{2\pi}\ln\left|\frac{\mu - \delta E_{\text{b}} - E_{\text{d}-}}{|D|}\right|\label{eq_lamb_hot_electron_coherent_plus}\\
 F_{-\ell}(E_{+0},E_{0-}) &\equalby{T\rightarrow 0} \mathcal{P}\int^{\infty}_{\mu - E_-} d\omega\frac{\prod_{\chi=\pm}\sqrt{\Gamma_{\ell}(\omega + E_\chi)}}{2\pi\omega}\notag\\
 &= \mathcal{P}\int^{\Eb}_{V_0} d\omega\frac{\sqrt{\Gamma_{\ell}(\omega)}\sqrt{\Gamma_{\ell}(\omega + E_{+-})}}{2\pi(\omega - E_-)}\notag\\
 &\phantom{=} -\lim_{|D|\rightarrow\infty}\frac{\Gamma_\ell}{2\pi}\ln\left|\frac{\Eb - E_-}{|D|}\right|\label{eq_lamb_hot_electron_coherent_minus}.
\end{align}
It is typical for such principal value $(\mathcal{P})$ integrals \eq{eq_lamb_hot_electron_plus}-\eq{eq_lamb_hot_electron_coherent_minus} to have a logarithmic divergence for infinite bandwidth $|D|\rightarrow\infty$. We stress, however, that the shifted splitting and coupling \BrackEq{eq_lamb_shifted_energies_expansion} only involve \emph{differences} of these integrals in which all divergent terms cancel out. Furthermore, the poles $E_\chi,E_{\text{d}\chi}$ avoided by principal value integration are either outside the finite support of the respective $\Gamma_\ell(E)$ or fully enclosed by the integration interval; their contribution thus mostly cancels out due to the sign switch. Indeed, as confirmed by our numerical results, the effective contribution to $\tilde{\tau},\Delta\tilde{\epsilon}$ is then $\sim\Gamma = \sum_{\ell}\Gamma_\ell$.

Note that while we evaluate the functions \eq{eq_lamb_hot_electron_plus}-\eq{eq_lamb_hot_electron_coherent_minus} numerically as described below, the above analytical expressions already highlight a few special cases for $\Lambda$. 
First, we consider $\tau = 0$, which means that the eigenvectors $\ket{\pm}$ are the bare dot states $\ket{\text{L/R}}$. We then have $C^\ell_+C^\ell_- = C^\text{L}_+C^{\text{L}}_+ = 0$, which according to \Eqs{eq_lamb_shifted_energies_expansion},\eq{eq_lamb_shift_matrix_elements} results in $\tilde{\tau} = \tau = 0$. Analogously, $\Deps = 0$ and $\tau > 0$ implies perfectly L-R (anti-)symmetric states $|C^\text{L}_\chi| = |C^\text{R}_\chi|$ and therefore $\Delta\tilde{\epsilon} = \Deps = 0$ according to \Eq{eq_lamb_shifted_energies_expansion}. 
This, importantly, proves that $\Lambda$ cannot rotate the single-particle eigenbasis if the latter is given by the dot states L,R or by the corresponding bonding and antibonding states. In case of $\tau = 0$, \Eq{eq_lamb_shift_matrix_elements} furthermore predicts the proportionality $(\Delta\tilde{\epsilon} - \Deps) \sim (\Lambda_{++} - \Lambda_{--}) \sim (\Gamma_{\text{L}} - \Gamma_{\text{R}})$, meaning that the Lamb shift then scales with the environment coupling asymmetry between the two bare dot states. For the setup studied in \Fig{fig_en_diffs_barrier_potential_dependent} and for all curves of \Fig{fig_etr} except the light grey one, the Lamb shift therefore vanishes exactly, $\Delta\tilde{\epsilon} = \Deps, \tilde{\tau} = \tau$.

We emphasize again that all system parameters entering \Eqs{eq_energies}-\eq{eq_rates_functions} can in principle be time-dependent in the way discussed in \Sec{sec_low_energy_theory}. We therefore need to solve the master equation \eq{eq_master_vector} numerically, which we achieve with a standard fourth-order Runge-Kutta integration scheme~\cite{SchulenborgZen2023Apr}. We use $\Nt = 2^{22} = 4194304$ time steps over a single period $2\pi/\Omega$ to create a single current trace as in \FigPart{fig_setting_0d}{a}, and substitute the result into \Eq{eq_averages} to obtain the emission times and energies $(t_x,\Emitx)$ for $x = 1,2$. This procedure is repeated for $\Nsw = 2^{9} = 512$ equidistant $\epso$-points in the closed range $[0,35U]$ to determine the energy-time relations $\Emitot(t)$ and $\Emitzo(t)$ as sketched in \FigPart{fig_setting_0d}{b}. Note that equidistant $\epso$-steps do not translate into equidistant time steps for these $\Emitx(t)$-curves. To determine the apparent transition energy $\Delta\Etr$ \BrackEq{eq_apparent_transition_diff}, we hence first pick $\Emitot$ at a given time $t$, and then sample $\Emitzo$ by interpolating between $\Emitzo(t_<)$ and $\Emitzo(t_>)$ with $t_<$ and $t_>$ being the closest times before and after $t$, so that $t_< \leq t \leq t_>$. 

Finally, to obtain the Lamb shift $\Lambda$, we use Simpson's 1/3 rule with $N_\text{Simp} = 2^{17} + 1 = 131073$ equidistant points to evaluate the principal value integrals \eq{eq_lamb_hot_electron_plus}-\eq{eq_lamb_hot_electron_coherent_minus}. Importantly, if any resonance lies within the integration interval, we implement the principal-value integration to achieve stable convergence by placing these points symmetrically around this resonance. Since the integrals need to be obtained for a large range of $E_{\chi}$ or $E_{d\chi}$ in many different driving protocols, we reduce the number of required computations by precomputing the integrals for $N_{\text{Int}} = 2^{23} = 8388608$ equidistant energy points in a range $[-2.5A,2.5A]$. The matrix elements \Eq{eq_lamb_shift_matrix_elements} are then obtained by linearly interpolating between these points. A further reduction of the integral computation time is achieved by likewise linearly interpolating between $N_{\Gamma} = 2^{18} = 262144$ precomputed points of the $\Gamma(E)$ functions. We have checked for convergence by individually increasing the above stated number of points and comparing the results.

\section{Energy current kernel} \label{sec_current_kernel}

The matrix representation $\mathbf{W}^{I,E}$ of the energy current kernel $W^{I,E}$ yielding \Eq{eq_currents} is derived from the full matrix $\mathbf{W}$ \BrackEq{eq_master_vector}. It turns out that in the regimes studied here, this can be done by neglecting all transitions between (instantaneous) dot energy eigenstates of $H(t)$ \BrackEq{eq_hamiltonian} with equal particle number. We calculate $\mathbf{W}^{I,E}$ in three steps: first we rotate the single-particle sector of $\mathbf{W}$ from the L,R basis to the single-particle eigenbasis \eq{eq_states}:
\begin{equation}
\mathbf{V} = \mathbf{O}\times\mathbf{W}\times\mathbf{O}^{-1}
\end{equation}
with the transformation matrix elements
\begin{equation}
O_{\chi\chi',\ell\ell'} = \sqbrack{\mathbf{O}}_{\chi\chi',\ell\ell'}  = \braket{\chi}{\ell}\braket{\ell'}{\chi'},
\end{equation}
between the basis states $\ell,\ell'\in\cbrack{0,\text{L},\text{R},\text{d}}$  and the eigenstates $\chi,\chi'\in\cbrack{0,-,+,\text{d}}$. Second, we set rates for transitions with constant particle number to zero:
\begin{align}
V_{\chi_1\chi_2,\chi_3\chi_4} \rightarrow 0 \lrsepa \chi_{1,2,3,4} \in \cbrack{-,+},\label{eq_set_to_zero}
\end{align}
where either $\chi_1 \neq \chi_3$ or $\chi_2 \neq \chi_4$.
The third and final step is to rotate back to the left-right located basis:
\begin{equation}
\mathbf{W}^{I,E} = \mathbf{O}^{-1}\times\mathbf{V}\times\mathbf{O}.
\end{equation}
Strictly speaking, \Eq{eq_set_to_zero} may introduce deviations from the trace preservation rule $\sum_i V_{i,\chi\chi'} = 0$ for the probability-coherence couplings\footnote{The rates $V_{-,+} = V_{+,-}$ already vanish due to weak coupling.}. However, since our manuscript always considers either single-particle degeneracy $\Deps = 0$ or large energy splittings $|E_+ - E_-| \gg |\Eb - V_0|$ compared to the barrier broadening, these transition rates are always negligible: we observe several orders of magnitude smaller rates $V_{\chi_1\chi_2,\chi_3\chi_4}$ compared to the transitions $V_{0,\chi\chi'},V_{\chi\chi',0},V_{\text{d},\chi\chi'},V_{\chi\chi',\text{d}}$ involving a particle number change. This justifies \Eq{eq_set_to_zero} for all emission processes studied in this paper, and shows that energy transfer in these situations is always tied to particle transfer. This also implies $W^{I,E}\times\rho \approx W\times\rho = \partial_t\rho$ to a very good approximation.


\begin{thebibliography}{67}%
\makeatletter
\providecommand \@ifxundefined [1]{%
 \@ifx{#1\undefined}
}%
\providecommand \@ifnum [1]{%
 \ifnum #1\expandafter \@firstoftwo
 \else \expandafter \@secondoftwo
 \fi
}%
\providecommand \@ifx [1]{%
 \ifx #1\expandafter \@firstoftwo
 \else \expandafter \@secondoftwo
 \fi
}%
\providecommand \natexlab [1]{#1}%
\providecommand \enquote  [1]{``#1''}%
\providecommand \bibnamefont  [1]{#1}%
\providecommand \bibfnamefont [1]{#1}%
\providecommand \citenamefont [1]{#1}%
\providecommand \href@noop [0]{\@secondoftwo}%
\providecommand \href [0]{\begingroup \@sanitize@url \@href}%
\providecommand \@href[1]{\@@startlink{#1}\@@href}%
\providecommand \@@href[1]{\endgroup#1\@@endlink}%
\providecommand \@sanitize@url [0]{\catcode `\\12\catcode `\$12\catcode
  `\&12\catcode `\#12\catcode `\^12\catcode `\_12\catcode `\%12\relax}%
\providecommand \@@startlink[1]{}%
\providecommand \@@endlink[0]{}%
\providecommand \url  [0]{\begingroup\@sanitize@url \@url }%
\providecommand \@url [1]{\endgroup\@href {#1}{\urlprefix }}%
\providecommand \urlprefix  [0]{URL }%
\providecommand \Eprint [0]{\href }%
\providecommand \doibase [0]{https://doi.org/}%
\providecommand \selectlanguage [0]{\@gobble}%
\providecommand \bibinfo  [0]{\@secondoftwo}%
\providecommand \bibfield  [0]{\@secondoftwo}%
\providecommand \translation [1]{[#1]}%
\providecommand \BibitemOpen [0]{}%
\providecommand \bibitemStop [0]{}%
\providecommand \bibitemNoStop [0]{.\EOS\space}%
\providecommand \EOS [0]{\spacefactor3000\relax}%
\providecommand \BibitemShut  [1]{\csname bibitem#1\endcsname}%
\let\auto@bib@innerbib\@empty
\bibitem [{\citenamefont {F{\ifmmode\grave{e}\else\`{e}\fi}ve}\ \emph
  {et~al.}(2007)\citenamefont {F{\ifmmode\grave{e}\else\`{e}\fi}ve},
  \citenamefont {Mah{\'{e}}}, \citenamefont {Berroir}, \citenamefont {Kontos},
  \citenamefont {Pla{\c{c}}ais}, \citenamefont {Glattli}, \citenamefont
  {Cavanna}, \citenamefont {Etienne},\ and\ \citenamefont {Jin}}]{Feve2007May}%
  \BibitemOpen
  \bibfield  {author} {\bibinfo {author} {\bibfnamefont {G.}~\bibnamefont
  {F{\ifmmode\grave{e}\else\`{e}\fi}ve}}, \bibinfo {author} {\bibfnamefont
  {A.}~\bibnamefont {Mah{\'{e}}}}, \bibinfo {author} {\bibfnamefont {J.-M.}\
  \bibnamefont {Berroir}}, \bibinfo {author} {\bibfnamefont {T.}~\bibnamefont
  {Kontos}}, \bibinfo {author} {\bibfnamefont {B.}~\bibnamefont
  {Pla{\c{c}}ais}}, \bibinfo {author} {\bibfnamefont {D.~C.}\ \bibnamefont
  {Glattli}}, \bibinfo {author} {\bibfnamefont {A.}~\bibnamefont {Cavanna}},
  \bibinfo {author} {\bibfnamefont {B.}~\bibnamefont {Etienne}},\ and\ \bibinfo
  {author} {\bibfnamefont {Y.}~\bibnamefont {Jin}},\ }\bibfield  {title}
  {\bibinfo {title} {{An On-Demand Coherent Single-Electron Source}},\ }\href
  {https://doi.org/10.1126/science.1141243} {\bibfield  {journal} {\bibinfo
  {journal} {Science}\ }\textbf {\bibinfo {volume} {316}},\ \bibinfo {pages}
  {1169--1172} (\bibinfo {year} {2007})}\BibitemShut {NoStop}%
\bibitem [{\citenamefont {Blumenthal}\ \emph {et~al.}(2007)\citenamefont
  {Blumenthal}, \citenamefont {Kaestner}, \citenamefont {Li}, \citenamefont
  {Giblin}, \citenamefont {Janssen}, \citenamefont {Pepper}, \citenamefont
  {Anderson}, \citenamefont {Jones},\ and\ \citenamefont
  {Ritchie}}]{Blumenthal2007May}%
  \BibitemOpen
  \bibfield  {author} {\bibinfo {author} {\bibfnamefont {M.~D.}\ \bibnamefont
  {Blumenthal}}, \bibinfo {author} {\bibfnamefont {B.}~\bibnamefont
  {Kaestner}}, \bibinfo {author} {\bibfnamefont {L.}~\bibnamefont {Li}},
  \bibinfo {author} {\bibfnamefont {S.}~\bibnamefont {Giblin}}, \bibinfo
  {author} {\bibfnamefont {T.~J. B.~M.}\ \bibnamefont {Janssen}}, \bibinfo
  {author} {\bibfnamefont {M.}~\bibnamefont {Pepper}}, \bibinfo {author}
  {\bibfnamefont {D.}~\bibnamefont {Anderson}}, \bibinfo {author}
  {\bibfnamefont {G.}~\bibnamefont {Jones}},\ and\ \bibinfo {author}
  {\bibfnamefont {D.~A.}\ \bibnamefont {Ritchie}},\ }\bibfield  {title}
  {\bibinfo {title} {{Gigahertz quantized charge pumping}},\ }\href
  {https://doi.org/10.1038/nphys582} {\bibfield  {journal} {\bibinfo  {journal}
  {Nat. Phys.}\ }\textbf {\bibinfo {volume} {3}},\ \bibinfo {pages} {343--347}
  (\bibinfo {year} {2007})}\BibitemShut {NoStop}%
\bibitem [{\citenamefont {Dubois}\ \emph {et~al.}(2013)\citenamefont {Dubois},
  \citenamefont {Jullien}, \citenamefont {Portier}, \citenamefont {Roche},
  \citenamefont {Cavanna}, \citenamefont {Jin}, \citenamefont {Wegscheider},
  \citenamefont {Roulleau},\ and\ \citenamefont {Glattli}}]{Dubois2013Oct}%
  \BibitemOpen
  \bibfield  {author} {\bibinfo {author} {\bibfnamefont {J.}~\bibnamefont
  {Dubois}}, \bibinfo {author} {\bibfnamefont {T.}~\bibnamefont {Jullien}},
  \bibinfo {author} {\bibfnamefont {F.}~\bibnamefont {Portier}}, \bibinfo
  {author} {\bibfnamefont {P.}~\bibnamefont {Roche}}, \bibinfo {author}
  {\bibfnamefont {A.}~\bibnamefont {Cavanna}}, \bibinfo {author} {\bibfnamefont
  {Y.}~\bibnamefont {Jin}}, \bibinfo {author} {\bibfnamefont {W.}~\bibnamefont
  {Wegscheider}}, \bibinfo {author} {\bibfnamefont {P.}~\bibnamefont
  {Roulleau}},\ and\ \bibinfo {author} {\bibfnamefont {D.~C.}\ \bibnamefont
  {Glattli}},\ }\bibfield  {title} {\bibinfo {title} {{Minimal-excitation
  states for electron quantum optics using levitons}},\ }\href
  {https://doi.org/10.1038/nature12713} {\bibfield  {journal} {\bibinfo
  {journal} {Nature}\ }\textbf {\bibinfo {volume} {502}},\ \bibinfo {pages}
  {659--663} (\bibinfo {year} {2013})}\BibitemShut {NoStop}%
\bibitem [{\citenamefont {Edlbauer}\ \emph {et~al.}(2022)\citenamefont
  {Edlbauer}, \citenamefont {Wang}, \citenamefont {Crozes}, \citenamefont
  {Perrier}, \citenamefont {Ouacel}, \citenamefont {Geffroy}, \citenamefont
  {Georgiou}, \citenamefont {Chatzikyriakou}, \citenamefont {Lacerda-Santos},
  \citenamefont {Waintal}, \citenamefont {Glattli}, \citenamefont {Roulleau},
  \citenamefont {Nath}, \citenamefont {Kataoka}, \citenamefont
  {Splettstoesser}, \citenamefont {Acciai}, \citenamefont {da~Silva~Figueira},
  \citenamefont {{\ifmmode\ddot{O}\else\"{O}\fi}ztas}, \citenamefont
  {Trellakis}, \citenamefont {Grange}, \citenamefont {Yevtushenko},
  \citenamefont {Birner},\ and\ \citenamefont
  {B{\ifmmode\ddot{a}\else\"{a}\fi}uerle}}]{Edlbauer2022Dec}%
  \BibitemOpen
  \bibfield  {author} {\bibinfo {author} {\bibfnamefont {H.}~\bibnamefont
  {Edlbauer}}, \bibinfo {author} {\bibfnamefont {J.}~\bibnamefont {Wang}},
  \bibinfo {author} {\bibfnamefont {T.}~\bibnamefont {Crozes}}, \bibinfo
  {author} {\bibfnamefont {P.}~\bibnamefont {Perrier}}, \bibinfo {author}
  {\bibfnamefont {S.}~\bibnamefont {Ouacel}}, \bibinfo {author} {\bibfnamefont
  {C.}~\bibnamefont {Geffroy}}, \bibinfo {author} {\bibfnamefont
  {G.}~\bibnamefont {Georgiou}}, \bibinfo {author} {\bibfnamefont
  {E.}~\bibnamefont {Chatzikyriakou}}, \bibinfo {author} {\bibfnamefont
  {A.}~\bibnamefont {Lacerda-Santos}}, \bibinfo {author} {\bibfnamefont
  {X.}~\bibnamefont {Waintal}}, \bibinfo {author} {\bibfnamefont {D.~C.}\
  \bibnamefont {Glattli}}, \bibinfo {author} {\bibfnamefont {P.}~\bibnamefont
  {Roulleau}}, \bibinfo {author} {\bibfnamefont {J.}~\bibnamefont {Nath}},
  \bibinfo {author} {\bibfnamefont {M.}~\bibnamefont {Kataoka}}, \bibinfo
  {author} {\bibfnamefont {J.}~\bibnamefont {Splettstoesser}}, \bibinfo
  {author} {\bibfnamefont {M.}~\bibnamefont {Acciai}}, \bibinfo {author}
  {\bibfnamefont {M.~C.}\ \bibnamefont {da~Silva~Figueira}}, \bibinfo {author}
  {\bibfnamefont {K.}~\bibnamefont {{\ifmmode\ddot{O}\else\"{O}\fi}ztas}},
  \bibinfo {author} {\bibfnamefont {A.}~\bibnamefont {Trellakis}}, \bibinfo
  {author} {\bibfnamefont {T.}~\bibnamefont {Grange}}, \bibinfo {author}
  {\bibfnamefont {O.~M.}\ \bibnamefont {Yevtushenko}}, \bibinfo {author}
  {\bibfnamefont {S.}~\bibnamefont {Birner}},\ and\ \bibinfo {author}
  {\bibfnamefont {C.}~\bibnamefont {B{\ifmmode\ddot{a}\else\"{a}\fi}uerle}},\
  }\bibfield  {title} {\bibinfo {title} {{Semiconductor-based electron flying
  qubits: review on recent progress accelerated by numerical modelling}},\
  }\href {https://doi.org/10.1140/epjqt/s40507-022-00139-w} {\bibfield
  {journal} {\bibinfo  {journal} {EPJ Quantum Technol.}\ }\textbf {\bibinfo
  {volume} {9}},\ \bibinfo {pages} {1--36} (\bibinfo {year}
  {2022})}\BibitemShut {NoStop}%
\bibitem [{\citenamefont {Bocquillon}\ \emph {et~al.}(2012)\citenamefont
  {Bocquillon}, \citenamefont {Parmentier}, \citenamefont {Grenier},
  \citenamefont {Berroir}, \citenamefont {Degiovanni}, \citenamefont {Glattli},
  \citenamefont {Pla{\ifmmode\mbox{\c{c}}\else\c{c}\fi}ais}, \citenamefont
  {Cavanna}, \citenamefont {Jin},\ and\ \citenamefont
  {F{\ifmmode\grave{e}\else\`{e}\fi}ve}}]{Bocquillon2012May}%
  \BibitemOpen
  \bibfield  {author} {\bibinfo {author} {\bibfnamefont {E.}~\bibnamefont
  {Bocquillon}}, \bibinfo {author} {\bibfnamefont {F.~D.}\ \bibnamefont
  {Parmentier}}, \bibinfo {author} {\bibfnamefont {C.}~\bibnamefont {Grenier}},
  \bibinfo {author} {\bibfnamefont {J.-M.}\ \bibnamefont {Berroir}}, \bibinfo
  {author} {\bibfnamefont {P.}~\bibnamefont {Degiovanni}}, \bibinfo {author}
  {\bibfnamefont {D.~C.}\ \bibnamefont {Glattli}}, \bibinfo {author}
  {\bibfnamefont {B.}~\bibnamefont
  {Pla{\ifmmode\mbox{\c{c}}\else\c{c}\fi}ais}}, \bibinfo {author}
  {\bibfnamefont {A.}~\bibnamefont {Cavanna}}, \bibinfo {author} {\bibfnamefont
  {Y.}~\bibnamefont {Jin}},\ and\ \bibinfo {author} {\bibfnamefont
  {G.}~\bibnamefont {F{\ifmmode\grave{e}\else\`{e}\fi}ve}},\ }\bibfield
  {title} {\bibinfo {title} {{Electron Quantum Optics: Partitioning Electrons
  One by One}},\ }\href {https://doi.org/10.1103/PhysRevLett.108.196803}
  {\bibfield  {journal} {\bibinfo  {journal} {Phys. Rev. Lett.}\ }\textbf
  {\bibinfo {volume} {108}},\ \bibinfo {pages} {196803} (\bibinfo {year}
  {2012})}\BibitemShut {NoStop}%
\bibitem [{\citenamefont {Bocquillon}\ \emph {et~al.}(2013)\citenamefont
  {Bocquillon}, \citenamefont {Freulon}, \citenamefont {Berroir}, \citenamefont
  {Degiovanni}, \citenamefont {Pla{\ifmmode\mbox{\c{c}}\else\c{c}\fi}ais},
  \citenamefont {Cavanna}, \citenamefont {Jin},\ and\ \citenamefont
  {F{\ifmmode\grave{e}\else\`{e}\fi}ve}}]{Bocquillon2013Mar}%
  \BibitemOpen
  \bibfield  {author} {\bibinfo {author} {\bibfnamefont {E.}~\bibnamefont
  {Bocquillon}}, \bibinfo {author} {\bibfnamefont {V.}~\bibnamefont {Freulon}},
  \bibinfo {author} {\bibfnamefont {J.-M.}\ \bibnamefont {Berroir}}, \bibinfo
  {author} {\bibfnamefont {P.}~\bibnamefont {Degiovanni}}, \bibinfo {author}
  {\bibfnamefont {B.}~\bibnamefont
  {Pla{\ifmmode\mbox{\c{c}}\else\c{c}\fi}ais}}, \bibinfo {author}
  {\bibfnamefont {A.}~\bibnamefont {Cavanna}}, \bibinfo {author} {\bibfnamefont
  {Y.}~\bibnamefont {Jin}},\ and\ \bibinfo {author} {\bibfnamefont
  {G.}~\bibnamefont {F{\ifmmode\grave{e}\else\`{e}\fi}ve}},\ }\bibfield
  {title} {\bibinfo {title} {{Coherence and Indistinguishability of Single
  Electrons Emitted by Independent Sources}},\ }\href
  {https://doi.org/10.1126/science.1232572} {\bibfield  {journal} {\bibinfo
  {journal} {Science}\ }\textbf {\bibinfo {volume} {339}},\ \bibinfo {pages}
  {1054--1057} (\bibinfo {year} {2013})}\BibitemShut {NoStop}%
\bibitem [{\citenamefont {Fletcher}\ \emph {et~al.}(2013)\citenamefont
  {Fletcher}, \citenamefont {See}, \citenamefont {Howe}, \citenamefont
  {Pepper}, \citenamefont {Giblin}, \citenamefont {Griffiths}, \citenamefont
  {Jones}, \citenamefont {Farrer}, \citenamefont {Ritchie}, \citenamefont
  {Janssen},\ and\ \citenamefont {Kataoka}}]{Fletcher2013Nov}%
  \BibitemOpen
  \bibfield  {author} {\bibinfo {author} {\bibfnamefont {J.~D.}\ \bibnamefont
  {Fletcher}}, \bibinfo {author} {\bibfnamefont {P.}~\bibnamefont {See}},
  \bibinfo {author} {\bibfnamefont {H.}~\bibnamefont {Howe}}, \bibinfo {author}
  {\bibfnamefont {M.}~\bibnamefont {Pepper}}, \bibinfo {author} {\bibfnamefont
  {S.~P.}\ \bibnamefont {Giblin}}, \bibinfo {author} {\bibfnamefont {J.~P.}\
  \bibnamefont {Griffiths}}, \bibinfo {author} {\bibfnamefont {G.~A.~C.}\
  \bibnamefont {Jones}}, \bibinfo {author} {\bibfnamefont {I.}~\bibnamefont
  {Farrer}}, \bibinfo {author} {\bibfnamefont {D.~A.}\ \bibnamefont {Ritchie}},
  \bibinfo {author} {\bibfnamefont {T.~J. B.~M.}\ \bibnamefont {Janssen}},\
  and\ \bibinfo {author} {\bibfnamefont {M.}~\bibnamefont {Kataoka}},\
  }\bibfield  {title} {\bibinfo {title} {{Clock-Controlled Emission of
  Single-Electron Wave Packets in a Solid-State Circuit}},\ }\href
  {https://doi.org/10.1103/PhysRevLett.111.216807} {\bibfield  {journal}
  {\bibinfo  {journal} {Phys. Rev. Lett.}\ }\textbf {\bibinfo {volume} {111}},\
  \bibinfo {pages} {216807} (\bibinfo {year} {2013})}\BibitemShut {NoStop}%
\bibitem [{\citenamefont {Ubbelohde}\ \emph {et~al.}(2015)\citenamefont
  {Ubbelohde}, \citenamefont {Hohls}, \citenamefont {Kashcheyevs},
  \citenamefont {Wagner}, \citenamefont {Fricke}, \citenamefont
  {K{\ifmmode\ddot{a}\else\"{a}\fi}stner}, \citenamefont {Pierz}, \citenamefont
  {Schumacher},\ and\ \citenamefont {Haug}}]{Ubbelohde2015Jan}%
  \BibitemOpen
  \bibfield  {author} {\bibinfo {author} {\bibfnamefont {N.}~\bibnamefont
  {Ubbelohde}}, \bibinfo {author} {\bibfnamefont {F.}~\bibnamefont {Hohls}},
  \bibinfo {author} {\bibfnamefont {V.}~\bibnamefont {Kashcheyevs}}, \bibinfo
  {author} {\bibfnamefont {T.}~\bibnamefont {Wagner}}, \bibinfo {author}
  {\bibfnamefont {L.}~\bibnamefont {Fricke}}, \bibinfo {author} {\bibfnamefont
  {B.}~\bibnamefont {K{\ifmmode\ddot{a}\else\"{a}\fi}stner}}, \bibinfo {author}
  {\bibfnamefont {K.}~\bibnamefont {Pierz}}, \bibinfo {author} {\bibfnamefont
  {H.~W.}\ \bibnamefont {Schumacher}},\ and\ \bibinfo {author} {\bibfnamefont
  {R.~J.}\ \bibnamefont {Haug}},\ }\bibfield  {title} {\bibinfo {title}
  {{Partitioning of on-demand electron pairs}},\ }\href
  {https://doi.org/10.1038/nnano.2014.275} {\bibfield  {journal} {\bibinfo
  {journal} {Nat. Nanotechnol.}\ }\textbf {\bibinfo {volume} {10}},\ \bibinfo
  {pages} {46--49} (\bibinfo {year} {2015})}\BibitemShut {NoStop}%
\bibitem [{\citenamefont {Waldie}\ \emph {et~al.}(2015)\citenamefont {Waldie},
  \citenamefont {See}, \citenamefont {Kashcheyevs}, \citenamefont {Griffiths},
  \citenamefont {Farrer}, \citenamefont {Jones}, \citenamefont {Ritchie},
  \citenamefont {Janssen},\ and\ \citenamefont {Kataoka}}]{Waldie2015Sep}%
  \BibitemOpen
  \bibfield  {author} {\bibinfo {author} {\bibfnamefont {J.}~\bibnamefont
  {Waldie}}, \bibinfo {author} {\bibfnamefont {P.}~\bibnamefont {See}},
  \bibinfo {author} {\bibfnamefont {V.}~\bibnamefont {Kashcheyevs}}, \bibinfo
  {author} {\bibfnamefont {J.~P.}\ \bibnamefont {Griffiths}}, \bibinfo {author}
  {\bibfnamefont {I.}~\bibnamefont {Farrer}}, \bibinfo {author} {\bibfnamefont
  {G.~A.~C.}\ \bibnamefont {Jones}}, \bibinfo {author} {\bibfnamefont {D.~A.}\
  \bibnamefont {Ritchie}}, \bibinfo {author} {\bibfnamefont {T.~J. B.~M.}\
  \bibnamefont {Janssen}},\ and\ \bibinfo {author} {\bibfnamefont
  {M.}~\bibnamefont {Kataoka}},\ }\bibfield  {title} {\bibinfo {title}
  {{Measurement and control of electron wave packets from a single-electron
  source}},\ }\href {https://doi.org/10.1103/PhysRevB.92.125305} {\bibfield
  {journal} {\bibinfo  {journal} {Phys. Rev. B}\ }\textbf {\bibinfo {volume}
  {92}},\ \bibinfo {pages} {125305} (\bibinfo {year} {2015})}\BibitemShut
  {NoStop}%
\bibitem [{\citenamefont {Fletcher}\ \emph {et~al.}(2019)\citenamefont
  {Fletcher}, \citenamefont {Johnson}, \citenamefont {Locane}, \citenamefont
  {See}, \citenamefont {Griffiths}, \citenamefont {Farrer}, \citenamefont
  {Ritchie}, \citenamefont {Brouwer}, \citenamefont {Kashcheyevs},\ and\
  \citenamefont {Kataoka}}]{Fletcher2019Nov}%
  \BibitemOpen
  \bibfield  {author} {\bibinfo {author} {\bibfnamefont {J.~D.}\ \bibnamefont
  {Fletcher}}, \bibinfo {author} {\bibfnamefont {N.}~\bibnamefont {Johnson}},
  \bibinfo {author} {\bibfnamefont {E.}~\bibnamefont {Locane}}, \bibinfo
  {author} {\bibfnamefont {P.}~\bibnamefont {See}}, \bibinfo {author}
  {\bibfnamefont {J.~P.}\ \bibnamefont {Griffiths}}, \bibinfo {author}
  {\bibfnamefont {I.}~\bibnamefont {Farrer}}, \bibinfo {author} {\bibfnamefont
  {D.~A.}\ \bibnamefont {Ritchie}}, \bibinfo {author} {\bibfnamefont {P.~W.}\
  \bibnamefont {Brouwer}}, \bibinfo {author} {\bibfnamefont {V.}~\bibnamefont
  {Kashcheyevs}},\ and\ \bibinfo {author} {\bibfnamefont {M.}~\bibnamefont
  {Kataoka}},\ }\bibfield  {title} {\bibinfo {title} {{Continuous-variable
  tomography of solitary electrons}},\ }\href
  {https://doi.org/10.1038/s41467-019-13222-1} {\bibfield  {journal} {\bibinfo
  {journal} {Nat. Commun.}\ }\textbf {\bibinfo {volume} {10}},\ \bibinfo
  {pages} {1--7} (\bibinfo {year} {2019})}\BibitemShut {NoStop}%
\bibitem [{\citenamefont {Jullien}\ \emph {et~al.}(2014)\citenamefont
  {Jullien}, \citenamefont {Roulleau}, \citenamefont {Roche}, \citenamefont
  {Cavanna}, \citenamefont {Jin},\ and\ \citenamefont
  {Glattli}}]{Jullien2014Oct}%
  \BibitemOpen
  \bibfield  {author} {\bibinfo {author} {\bibfnamefont {T.}~\bibnamefont
  {Jullien}}, \bibinfo {author} {\bibfnamefont {P.}~\bibnamefont {Roulleau}},
  \bibinfo {author} {\bibfnamefont {B.}~\bibnamefont {Roche}}, \bibinfo
  {author} {\bibfnamefont {A.}~\bibnamefont {Cavanna}}, \bibinfo {author}
  {\bibfnamefont {Y.}~\bibnamefont {Jin}},\ and\ \bibinfo {author}
  {\bibfnamefont {D.~C.}\ \bibnamefont {Glattli}},\ }\bibfield  {title}
  {\bibinfo {title} {{Quantum tomography of an electron}},\ }\href
  {https://doi.org/10.1038/nature13821} {\bibfield  {journal} {\bibinfo
  {journal} {Nature}\ }\textbf {\bibinfo {volume} {514}},\ \bibinfo {pages}
  {603--607} (\bibinfo {year} {2014})}\BibitemShut {NoStop}%
\bibitem [{\citenamefont {Thibierge}\ \emph {et~al.}(2016)\citenamefont
  {Thibierge}, \citenamefont {Ferraro}, \citenamefont {Roussel}, \citenamefont
  {Cabart}, \citenamefont {Marguerite}, \citenamefont
  {F{\ifmmode\grave{e}\else\`{e}\fi}ve},\ and\ \citenamefont
  {Degiovanni}}]{Thibierge2016Feb}%
  \BibitemOpen
  \bibfield  {author} {\bibinfo {author} {\bibfnamefont
  {{\ifmmode\acute{E}\else\'{E}\fi}.}~\bibnamefont {Thibierge}}, \bibinfo
  {author} {\bibfnamefont {D.}~\bibnamefont {Ferraro}}, \bibinfo {author}
  {\bibfnamefont {B.}~\bibnamefont {Roussel}}, \bibinfo {author} {\bibfnamefont
  {C.}~\bibnamefont {Cabart}}, \bibinfo {author} {\bibfnamefont
  {A.}~\bibnamefont {Marguerite}}, \bibinfo {author} {\bibfnamefont
  {G.}~\bibnamefont {F{\ifmmode\grave{e}\else\`{e}\fi}ve}},\ and\ \bibinfo
  {author} {\bibfnamefont {P.}~\bibnamefont {Degiovanni}},\ }\bibfield  {title}
  {\bibinfo {title} {{Two-electron coherence and its measurement in electron
  quantum optics}},\ }\href {https://doi.org/10.1103/PhysRevB.93.081302}
  {\bibfield  {journal} {\bibinfo  {journal} {Phys. Rev. B}\ }\textbf {\bibinfo
  {volume} {93}},\ \bibinfo {pages} {081302} (\bibinfo {year}
  {2016})}\BibitemShut {NoStop}%
\bibitem [{\citenamefont {Bisognin}\ \emph {et~al.}(2019)\citenamefont
  {Bisognin}, \citenamefont {Marguerite}, \citenamefont {Roussel},
  \citenamefont {Kumar}, \citenamefont {Cabart}, \citenamefont {Chapdelaine},
  \citenamefont {Mohammad-Djafari}, \citenamefont {Berroir}, \citenamefont
  {Bocquillon}, \citenamefont {Pla{\ifmmode\mbox{\c{c}}\else\c{c}\fi}ais},
  \citenamefont {Cavanna}, \citenamefont {Gennser}, \citenamefont {Jin},
  \citenamefont {Degiovanni},\ and\ \citenamefont
  {F{\ifmmode\grave{e}\else\`{e}\fi}ve}}]{Bisognin2019Jul}%
  \BibitemOpen
  \bibfield  {author} {\bibinfo {author} {\bibfnamefont {R.}~\bibnamefont
  {Bisognin}}, \bibinfo {author} {\bibfnamefont {A.}~\bibnamefont
  {Marguerite}}, \bibinfo {author} {\bibfnamefont {B.}~\bibnamefont {Roussel}},
  \bibinfo {author} {\bibfnamefont {M.}~\bibnamefont {Kumar}}, \bibinfo
  {author} {\bibfnamefont {C.}~\bibnamefont {Cabart}}, \bibinfo {author}
  {\bibfnamefont {C.}~\bibnamefont {Chapdelaine}}, \bibinfo {author}
  {\bibfnamefont {A.}~\bibnamefont {Mohammad-Djafari}}, \bibinfo {author}
  {\bibfnamefont {J.-M.}\ \bibnamefont {Berroir}}, \bibinfo {author}
  {\bibfnamefont {E.}~\bibnamefont {Bocquillon}}, \bibinfo {author}
  {\bibfnamefont {B.}~\bibnamefont
  {Pla{\ifmmode\mbox{\c{c}}\else\c{c}\fi}ais}}, \bibinfo {author}
  {\bibfnamefont {A.}~\bibnamefont {Cavanna}}, \bibinfo {author} {\bibfnamefont
  {U.}~\bibnamefont {Gennser}}, \bibinfo {author} {\bibfnamefont
  {Y.}~\bibnamefont {Jin}}, \bibinfo {author} {\bibfnamefont {P.}~\bibnamefont
  {Degiovanni}},\ and\ \bibinfo {author} {\bibfnamefont {G.}~\bibnamefont
  {F{\ifmmode\grave{e}\else\`{e}\fi}ve}},\ }\bibfield  {title} {\bibinfo
  {title} {{Quantum tomography of electrical currents}},\ }\href
  {https://doi.org/10.1038/s41467-019-11369-5} {\bibfield  {journal} {\bibinfo
  {journal} {Nat. Commun.}\ }\textbf {\bibinfo {volume} {10}},\ \bibinfo
  {pages} {1--12} (\bibinfo {year} {2019})}\BibitemShut {NoStop}%
\bibitem [{\citenamefont {Locane}\ \emph {et~al.}(2019)\citenamefont {Locane},
  \citenamefont {Brouwer},\ and\ \citenamefont {Kashcheyevs}}]{Locane2019Sep}%
  \BibitemOpen
  \bibfield  {author} {\bibinfo {author} {\bibfnamefont {E.}~\bibnamefont
  {Locane}}, \bibinfo {author} {\bibfnamefont {P.~W.}\ \bibnamefont
  {Brouwer}},\ and\ \bibinfo {author} {\bibfnamefont {V.}~\bibnamefont
  {Kashcheyevs}},\ }\bibfield  {title} {\bibinfo {title} {{Time-energy
  filtering of single electrons in ballistic waveguides}},\ }\href
  {https://doi.org/10.1088/1367-2630/ab3fbb} {\bibfield  {journal} {\bibinfo
  {journal} {New J. Phys.}\ }\textbf {\bibinfo {volume} {21}},\ \bibinfo
  {pages} {093042} (\bibinfo {year} {2019})}\BibitemShut {NoStop}%
\bibitem [{\citenamefont {Fletcher}\ \emph {et~al.}(2022)\citenamefont
  {Fletcher}, \citenamefont {Park}, \citenamefont {Ryu}, \citenamefont {See},
  \citenamefont {Griffiths}, \citenamefont {Jones}, \citenamefont {Farrer},
  \citenamefont {Ritchie}, \citenamefont {Sim},\ and\ \citenamefont
  {Kataoka}}]{Fletcher2022Oct}%
  \BibitemOpen
  \bibfield  {author} {\bibinfo {author} {\bibfnamefont {J.~D.}\ \bibnamefont
  {Fletcher}}, \bibinfo {author} {\bibfnamefont {W.}~\bibnamefont {Park}},
  \bibinfo {author} {\bibfnamefont {S.}~\bibnamefont {Ryu}}, \bibinfo {author}
  {\bibfnamefont {P.}~\bibnamefont {See}}, \bibinfo {author} {\bibfnamefont
  {J.~P.}\ \bibnamefont {Griffiths}}, \bibinfo {author} {\bibfnamefont
  {G.~A.~C.}\ \bibnamefont {Jones}}, \bibinfo {author} {\bibfnamefont
  {I.}~\bibnamefont {Farrer}}, \bibinfo {author} {\bibfnamefont {D.~A.}\
  \bibnamefont {Ritchie}}, \bibinfo {author} {\bibfnamefont {H.-S.}\
  \bibnamefont {Sim}},\ and\ \bibinfo {author} {\bibfnamefont {M.}~\bibnamefont
  {Kataoka}},\ }\bibfield  {title} {\bibinfo {title} {{Time-resolved Coulomb
  collision of single electrons}},\ }\bibfield  {journal} {\bibinfo  {journal}
  {arXiv}\ }\href {https://doi.org/10.48550/arXiv.2210.03473}
  {10.48550/arXiv.2210.03473} (\bibinfo {year} {2022}),\ \Eprint
  {https://arxiv.org/abs/2210.03473} {2210.03473} \BibitemShut {NoStop}%
\bibitem [{\citenamefont {Wang}\ \emph {et~al.}(2022)\citenamefont {Wang},
  \citenamefont {Edlbauer}, \citenamefont {Richard}, \citenamefont {Ota},
  \citenamefont {Park}, \citenamefont {Shim}, \citenamefont {Ludwig},
  \citenamefont {Wieck}, \citenamefont {Sim}, \citenamefont {Urdampilleta},
  \citenamefont {Meunier}, \citenamefont {Kodera}, \citenamefont {Kaneko},
  \citenamefont {Sellier}, \citenamefont {Waintal}, \citenamefont {Takada},\
  and\ \citenamefont {Bäuerle}}]{WangSAW}%
  \BibitemOpen
  \bibfield  {author} {\bibinfo {author} {\bibfnamefont {J.}~\bibnamefont
  {Wang}}, \bibinfo {author} {\bibfnamefont {H.}~\bibnamefont {Edlbauer}},
  \bibinfo {author} {\bibfnamefont {A.}~\bibnamefont {Richard}}, \bibinfo
  {author} {\bibfnamefont {S.}~\bibnamefont {Ota}}, \bibinfo {author}
  {\bibfnamefont {W.}~\bibnamefont {Park}}, \bibinfo {author} {\bibfnamefont
  {J.}~\bibnamefont {Shim}}, \bibinfo {author} {\bibfnamefont {A.}~\bibnamefont
  {Ludwig}}, \bibinfo {author} {\bibfnamefont {A.}~\bibnamefont {Wieck}},
  \bibinfo {author} {\bibfnamefont {H.-S.}\ \bibnamefont {Sim}}, \bibinfo
  {author} {\bibfnamefont {M.}~\bibnamefont {Urdampilleta}}, \bibinfo {author}
  {\bibfnamefont {T.}~\bibnamefont {Meunier}}, \bibinfo {author} {\bibfnamefont
  {T.}~\bibnamefont {Kodera}}, \bibinfo {author} {\bibfnamefont {N.-H.}\
  \bibnamefont {Kaneko}}, \bibinfo {author} {\bibfnamefont {H.}~\bibnamefont
  {Sellier}}, \bibinfo {author} {\bibfnamefont {X.}~\bibnamefont {Waintal}},
  \bibinfo {author} {\bibfnamefont {S.}~\bibnamefont {Takada}},\ and\ \bibinfo
  {author} {\bibfnamefont {C.}~\bibnamefont {Bäuerle}},\ }\bibfield  {title}
  {\bibinfo {title} {Coulomb-mediated antibunching of an electron pair surfing
  on sound},\ }\bibfield  {journal} {\bibinfo  {journal} {arXiv}\ }\href
  {https://doi.org/10.48550/arXiv.2210.03452} {10.48550/arXiv.2210.03452}
  (\bibinfo {year} {2022})\BibitemShut {NoStop}%
\bibitem [{\citenamefont {Ubbelohde}\ \emph {et~al.}(2022)\citenamefont
  {Ubbelohde}, \citenamefont {Freise}, \citenamefont {Pavlovska}, \citenamefont
  {Silvestrov}, \citenamefont {Recher}, \citenamefont {Kokainis}, \citenamefont
  {Barinovs}, \citenamefont {Hohls}, \citenamefont {Weimann}, \citenamefont
  {Pierz},\ and\ \citenamefont {Kashcheyevs}}]{UbbelohdeChargeDet}%
  \BibitemOpen
  \bibfield  {author} {\bibinfo {author} {\bibfnamefont {N.}~\bibnamefont
  {Ubbelohde}}, \bibinfo {author} {\bibfnamefont {L.}~\bibnamefont {Freise}},
  \bibinfo {author} {\bibfnamefont {E.}~\bibnamefont {Pavlovska}}, \bibinfo
  {author} {\bibfnamefont {P.~G.}\ \bibnamefont {Silvestrov}}, \bibinfo
  {author} {\bibfnamefont {P.}~\bibnamefont {Recher}}, \bibinfo {author}
  {\bibfnamefont {M.}~\bibnamefont {Kokainis}}, \bibinfo {author}
  {\bibfnamefont {G.}~\bibnamefont {Barinovs}}, \bibinfo {author}
  {\bibfnamefont {F.}~\bibnamefont {Hohls}}, \bibinfo {author} {\bibfnamefont
  {T.}~\bibnamefont {Weimann}}, \bibinfo {author} {\bibfnamefont
  {K.}~\bibnamefont {Pierz}},\ and\ \bibinfo {author} {\bibfnamefont
  {V.}~\bibnamefont {Kashcheyevs}},\ }\bibfield  {title} {\bibinfo {title} {Two
  electrons interacting at a mesoscopic beam splitter},\ }\bibfield  {journal}
  {\bibinfo  {journal} {arXiv}\ }\href
  {https://doi.org/10.48550/arXiv.2210.03632} {10.48550/arXiv.2210.03632}
  (\bibinfo {year} {2022})\BibitemShut {NoStop}%
\bibitem [{\citenamefont {Freulon}\ \emph {et~al.}(2015)\citenamefont
  {Freulon}, \citenamefont {Marguerite}, \citenamefont {Berroir}, \citenamefont
  {Pla{\ifmmode\mbox{\c{c}}\else\c{c}\fi}ais}, \citenamefont {Cavanna},
  \citenamefont {Jin},\ and\ \citenamefont
  {F{\ifmmode\grave{e}\else\`{e}\fi}ve}}]{Freulon2015Apr}%
  \BibitemOpen
  \bibfield  {author} {\bibinfo {author} {\bibfnamefont {V.}~\bibnamefont
  {Freulon}}, \bibinfo {author} {\bibfnamefont {A.}~\bibnamefont {Marguerite}},
  \bibinfo {author} {\bibfnamefont {J.-M.}\ \bibnamefont {Berroir}}, \bibinfo
  {author} {\bibfnamefont {B.}~\bibnamefont
  {Pla{\ifmmode\mbox{\c{c}}\else\c{c}\fi}ais}}, \bibinfo {author}
  {\bibfnamefont {A.}~\bibnamefont {Cavanna}}, \bibinfo {author} {\bibfnamefont
  {Y.}~\bibnamefont {Jin}},\ and\ \bibinfo {author} {\bibfnamefont
  {G.}~\bibnamefont {F{\ifmmode\grave{e}\else\`{e}\fi}ve}},\ }\bibfield
  {title} {\bibinfo {title} {{Hong-Ou-Mandel experiment for temporal
  investigation of single-electron fractionalization}},\ }\href
  {https://doi.org/10.1038/ncomms7854} {\bibfield  {journal} {\bibinfo
  {journal} {Nat. Commun.}\ }\textbf {\bibinfo {volume} {6}},\ \bibinfo {pages}
  {1--6} (\bibinfo {year} {2015})}\BibitemShut {NoStop}%
\bibitem [{\citenamefont {Ryu}\ and\ \citenamefont {Sim}(2022)}]{Sungguen2022}%
  \BibitemOpen
  \bibfield  {author} {\bibinfo {author} {\bibfnamefont {S.}~\bibnamefont
  {Ryu}}\ and\ \bibinfo {author} {\bibfnamefont {H.-S.}\ \bibnamefont {Sim}},\
  }\bibfield  {title} {\bibinfo {title} {Partition of two interacting electrons
  by a potential barrier},\ }\href
  {https://doi.org/10.1103/PhysRevLett.129.166801} {\bibfield  {journal}
  {\bibinfo  {journal} {Phys. Rev. Lett.}\ }\textbf {\bibinfo {volume} {129}},\
  \bibinfo {pages} {166801} (\bibinfo {year} {2022})}\BibitemShut {NoStop}%
\bibitem [{\citenamefont {Pavlovska}\ \emph {et~al.}(2022)\citenamefont
  {Pavlovska}, \citenamefont {Silvestrov}, \citenamefont {Recher},
  \citenamefont {Barinovs},\ and\ \citenamefont
  {Kashcheyevs}}]{Pavlovska2022Jan}%
  \BibitemOpen
  \bibfield  {author} {\bibinfo {author} {\bibfnamefont {E.}~\bibnamefont
  {Pavlovska}}, \bibinfo {author} {\bibfnamefont {P.~G.}\ \bibnamefont
  {Silvestrov}}, \bibinfo {author} {\bibfnamefont {P.}~\bibnamefont {Recher}},
  \bibinfo {author} {\bibfnamefont {G.}~\bibnamefont {Barinovs}},\ and\
  \bibinfo {author} {\bibfnamefont {V.}~\bibnamefont {Kashcheyevs}},\
  }\bibfield  {title} {\bibinfo {title} {{Collision of two interacting
  electrons on a mesoscopic beamsplitter: exact solution in the classical
  limit}},\ }\bibfield  {journal} {\bibinfo  {journal} {arXiv}\ }\href
  {https://doi.org/10.48550/arXiv.2201.13439} {10.48550/arXiv.2201.13439}
  (\bibinfo {year} {2022}),\ \Eprint {https://arxiv.org/abs/2201.13439}
  {2201.13439} \BibitemShut {NoStop}%
\bibitem [{\citenamefont {Leicht}\ \emph {et~al.}(2011)\citenamefont {Leicht},
  \citenamefont {Mirovsky}, \citenamefont {Kaestner}, \citenamefont {Hohls},
  \citenamefont {Kashcheyevs}, \citenamefont {Kurganova}, \citenamefont
  {Zeitler}, \citenamefont {Weimann}, \citenamefont {Pierz},\ and\
  \citenamefont {Schumacher}}]{Leicht2011Mar}%
  \BibitemOpen
  \bibfield  {author} {\bibinfo {author} {\bibfnamefont {C.}~\bibnamefont
  {Leicht}}, \bibinfo {author} {\bibfnamefont {P.}~\bibnamefont {Mirovsky}},
  \bibinfo {author} {\bibfnamefont {B.}~\bibnamefont {Kaestner}}, \bibinfo
  {author} {\bibfnamefont {F.}~\bibnamefont {Hohls}}, \bibinfo {author}
  {\bibfnamefont {V.}~\bibnamefont {Kashcheyevs}}, \bibinfo {author}
  {\bibfnamefont {E.~V.}\ \bibnamefont {Kurganova}}, \bibinfo {author}
  {\bibfnamefont {U.}~\bibnamefont {Zeitler}}, \bibinfo {author} {\bibfnamefont
  {T.}~\bibnamefont {Weimann}}, \bibinfo {author} {\bibfnamefont
  {K.}~\bibnamefont {Pierz}},\ and\ \bibinfo {author} {\bibfnamefont {H.~W.}\
  \bibnamefont {Schumacher}},\ }\bibfield  {title} {\bibinfo {title}
  {{Generation of energy selective excitations in quantum Hall edge states}},\
  }\href {https://doi.org/10.1088/0268-1242/26/5/055010} {\bibfield  {journal}
  {\bibinfo  {journal} {Semicond. Sci. Technol.}\ }\textbf {\bibinfo {volume}
  {26}},\ \bibinfo {pages} {055010} (\bibinfo {year} {2011})}\BibitemShut
  {NoStop}%
\bibitem [{\citenamefont {Cho}\ \emph {et~al.}(2022)\citenamefont {Cho},
  \citenamefont {Park}, \citenamefont {Kim}, \citenamefont {Seo}, \citenamefont
  {Park}, \citenamefont {Choi}, \citenamefont {Kim}, \citenamefont {Sim},\ and\
  \citenamefont {Bae}}]{Cho2022Dec}%
  \BibitemOpen
  \bibfield  {author} {\bibinfo {author} {\bibfnamefont {S.~U.}\ \bibnamefont
  {Cho}}, \bibinfo {author} {\bibfnamefont {W.}~\bibnamefont {Park}}, \bibinfo
  {author} {\bibfnamefont {B.-K.}\ \bibnamefont {Kim}}, \bibinfo {author}
  {\bibfnamefont {M.}~\bibnamefont {Seo}}, \bibinfo {author} {\bibfnamefont
  {D.~T.}\ \bibnamefont {Park}}, \bibinfo {author} {\bibfnamefont
  {H.}~\bibnamefont {Choi}}, \bibinfo {author} {\bibfnamefont {N.}~\bibnamefont
  {Kim}}, \bibinfo {author} {\bibfnamefont {H.-S.}\ \bibnamefont {Sim}},\ and\
  \bibinfo {author} {\bibfnamefont {M.-H.}\ \bibnamefont {Bae}},\ }\bibfield
  {title} {\bibinfo {title} {{One-Lead Single-Electron Source with Charging
  Energy}},\ }\href {https://doi.org/10.1021/acs.nanolett.2c02893} {\bibfield
  {journal} {\bibinfo  {journal} {Nano Lett.}\ }\textbf {\bibinfo {volume}
  {22}},\ \bibinfo {pages} {9313--9318} (\bibinfo {year} {2022})}\BibitemShut
  {NoStop}%
\bibitem [{\citenamefont {Pekola}\ \emph {et~al.}(2013)\citenamefont {Pekola},
  \citenamefont {Saira}, \citenamefont {Maisi}, \citenamefont {Kemppinen},
  \citenamefont
  {M{\ifmmode\ddot{o}\else\"{o}\fi}tt{\ifmmode\ddot{o}\else\"{o}\fi}nen},
  \citenamefont {Pashkin},\ and\ \citenamefont {Averin}}]{Pekola2013Oct}%
  \BibitemOpen
  \bibfield  {author} {\bibinfo {author} {\bibfnamefont {J.~P.}\ \bibnamefont
  {Pekola}}, \bibinfo {author} {\bibfnamefont {O.-P.}\ \bibnamefont {Saira}},
  \bibinfo {author} {\bibfnamefont {V.~F.}\ \bibnamefont {Maisi}}, \bibinfo
  {author} {\bibfnamefont {A.}~\bibnamefont {Kemppinen}}, \bibinfo {author}
  {\bibfnamefont {M.}~\bibnamefont
  {M{\ifmmode\ddot{o}\else\"{o}\fi}tt{\ifmmode\ddot{o}\else\"{o}\fi}nen}},
  \bibinfo {author} {\bibfnamefont {Y.~A.}\ \bibnamefont {Pashkin}},\ and\
  \bibinfo {author} {\bibfnamefont {D.~V.}\ \bibnamefont {Averin}},\ }\bibfield
   {title} {\bibinfo {title} {{Single-electron current sources: Toward a
  refined definition of the ampere}},\ }\href
  {https://doi.org/10.1103/RevModPhys.85.1421} {\bibfield  {journal} {\bibinfo
  {journal} {Rev. Mod. Phys.}\ }\textbf {\bibinfo {volume} {85}},\ \bibinfo
  {pages} {1421--1472} (\bibinfo {year} {2013})}\BibitemShut {NoStop}%
\bibitem [{\citenamefont {Schulenborg}\ \emph {et~al.}(2016)\citenamefont
  {Schulenborg}, \citenamefont {Saptsov}, \citenamefont {Haupt}, \citenamefont
  {Splettstoesser},\ and\ \citenamefont {Wegewijs}}]{Schulenborg2016Feb}%
  \BibitemOpen
  \bibfield  {author} {\bibinfo {author} {\bibfnamefont {J.}~\bibnamefont
  {Schulenborg}}, \bibinfo {author} {\bibfnamefont {R.~B.}\ \bibnamefont
  {Saptsov}}, \bibinfo {author} {\bibfnamefont {F.}~\bibnamefont {Haupt}},
  \bibinfo {author} {\bibfnamefont {J.}~\bibnamefont {Splettstoesser}},\ and\
  \bibinfo {author} {\bibfnamefont {M.~R.}\ \bibnamefont {Wegewijs}},\
  }\bibfield  {title} {\bibinfo {title} {{Fermion-parity duality and energy
  relaxation in interacting open systems}},\ }\href
  {https://doi.org/10.1103/PhysRevB.93.081411} {\bibfield  {journal} {\bibinfo
  {journal} {Phys. Rev. B}\ }\textbf {\bibinfo {volume} {93}},\ \bibinfo
  {pages} {081411} (\bibinfo {year} {2016})}\BibitemShut {NoStop}%
\bibitem [{\citenamefont {Vanherck}\ \emph {et~al.}(2017)\citenamefont
  {Vanherck}, \citenamefont {Schulenborg}, \citenamefont {Saptsov},
  \citenamefont {Splettstoesser},\ and\ \citenamefont
  {Wegewijs}}]{Vanherck2017Mar}%
  \BibitemOpen
  \bibfield  {author} {\bibinfo {author} {\bibfnamefont {J.}~\bibnamefont
  {Vanherck}}, \bibinfo {author} {\bibfnamefont {J.}~\bibnamefont
  {Schulenborg}}, \bibinfo {author} {\bibfnamefont {R.~B.}\ \bibnamefont
  {Saptsov}}, \bibinfo {author} {\bibfnamefont {J.}~\bibnamefont
  {Splettstoesser}},\ and\ \bibinfo {author} {\bibfnamefont {M.~R.}\
  \bibnamefont {Wegewijs}},\ }\bibfield  {title} {\bibinfo {title} {{Relaxation
  of quantum dots in a magnetic field at finite bias {\textendash} Charge,
  spin, and heat currents}},\ }\href {https://doi.org/10.1002/pssb.201600614}
  {\bibfield  {journal} {\bibinfo  {journal} {Phys. Status Solidi B}\ }\textbf
  {\bibinfo {volume} {254}},\ \bibinfo {pages} {1600614} (\bibinfo {year}
  {2017})}\BibitemShut {NoStop}%
\bibitem [{\citenamefont {Schulenborg}\ \emph {et~al.}(2018)\citenamefont
  {Schulenborg}, \citenamefont {Splettstoesser},\ and\ \citenamefont
  {Wegewijs}}]{Schulenborg2018Dec}%
  \BibitemOpen
  \bibfield  {author} {\bibinfo {author} {\bibfnamefont {J.}~\bibnamefont
  {Schulenborg}}, \bibinfo {author} {\bibfnamefont {J.}~\bibnamefont
  {Splettstoesser}},\ and\ \bibinfo {author} {\bibfnamefont {M.~R.}\
  \bibnamefont {Wegewijs}},\ }\bibfield  {title} {\bibinfo {title} {{Duality
  for open fermion systems: Energy-dependent weak coupling and quantum master
  equations}},\ }\href {https://doi.org/10.1103/PhysRevB.98.235405} {\bibfield
  {journal} {\bibinfo  {journal} {Phys. Rev. B}\ }\textbf {\bibinfo {volume}
  {98}},\ \bibinfo {pages} {235405} (\bibinfo {year} {2018})}\BibitemShut
  {NoStop}%
\bibitem [{\citenamefont {Kashcheyevs}\ and\ \citenamefont
  {Samuelsson}(2017)}]{Kashcheyevs2017Jun}%
  \BibitemOpen
  \bibfield  {author} {\bibinfo {author} {\bibfnamefont {V.}~\bibnamefont
  {Kashcheyevs}}\ and\ \bibinfo {author} {\bibfnamefont {P.}~\bibnamefont
  {Samuelsson}},\ }\bibfield  {title} {\bibinfo {title} {{Classical-to-quantum
  crossover in electron on-demand emission}},\ }\href
  {https://doi.org/10.1103/PhysRevB.95.245424} {\bibfield  {journal} {\bibinfo
  {journal} {Phys. Rev. B}\ }\textbf {\bibinfo {volume} {95}},\ \bibinfo
  {pages} {245424} (\bibinfo {year} {2017})}\BibitemShut {NoStop}%
\bibitem [{\citenamefont {Splettstoesser}\ \emph {et~al.}(2010)\citenamefont
  {Splettstoesser}, \citenamefont {Governale}, \citenamefont
  {K{\ifmmode\ddot{o}\else\"{o}\fi}nig},\ and\ \citenamefont
  {B{\ifmmode\ddot{u}\else\"{u}\fi}ttiker}}]{Splettstoesser2010Apr}%
  \BibitemOpen
  \bibfield  {author} {\bibinfo {author} {\bibfnamefont {J.}~\bibnamefont
  {Splettstoesser}}, \bibinfo {author} {\bibfnamefont {M.}~\bibnamefont
  {Governale}}, \bibinfo {author} {\bibfnamefont {J.}~\bibnamefont
  {K{\ifmmode\ddot{o}\else\"{o}\fi}nig}},\ and\ \bibinfo {author}
  {\bibfnamefont {M.}~\bibnamefont {B{\ifmmode\ddot{u}\else\"{u}\fi}ttiker}},\
  }\bibfield  {title} {\bibinfo {title} {{Charge and spin dynamics in
  interacting quantum dots}},\ }\href
  {https://doi.org/10.1103/PhysRevB.81.165318} {\bibfield  {journal} {\bibinfo
  {journal} {Phys. Rev. B}\ }\textbf {\bibinfo {volume} {81}},\ \bibinfo
  {pages} {165318} (\bibinfo {year} {2010})}\BibitemShut {NoStop}%
\bibitem [{\citenamefont {Filippone}\ \emph {et~al.}(2011)\citenamefont
  {Filippone}, \citenamefont {Le~Hur},\ and\ \citenamefont
  {Mora}}]{Filippone2011Oct}%
  \BibitemOpen
  \bibfield  {author} {\bibinfo {author} {\bibfnamefont {M.}~\bibnamefont
  {Filippone}}, \bibinfo {author} {\bibfnamefont {K.}~\bibnamefont {Le~Hur}},\
  and\ \bibinfo {author} {\bibfnamefont {C.}~\bibnamefont {Mora}},\ }\bibfield
  {title} {\bibinfo {title} {{Giant Charge Relaxation Resistance in the
  Anderson Model}},\ }\href {https://doi.org/10.1103/PhysRevLett.107.176601}
  {\bibfield  {journal} {\bibinfo  {journal} {Phys. Rev. Lett.}\ }\textbf
  {\bibinfo {volume} {107}},\ \bibinfo {pages} {176601} (\bibinfo {year}
  {2011})}\BibitemShut {NoStop}%
\bibitem [{\citenamefont {Kashuba}\ \emph {et~al.}(2012)\citenamefont
  {Kashuba}, \citenamefont {Schoeller},\ and\ \citenamefont
  {Splettstoesser}}]{Kashuba2012Jun}%
  \BibitemOpen
  \bibfield  {author} {\bibinfo {author} {\bibfnamefont {O.}~\bibnamefont
  {Kashuba}}, \bibinfo {author} {\bibfnamefont {H.}~\bibnamefont {Schoeller}},\
  and\ \bibinfo {author} {\bibfnamefont {J.}~\bibnamefont {Splettstoesser}},\
  }\bibfield  {title} {\bibinfo {title} {{Nonlinear adiabatic response of
  interacting quantum dots}},\ }\href
  {https://doi.org/10.1209/0295-5075/98/57003} {\bibfield  {journal} {\bibinfo
  {journal} {Europhys. Lett.}\ }\textbf {\bibinfo {volume} {98}},\ \bibinfo
  {pages} {57003} (\bibinfo {year} {2012})}\BibitemShut {NoStop}%
\bibitem [{\citenamefont {Schulenborg}\ \emph {et~al.}(2014)\citenamefont
  {Schulenborg}, \citenamefont {Splettstoesser}, \citenamefont {Governale},\
  and\ \citenamefont {Contreras-Pulido}}]{Schulenborg2014May}%
  \BibitemOpen
  \bibfield  {author} {\bibinfo {author} {\bibfnamefont {J.}~\bibnamefont
  {Schulenborg}}, \bibinfo {author} {\bibfnamefont {J.}~\bibnamefont
  {Splettstoesser}}, \bibinfo {author} {\bibfnamefont {M.}~\bibnamefont
  {Governale}},\ and\ \bibinfo {author} {\bibfnamefont {L.~D.}\ \bibnamefont
  {Contreras-Pulido}},\ }\bibfield  {title} {\bibinfo {title} {{Detection of
  the relaxation rates of an interacting quantum dot by a capacitively coupled
  sensor dot}},\ }\href {https://doi.org/10.1103/PhysRevB.89.195305} {\bibfield
   {journal} {\bibinfo  {journal} {Phys. Rev. B}\ }\textbf {\bibinfo {volume}
  {89}},\ \bibinfo {pages} {195305} (\bibinfo {year} {2014})}\BibitemShut
  {NoStop}%
\bibitem [{\citenamefont {Alomar}\ \emph {et~al.}(2016)\citenamefont {Alomar},
  \citenamefont {Lim},\ and\ \citenamefont
  {S{\ifmmode\acute{a}\else\'{a}\fi}nchez}}]{Alomar2016Oct}%
  \BibitemOpen
  \bibfield  {author} {\bibinfo {author} {\bibfnamefont {M.~I.}\ \bibnamefont
  {Alomar}}, \bibinfo {author} {\bibfnamefont {J.~S.}\ \bibnamefont {Lim}},\
  and\ \bibinfo {author} {\bibfnamefont {D.}~\bibnamefont
  {S{\ifmmode\acute{a}\else\'{a}\fi}nchez}},\ }\bibfield  {title} {\bibinfo
  {title} {{Coulomb-blockade effect in nonlinear mesoscopic capacitors}},\
  }\href {https://doi.org/10.1103/PhysRevB.94.165425} {\bibfield  {journal}
  {\bibinfo  {journal} {Phys. Rev. B}\ }\textbf {\bibinfo {volume} {94}},\
  \bibinfo {pages} {165425} (\bibinfo {year} {2016})}\BibitemShut {NoStop}%
\bibitem [{\citenamefont {Filippone}\ \emph {et~al.}(2020)\citenamefont
  {Filippone}, \citenamefont {Marguerite}, \citenamefont {Le~Hur},
  \citenamefont {F{\ifmmode\grave{e}\else\`{e}\fi}ve},\ and\ \citenamefont
  {Mora}}]{Filippone2020Jul}%
  \BibitemOpen
  \bibfield  {author} {\bibinfo {author} {\bibfnamefont {M.}~\bibnamefont
  {Filippone}}, \bibinfo {author} {\bibfnamefont {A.}~\bibnamefont
  {Marguerite}}, \bibinfo {author} {\bibfnamefont {K.}~\bibnamefont {Le~Hur}},
  \bibinfo {author} {\bibfnamefont {G.}~\bibnamefont
  {F{\ifmmode\grave{e}\else\`{e}\fi}ve}},\ and\ \bibinfo {author}
  {\bibfnamefont {C.}~\bibnamefont {Mora}},\ }\bibfield  {title} {\bibinfo
  {title} {{Phase-Coherent Dynamics of Quantum Devices with Local
  Interactions}},\ }\href {https://doi.org/10.3390/e22080847} {\bibfield
  {journal} {\bibinfo  {journal} {Entropy}\ }\textbf {\bibinfo {volume} {22}},\
  \bibinfo {pages} {847} (\bibinfo {year} {2020})}\BibitemShut {NoStop}%
\bibitem [{\citenamefont {Kaestner}\ and\ \citenamefont
  {Kashcheyevs}(2015)}]{Kaestner2015Sep}%
  \BibitemOpen
  \bibfield  {author} {\bibinfo {author} {\bibfnamefont {B.}~\bibnamefont
  {Kaestner}}\ and\ \bibinfo {author} {\bibfnamefont {V.}~\bibnamefont
  {Kashcheyevs}},\ }\bibfield  {title} {\bibinfo {title} {{Non-adiabatic
  quantized charge pumping with tunable-barrier quantum dots: a review of
  current progress}},\ }\href {https://doi.org/10.1088/0034-4885/78/10/103901}
  {\bibfield  {journal} {\bibinfo  {journal} {Rep. Prog. Phys.}\ }\textbf
  {\bibinfo {volume} {78}},\ \bibinfo {pages} {103901} (\bibinfo {year}
  {2015})}\BibitemShut {NoStop}%
\bibitem [{\citenamefont {Hofmann}\ \emph {et~al.}(2016)\citenamefont
  {Hofmann}, \citenamefont {Maisi}, \citenamefont {Gold}, \citenamefont
  {Kr{\ifmmode\ddot{a}\else\"{a}\fi}henmann}, \citenamefont
  {R{\ifmmode\ddot{o}\else\"{o}\fi}ssler}, \citenamefont {Basset},
  \citenamefont {M{\ifmmode\ddot{a}\else\"{a}\fi}rki}, \citenamefont {Reichl},
  \citenamefont {Wegscheider}, \citenamefont {Ensslin},\ and\ \citenamefont
  {Ihn}}]{Hofmann2016Nov}%
  \BibitemOpen
  \bibfield  {author} {\bibinfo {author} {\bibfnamefont {A.}~\bibnamefont
  {Hofmann}}, \bibinfo {author} {\bibfnamefont {V.~F.}\ \bibnamefont {Maisi}},
  \bibinfo {author} {\bibfnamefont {C.}~\bibnamefont {Gold}}, \bibinfo {author}
  {\bibfnamefont {T.}~\bibnamefont {Kr{\ifmmode\ddot{a}\else\"{a}\fi}henmann}},
  \bibinfo {author} {\bibfnamefont {C.}~\bibnamefont
  {R{\ifmmode\ddot{o}\else\"{o}\fi}ssler}}, \bibinfo {author} {\bibfnamefont
  {J.}~\bibnamefont {Basset}}, \bibinfo {author} {\bibfnamefont
  {P.}~\bibnamefont {M{\ifmmode\ddot{a}\else\"{a}\fi}rki}}, \bibinfo {author}
  {\bibfnamefont {C.}~\bibnamefont {Reichl}}, \bibinfo {author} {\bibfnamefont
  {W.}~\bibnamefont {Wegscheider}}, \bibinfo {author} {\bibfnamefont
  {K.}~\bibnamefont {Ensslin}},\ and\ \bibinfo {author} {\bibfnamefont
  {T.}~\bibnamefont {Ihn}},\ }\bibfield  {title} {\bibinfo {title} {{Measuring
  the Degeneracy of Discrete Energy Levels Using a
  $\mathrm{GaAs}/\mathrm{AlGaAs}$ Quantum Dot}},\ }\href
  {https://doi.org/10.1103/PhysRevLett.117.206803} {\bibfield  {journal}
  {\bibinfo  {journal} {Phys. Rev. Lett.}\ }\textbf {\bibinfo {volume} {117}},\
  \bibinfo {pages} {206803} (\bibinfo {year} {2016})}\BibitemShut {NoStop}%
\bibitem [{\citenamefont {Beckel}\ \emph {et~al.}(2014)\citenamefont {Beckel},
  \citenamefont {Kurzmann}, \citenamefont {Geller}, \citenamefont {Ludwig},
  \citenamefont {Wieck}, \citenamefont {K{\ifmmode\ddot{o}\else\"{o}\fi}nig},\
  and\ \citenamefont {Lorke}}]{Beckel2014May}%
  \BibitemOpen
  \bibfield  {author} {\bibinfo {author} {\bibfnamefont {A.}~\bibnamefont
  {Beckel}}, \bibinfo {author} {\bibfnamefont {A.}~\bibnamefont {Kurzmann}},
  \bibinfo {author} {\bibfnamefont {M.}~\bibnamefont {Geller}}, \bibinfo
  {author} {\bibfnamefont {A.}~\bibnamefont {Ludwig}}, \bibinfo {author}
  {\bibfnamefont {A.~D.}\ \bibnamefont {Wieck}}, \bibinfo {author}
  {\bibfnamefont {J.}~\bibnamefont {K{\ifmmode\ddot{o}\else\"{o}\fi}nig}},\
  and\ \bibinfo {author} {\bibfnamefont {A.}~\bibnamefont {Lorke}},\ }\bibfield
   {title} {\bibinfo {title} {{Asymmetry of charge relaxation times in quantum
  dots: The influence of degeneracy}},\ }\href
  {https://doi.org/10.1209/0295-5075/106/47002} {\bibfield  {journal} {\bibinfo
   {journal} {Europhys. Lett.}\ }\textbf {\bibinfo {volume} {106}},\ \bibinfo
  {pages} {47002} (\bibinfo {year} {2014})}\BibitemShut {NoStop}%
\bibitem [{\citenamefont {Skinner}\ and\ \citenamefont
  {Shklovskii}(2010)}]{Skinner2010Oct}%
  \BibitemOpen
  \bibfield  {author} {\bibinfo {author} {\bibfnamefont {B.}~\bibnamefont
  {Skinner}}\ and\ \bibinfo {author} {\bibfnamefont {B.~I.}\ \bibnamefont
  {Shklovskii}},\ }\bibfield  {title} {\bibinfo {title} {{Anomalously large
  capacitance of a plane capacitor with a two-dimensional electron gas}},\
  }\href {https://doi.org/10.1103/PhysRevB.82.155111} {\bibfield  {journal}
  {\bibinfo  {journal} {Phys. Rev. B}\ }\textbf {\bibinfo {volume} {82}},\
  \bibinfo {pages} {155111} (\bibinfo {year} {2010})}\BibitemShut {NoStop}%
\bibitem [{\citenamefont {Akmentinsh}\ \emph {et~al.}(2023)\citenamefont
  {Akmentinsh}, \citenamefont {Reifert}, \citenamefont {Weimann}, \citenamefont
  {Pierz}, \citenamefont {Kashcheyevs},\ and\ \citenamefont
  {Ubbelohde}}]{Akmentinsh2023Jan}%
  \BibitemOpen
  \bibfield  {author} {\bibinfo {author} {\bibfnamefont {A.}~\bibnamefont
  {Akmentinsh}}, \bibinfo {author} {\bibfnamefont {D.}~\bibnamefont {Reifert}},
  \bibinfo {author} {\bibfnamefont {T.}~\bibnamefont {Weimann}}, \bibinfo
  {author} {\bibfnamefont {K.}~\bibnamefont {Pierz}}, \bibinfo {author}
  {\bibfnamefont {V.}~\bibnamefont {Kashcheyevs}},\ and\ \bibinfo {author}
  {\bibfnamefont {N.}~\bibnamefont {Ubbelohde}},\ }\bibfield  {title} {\bibinfo
  {title} {{Universal scaling of adiabatic tunneling out of a shallow
  confinement potential}},\ }\bibfield  {journal} {\bibinfo  {journal} {arXiv}\
  }\href {https://doi.org/10.48550/arXiv.2301.11295}
  {10.48550/arXiv.2301.11295} (\bibinfo {year} {2023}),\ \Eprint
  {https://arxiv.org/abs/2301.11295} {2301.11295} \BibitemShut {NoStop}%
\bibitem [{\citenamefont {Kashcheyevs}\ and\ \citenamefont
  {Kaestner}(2010)}]{Kashcheyevs2010May}%
  \BibitemOpen
  \bibfield  {author} {\bibinfo {author} {\bibfnamefont {V.}~\bibnamefont
  {Kashcheyevs}}\ and\ \bibinfo {author} {\bibfnamefont {B.}~\bibnamefont
  {Kaestner}},\ }\bibfield  {title} {\bibinfo {title} {{Universal Decay Cascade
  Model for Dynamic Quantum Dot Initialization}},\ }\href
  {https://doi.org/10.1103/PhysRevLett.104.186805} {\bibfield  {journal}
  {\bibinfo  {journal} {Phys. Rev. Lett.}\ }\textbf {\bibinfo {volume} {104}},\
  \bibinfo {pages} {186805} (\bibinfo {year} {2010})}\BibitemShut {NoStop}%
\bibitem [{\citenamefont {Fricke}\ \emph {et~al.}(2013)\citenamefont {Fricke},
  \citenamefont {Wulf}, \citenamefont {Kaestner}, \citenamefont {Kashcheyevs},
  \citenamefont {Timoshenko}, \citenamefont {Nazarov}, \citenamefont {Hohls},
  \citenamefont {Mirovsky}, \citenamefont {Mackrodt}, \citenamefont {Dolata},
  \citenamefont {Weimann}, \citenamefont {Pierz},\ and\ \citenamefont
  {Schumacher}}]{Fricke2013Mar}%
  \BibitemOpen
  \bibfield  {author} {\bibinfo {author} {\bibfnamefont {L.}~\bibnamefont
  {Fricke}}, \bibinfo {author} {\bibfnamefont {M.}~\bibnamefont {Wulf}},
  \bibinfo {author} {\bibfnamefont {B.}~\bibnamefont {Kaestner}}, \bibinfo
  {author} {\bibfnamefont {V.}~\bibnamefont {Kashcheyevs}}, \bibinfo {author}
  {\bibfnamefont {J.}~\bibnamefont {Timoshenko}}, \bibinfo {author}
  {\bibfnamefont {P.}~\bibnamefont {Nazarov}}, \bibinfo {author} {\bibfnamefont
  {F.}~\bibnamefont {Hohls}}, \bibinfo {author} {\bibfnamefont
  {P.}~\bibnamefont {Mirovsky}}, \bibinfo {author} {\bibfnamefont
  {B.}~\bibnamefont {Mackrodt}}, \bibinfo {author} {\bibfnamefont
  {R.}~\bibnamefont {Dolata}}, \bibinfo {author} {\bibfnamefont
  {T.}~\bibnamefont {Weimann}}, \bibinfo {author} {\bibfnamefont
  {K.}~\bibnamefont {Pierz}},\ and\ \bibinfo {author} {\bibfnamefont {H.~W.}\
  \bibnamefont {Schumacher}},\ }\bibfield  {title} {\bibinfo {title} {{Counting
  Statistics for Electron Capture in a Dynamic Quantum Dot}},\ }\href
  {https://doi.org/10.1103/PhysRevLett.110.126803} {\bibfield  {journal}
  {\bibinfo  {journal} {Phys. Rev. Lett.}\ }\textbf {\bibinfo {volume} {110}},\
  \bibinfo {pages} {126803} (\bibinfo {year} {2013})}\BibitemShut {NoStop}%
\bibitem [{\citenamefont {Wenz}\ \emph {et~al.}(2019)\citenamefont {Wenz},
  \citenamefont {Klochan}, \citenamefont {Hohls}, \citenamefont {Gerster},
  \citenamefont {Kashcheyevs},\ and\ \citenamefont {Schumacher}}]{Wenz2019May}%
  \BibitemOpen
  \bibfield  {author} {\bibinfo {author} {\bibfnamefont {T.}~\bibnamefont
  {Wenz}}, \bibinfo {author} {\bibfnamefont {J.}~\bibnamefont {Klochan}},
  \bibinfo {author} {\bibfnamefont {F.}~\bibnamefont {Hohls}}, \bibinfo
  {author} {\bibfnamefont {T.}~\bibnamefont {Gerster}}, \bibinfo {author}
  {\bibfnamefont {V.}~\bibnamefont {Kashcheyevs}},\ and\ \bibinfo {author}
  {\bibfnamefont {H.~W.}\ \bibnamefont {Schumacher}},\ }\bibfield  {title}
  {\bibinfo {title} {{Quantum dot state initialization by control of tunneling
  rates}},\ }\href {https://doi.org/10.1103/PhysRevB.99.201409} {\bibfield
  {journal} {\bibinfo  {journal} {Phys. Rev. B}\ }\textbf {\bibinfo {volume}
  {99}},\ \bibinfo {pages} {201409} (\bibinfo {year} {2019})}\BibitemShut
  {NoStop}%
\bibitem [{\citenamefont {Wei{\ss}e}\ \emph {et~al.}(2006)\citenamefont
  {Wei{\ss}e}, \citenamefont {Wellein}, \citenamefont {Alvermann},\ and\
  \citenamefont {Fehske}}]{Weisse2006Mar}%
  \BibitemOpen
  \bibfield  {author} {\bibinfo {author} {\bibfnamefont {A.}~\bibnamefont
  {Wei{\ss}e}}, \bibinfo {author} {\bibfnamefont {G.}~\bibnamefont {Wellein}},
  \bibinfo {author} {\bibfnamefont {A.}~\bibnamefont {Alvermann}},\ and\
  \bibinfo {author} {\bibfnamefont {H.}~\bibnamefont {Fehske}},\ }\bibfield
  {title} {\bibinfo {title} {{The kernel polynomial method}},\ }\href
  {https://doi.org/10.1103/RevModPhys.78.275} {\bibfield  {journal} {\bibinfo
  {journal} {Rev. Mod. Phys.}\ }\textbf {\bibinfo {volume} {78}},\ \bibinfo
  {pages} {275--306} (\bibinfo {year} {2006})}\BibitemShut {NoStop}%
\bibitem [{\citenamefont
  {Bj{\ifmmode\ddot{o}\else\"{o}\fi}rnson}(2016)}]{Bjornson2016}%
  \BibitemOpen
  \bibfield  {author} {\bibinfo {author} {\bibfnamefont {K.}~\bibnamefont
  {Bj{\ifmmode\ddot{o}\else\"{o}\fi}rnson}},\ }\emph {\bibinfo {title}
  {{Topological band theory and Majorana fermions : With focus on
  self-consistent lattice models}}},\ \href
  {http://www.diva-portal.org/smash/record.jsf?pid=diva2%3A1034719&dswid=-9098}
  {Ph.D. thesis},\ \bibinfo  {school} {Acta Universitatis Upsaliensis}
  (\bibinfo {year} {2016})\BibitemShut {NoStop}%
\bibitem [{\citenamefont {Schulenborg}\ \emph {et~al.}(2024)\citenamefont
  {Schulenborg}, \citenamefont {Fletcher}, \citenamefont {Kataoka},\ and\
  \citenamefont {Splettstoesser}}]{SchulenborgZen2023Apr}%
  \BibitemOpen
  \bibfield  {author} {\bibinfo {author} {\bibfnamefont {J.}~\bibnamefont
  {Schulenborg}}, \bibinfo {author} {\bibfnamefont {J.~D.}\ \bibnamefont
  {Fletcher}}, \bibinfo {author} {\bibfnamefont {M.}~\bibnamefont {Kataoka}},\
  and\ \bibinfo {author} {\bibfnamefont {J.}~\bibnamefont {Splettstoesser}},\
  }\bibfield  {title} {\bibinfo {title} {{Zenodo repository containing raw data
  and code}},\ }\href {https://doi.org/10.5281/zenodo.10463201}
  {10.5281/zenodo.10463201} (\bibinfo {year} {2024})\BibitemShut {NoStop}%
\bibitem [{\citenamefont {Jauho}\ \emph {et~al.}(1994)\citenamefont {Jauho},
  \citenamefont {Wingreen},\ and\ \citenamefont {Meir}}]{Jauho1994Aug}%
  \BibitemOpen
  \bibfield  {author} {\bibinfo {author} {\bibfnamefont {A.-P.}\ \bibnamefont
  {Jauho}}, \bibinfo {author} {\bibfnamefont {N.~S.}\ \bibnamefont
  {Wingreen}},\ and\ \bibinfo {author} {\bibfnamefont {Y.}~\bibnamefont
  {Meir}},\ }\bibfield  {title} {\bibinfo {title} {{Time-dependent transport in
  interacting and noninteracting resonant-tunneling systems}},\ }\href
  {https://doi.org/10.1103/PhysRevB.50.5528} {\bibfield  {journal} {\bibinfo
  {journal} {Phys. Rev. B}\ }\textbf {\bibinfo {volume} {50}},\ \bibinfo
  {pages} {5528--5544} (\bibinfo {year} {1994})}\BibitemShut {NoStop}%
\bibitem [{\citenamefont {Breuer}\ \emph {et~al.}(2007)\citenamefont {Breuer},
  \citenamefont {Petruccione}, \citenamefont {Breuer},\ and\ \citenamefont
  {Petruccione}}]{Breuer2007Jan}%
  \BibitemOpen
  \bibfield  {author} {\bibinfo {author} {\bibfnamefont {H.-P.}\ \bibnamefont
  {Breuer}}, \bibinfo {author} {\bibfnamefont {F.}~\bibnamefont {Petruccione}},
  \bibinfo {author} {\bibfnamefont {H.-P.}\ \bibnamefont {Breuer}},\ and\
  \bibinfo {author} {\bibfnamefont {F.}~\bibnamefont {Petruccione}},\ }\href
  {https://global.oup.com/academic/product/the-theory-of-open-quantum-systems-9780199213900}
  {\emph {\bibinfo {title} {{The Theory of Open Quantum Systems}}}}\ (\bibinfo
  {publisher} {Oxford University Press},\ \bibinfo {address} {Oxford, England,
  UK},\ \bibinfo {year} {2007})\BibitemShut {NoStop}%
\bibitem [{\citenamefont {Koenig}(1999)}]{Koenig1999}%
  \BibitemOpen
  \bibfield  {author} {\bibinfo {author} {\bibfnamefont {J.}~\bibnamefont
  {Koenig}},\ }\href {https://publikationen.bibliothek.kit.edu/1299} {\bibinfo
  {title} {{Quantum fluctuations in the single-electron transistor}}} (\bibinfo
  {year} {1999}),\ \bibinfo {note} {[Online; accessed 10. Mar.
  2023]}\BibitemShut {NoStop}%
\bibitem [{\citenamefont {Gurvitz}\ and\ \citenamefont
  {Prager}(1996)}]{Gurvitz1996Jun}%
  \BibitemOpen
  \bibfield  {author} {\bibinfo {author} {\bibfnamefont {S.~A.}\ \bibnamefont
  {Gurvitz}}\ and\ \bibinfo {author} {\bibfnamefont {{\relax Ya}.~S.}\
  \bibnamefont {Prager}},\ }\bibfield  {title} {\bibinfo {title} {{Microscopic
  derivation of rate equations for quantum transport}},\ }\href
  {https://doi.org/10.1103/PhysRevB.53.15932} {\bibfield  {journal} {\bibinfo
  {journal} {Phys. Rev. B}\ }\textbf {\bibinfo {volume} {53}},\ \bibinfo
  {pages} {15932--15943} (\bibinfo {year} {1996})}\BibitemShut {NoStop}%
\bibitem [{\citenamefont {Gurvitz}(1997)}]{Gurvitz1997Dec}%
  \BibitemOpen
  \bibfield  {author} {\bibinfo {author} {\bibfnamefont {S.~A.}\ \bibnamefont
  {Gurvitz}},\ }\bibfield  {title} {\bibinfo {title} {{Measurements with a
  noninvasive detector and dephasing mechanism}},\ }\href
  {https://doi.org/10.1103/PhysRevB.56.15215} {\bibfield  {journal} {\bibinfo
  {journal} {Phys. Rev. B}\ }\textbf {\bibinfo {volume} {56}},\ \bibinfo
  {pages} {15215--15223} (\bibinfo {year} {1997})}\BibitemShut {NoStop}%
\bibitem [{\citenamefont {Oguri}\ and\ \citenamefont
  {Sakano}(2013)}]{Oguri2013Oct}%
  \BibitemOpen
  \bibfield  {author} {\bibinfo {author} {\bibfnamefont {A.}~\bibnamefont
  {Oguri}}\ and\ \bibinfo {author} {\bibfnamefont {R.}~\bibnamefont {Sakano}},\
  }\bibfield  {title} {\bibinfo {title} {{Exact interacting Green's function
  for the Anderson impurity at high bias voltages}},\ }\href
  {https://doi.org/10.1103/PhysRevB.88.155424} {\bibfield  {journal} {\bibinfo
  {journal} {Phys. Rev. B}\ }\textbf {\bibinfo {volume} {88}},\ \bibinfo
  {pages} {155424} (\bibinfo {year} {2013})}\BibitemShut {NoStop}%
\bibitem [{\citenamefont {Nathan}\ and\ \citenamefont
  {Rudner}(2020)}]{Nathan2020Sep}%
  \BibitemOpen
  \bibfield  {author} {\bibinfo {author} {\bibfnamefont {F.}~\bibnamefont
  {Nathan}}\ and\ \bibinfo {author} {\bibfnamefont {M.~S.}\ \bibnamefont
  {Rudner}},\ }\bibfield  {title} {\bibinfo {title} {{Universal Lindblad
  equation for open quantum systems}},\ }\href
  {https://doi.org/10.1103/PhysRevB.102.115109} {\bibfield  {journal} {\bibinfo
   {journal} {Phys. Rev. B}\ }\textbf {\bibinfo {volume} {102}},\ \bibinfo
  {pages} {115109} (\bibinfo {year} {2020})}\BibitemShut {NoStop}%
\bibitem [{\citenamefont {Splettstoesser}\ \emph {et~al.}(2006)\citenamefont
  {Splettstoesser}, \citenamefont {Governale}, \citenamefont
  {K{\ifmmode\ddot{o}\else\"{o}\fi}nig},\ and\ \citenamefont
  {Fazio}}]{Splettstoesser2006Aug}%
  \BibitemOpen
  \bibfield  {author} {\bibinfo {author} {\bibfnamefont {J.}~\bibnamefont
  {Splettstoesser}}, \bibinfo {author} {\bibfnamefont {M.}~\bibnamefont
  {Governale}}, \bibinfo {author} {\bibfnamefont {J.}~\bibnamefont
  {K{\ifmmode\ddot{o}\else\"{o}\fi}nig}},\ and\ \bibinfo {author}
  {\bibfnamefont {R.}~\bibnamefont {Fazio}},\ }\bibfield  {title} {\bibinfo
  {title} {{Adiabatic pumping through a quantum dot with coulomb interactions:
  A perturbation expansion in the tunnel coupling}},\ }\href
  {https://doi.org/10.1103/PhysRevB.74.085305} {\bibfield  {journal} {\bibinfo
  {journal} {Phys. Rev. B}\ }\textbf {\bibinfo {volume} {74}},\ \bibinfo
  {pages} {085305} (\bibinfo {year} {2006})}\BibitemShut {NoStop}%
\bibitem [{\citenamefont {Reckermann}\ \emph {et~al.}(2010)\citenamefont
  {Reckermann}, \citenamefont {Splettstoesser},\ and\ \citenamefont
  {Wegewijs}}]{Reckermann2010Jun}%
  \BibitemOpen
  \bibfield  {author} {\bibinfo {author} {\bibfnamefont {F.}~\bibnamefont
  {Reckermann}}, \bibinfo {author} {\bibfnamefont {J.}~\bibnamefont
  {Splettstoesser}},\ and\ \bibinfo {author} {\bibfnamefont {M.~R.}\
  \bibnamefont {Wegewijs}},\ }\bibfield  {title} {\bibinfo {title}
  {{Interaction-Induced Adiabatic Nonlinear Transport}},\ }\href
  {https://doi.org/10.1103/PhysRevLett.104.226803} {\bibfield  {journal}
  {\bibinfo  {journal} {Phys. Rev. Lett.}\ }\textbf {\bibinfo {volume} {104}},\
  \bibinfo {pages} {226803} (\bibinfo {year} {2010})}\BibitemShut {NoStop}%
\bibitem [{\citenamefont {Riwar}(2014)}]{Riwar2014}%
  \BibitemOpen
  \bibfield  {author} {\bibinfo {author} {\bibfnamefont {R.-P.}\ \bibnamefont
  {Riwar}},\ }\emph {\bibinfo {title} {{Current and noise in interacting
  quantum pumps}}},\ \href {https://publications.rwth-aachen.de/record/229331}
  {Ph.D. thesis},\ \bibinfo  {school} {RWTH Aachen University}, \bibinfo
  {address} {Aachen, Germany} (\bibinfo {year} {2014})\BibitemShut {NoStop}%
\bibitem [{\citenamefont {Kashcheyevs}\ and\ \citenamefont
  {Timoshenko}(2012)}]{Kashcheyevs2012Nov}%
  \BibitemOpen
  \bibfield  {author} {\bibinfo {author} {\bibfnamefont {V.}~\bibnamefont
  {Kashcheyevs}}\ and\ \bibinfo {author} {\bibfnamefont {J.}~\bibnamefont
  {Timoshenko}},\ }\bibfield  {title} {\bibinfo {title} {{Quantum Fluctuations
  and Coherence in High-Precision Single-Electron Capture}},\ }\href
  {https://doi.org/10.1103/PhysRevLett.109.216801} {\bibfield  {journal}
  {\bibinfo  {journal} {Phys. Rev. Lett.}\ }\textbf {\bibinfo {volume} {109}},\
  \bibinfo {pages} {216801} (\bibinfo {year} {2012})}\BibitemShut {NoStop}%
\bibitem [{\citenamefont {Schulenborg}(2016)}]{SchulenborgLic}%
  \BibitemOpen
  \bibfield  {author} {\bibinfo {author} {\bibfnamefont {J.}~\bibnamefont
  {Schulenborg}},\ }\href
  {http://publications.lib.chalmers.se/publication/232302-time-dependent-relaxation-of-charge-and-energy-in-electronic-nanosystems}
  {\bibinfo {title} {{Time-dependent relaxation of charge and energy in
  electronic nanosystems}}} (\bibinfo {year} {2016}),\ \bibinfo {note}
  {{Licentiate thesis}}\BibitemShut {NoStop}%
\bibitem [{Jon()}]{Jonunpublished}%
  \BibitemOpen
  \href@noop {} {}\bibinfo {note} {Unpublished data}\BibitemShut {NoStop}%
\bibitem [{\citenamefont {Bonet}\ \emph {et~al.}(2002)\citenamefont {Bonet},
  \citenamefont {Deshmukh},\ and\ \citenamefont {Ralph}}]{Bonet2002Jan}%
  \BibitemOpen
  \bibfield  {author} {\bibinfo {author} {\bibfnamefont {E.}~\bibnamefont
  {Bonet}}, \bibinfo {author} {\bibfnamefont {M.~M.}\ \bibnamefont
  {Deshmukh}},\ and\ \bibinfo {author} {\bibfnamefont {D.~C.}\ \bibnamefont
  {Ralph}},\ }\bibfield  {title} {\bibinfo {title} {{Solving rate equations for
  electron tunneling via discrete quantum states}},\ }\href
  {https://doi.org/10.1103/PhysRevB.65.045317} {\bibfield  {journal} {\bibinfo
  {journal} {Phys. Rev. B}\ }\textbf {\bibinfo {volume} {65}},\ \bibinfo
  {pages} {045317} (\bibinfo {year} {2002})}\BibitemShut {NoStop}%
\bibitem [{\citenamefont {Hartman}\ \emph {et~al.}(2018)\citenamefont
  {Hartman}, \citenamefont {Olsen}, \citenamefont
  {L{\ifmmode\ddot{u}\else\"{u}\fi}scher}, \citenamefont {Samani},
  \citenamefont {Fallahi}, \citenamefont {Gardner}, \citenamefont {Manfra},\
  and\ \citenamefont {Folk}}]{Hartman2018Nov}%
  \BibitemOpen
  \bibfield  {author} {\bibinfo {author} {\bibfnamefont {N.}~\bibnamefont
  {Hartman}}, \bibinfo {author} {\bibfnamefont {C.}~\bibnamefont {Olsen}},
  \bibinfo {author} {\bibfnamefont {S.}~\bibnamefont
  {L{\ifmmode\ddot{u}\else\"{u}\fi}scher}}, \bibinfo {author} {\bibfnamefont
  {M.}~\bibnamefont {Samani}}, \bibinfo {author} {\bibfnamefont
  {S.}~\bibnamefont {Fallahi}}, \bibinfo {author} {\bibfnamefont {G.~C.}\
  \bibnamefont {Gardner}}, \bibinfo {author} {\bibfnamefont {M.}~\bibnamefont
  {Manfra}},\ and\ \bibinfo {author} {\bibfnamefont {J.}~\bibnamefont {Folk}},\
  }\bibfield  {title} {\bibinfo {title} {{Direct entropy measurement in a
  mesoscopic quantum system}},\ }\href
  {https://doi.org/10.1038/s41567-018-0250-5} {\bibfield  {journal} {\bibinfo
  {journal} {Nat. Phys.}\ }\textbf {\bibinfo {volume} {14}},\ \bibinfo {pages}
  {1083--1086} (\bibinfo {year} {2018})}\BibitemShut {NoStop}%
\bibitem [{\citenamefont {Kleeorin}\ \emph {et~al.}(2019)\citenamefont
  {Kleeorin}, \citenamefont {Thierschmann}, \citenamefont {Buhmann},
  \citenamefont {Georges}, \citenamefont {Molenkamp},\ and\ \citenamefont
  {Meir}}]{Kleeorin2019Dec}%
  \BibitemOpen
  \bibfield  {author} {\bibinfo {author} {\bibfnamefont {Y.}~\bibnamefont
  {Kleeorin}}, \bibinfo {author} {\bibfnamefont {H.}~\bibnamefont
  {Thierschmann}}, \bibinfo {author} {\bibfnamefont {H.}~\bibnamefont
  {Buhmann}}, \bibinfo {author} {\bibfnamefont {A.}~\bibnamefont {Georges}},
  \bibinfo {author} {\bibfnamefont {L.~W.}\ \bibnamefont {Molenkamp}},\ and\
  \bibinfo {author} {\bibfnamefont {Y.}~\bibnamefont {Meir}},\ }\bibfield
  {title} {\bibinfo {title} {{How to measure the entropy of a mesoscopic system
  via thermoelectric transport}},\ }\href
  {https://doi.org/10.1038/s41467-019-13630-3} {\bibfield  {journal} {\bibinfo
  {journal} {Nat. Commun.}\ }\textbf {\bibinfo {volume} {10}},\ \bibinfo
  {pages} {1--8} (\bibinfo {year} {2019})}\BibitemShut {NoStop}%
\bibitem [{\citenamefont {Riwar}\ \emph {et~al.}(2016)\citenamefont {Riwar},
  \citenamefont {Roche}, \citenamefont {Jehl},\ and\ \citenamefont
  {Splettstoesser}}]{Riwar2016Jun}%
  \BibitemOpen
  \bibfield  {author} {\bibinfo {author} {\bibfnamefont {R.-P.}\ \bibnamefont
  {Riwar}}, \bibinfo {author} {\bibfnamefont {B.}~\bibnamefont {Roche}},
  \bibinfo {author} {\bibfnamefont {X.}~\bibnamefont {Jehl}},\ and\ \bibinfo
  {author} {\bibfnamefont {J.}~\bibnamefont {Splettstoesser}},\ }\bibfield
  {title} {\bibinfo {title} {{Readout of relaxation rates by nonadiabatic
  pumping spectroscopy}},\ }\href {https://doi.org/10.1103/PhysRevB.93.235401}
  {\bibfield  {journal} {\bibinfo  {journal} {Phys. Rev. B}\ }\textbf {\bibinfo
  {volume} {93}},\ \bibinfo {pages} {235401} (\bibinfo {year}
  {2016})}\BibitemShut {NoStop}%
\bibitem [{\citenamefont {Antoine}\ \emph {et~al.}(2017)\citenamefont
  {Antoine}, \citenamefont {Lorin},\ and\ \citenamefont
  {Tang}}]{Antoine2017Aug}%
  \BibitemOpen
  \bibfield  {author} {\bibinfo {author} {\bibfnamefont {X.}~\bibnamefont
  {Antoine}}, \bibinfo {author} {\bibfnamefont {E.}~\bibnamefont {Lorin}},\
  and\ \bibinfo {author} {\bibfnamefont {Q.}~\bibnamefont {Tang}},\ }\bibfield
  {title} {\bibinfo {title} {{A friendly review of absorbing boundary
  conditions and perfectly matched layers for classical and relativistic
  quantum waves equations}},\ }\href
  {https://doi.org/10.1080/00268976.2017.1290834} {\bibfield  {journal}
  {\bibinfo  {journal} {Mol. Phys.}\ }\textbf {\bibinfo {volume} {115}},\
  \bibinfo {pages} {1861--1879} (\bibinfo {year} {2017})}\BibitemShut {NoStop}%
\bibitem [{\citenamefont {Selst{\o}}\ and\ \citenamefont
  {Kvaal}(2010)}]{Selsto2010Mar}%
  \BibitemOpen
  \bibfield  {author} {\bibinfo {author} {\bibfnamefont {S.}~\bibnamefont
  {Selst{\o}}}\ and\ \bibinfo {author} {\bibfnamefont {S.}~\bibnamefont
  {Kvaal}},\ }\bibfield  {title} {\bibinfo {title} {{Absorbing boundary
  conditions for dynamical many-body quantum systems}},\ }\href
  {https://doi.org/10.1088/0953-4075/43/6/065004} {\bibfield  {journal}
  {\bibinfo  {journal} {J. Phys. B: At. Mol. Opt. Phys.}\ }\textbf {\bibinfo
  {volume} {43}},\ \bibinfo {pages} {065004} (\bibinfo {year}
  {2010})}\BibitemShut {NoStop}%
\bibitem [{\citenamefont {Park}\ and\ \citenamefont
  {Light}(1986)}]{Park1986Nov}%
  \BibitemOpen
  \bibfield  {author} {\bibinfo {author} {\bibfnamefont {T.~J.}\ \bibnamefont
  {Park}}\ and\ \bibinfo {author} {\bibfnamefont {J.~C.}\ \bibnamefont
  {Light}},\ }\bibfield  {title} {\bibinfo {title} {{Unitary quantum time
  evolution by iterative Lanczos reduction}},\ }\href
  {https://doi.org/10.1063/1.451548} {\bibfield  {journal} {\bibinfo  {journal}
  {J. Chem. Phys.}\ }\textbf {\bibinfo {volume} {85}},\ \bibinfo {pages}
  {5870--5876} (\bibinfo {year} {1986})}\BibitemShut {NoStop}%
\bibitem [{\citenamefont {Kosloff}(1988)}]{Kosloff1988Apr}%
  \BibitemOpen
  \bibfield  {author} {\bibinfo {author} {\bibfnamefont {R.}~\bibnamefont
  {Kosloff}},\ }\bibfield  {title} {\bibinfo {title} {{Time-dependent
  quantum-mechanical methods for molecular dynamics}},\ }\href
  {https://doi.org/10.1021/j100319a003} {\bibfield  {journal} {\bibinfo
  {journal} {J. Phys. Chem.}\ }\textbf {\bibinfo {volume} {92}},\ \bibinfo
  {pages} {2087--2100} (\bibinfo {year} {1988})}\BibitemShut {NoStop}%
\bibitem [{\citenamefont {Visscher}(1991)}]{Visscher1991Nov}%
  \BibitemOpen
  \bibfield  {author} {\bibinfo {author} {\bibfnamefont {P.~B.}\ \bibnamefont
  {Visscher}},\ }\bibfield  {title} {\bibinfo {title} {{A fast explicit
  algorithm for the time{-}dependent Schr{\ifmmode\ddot{o}\else\"{o}\fi}dinger
  equation}},\ }\href {https://doi.org/10.1063/1.168415} {\bibfield  {journal}
  {\bibinfo  {journal} {Comput. Phys.}\ }\textbf {\bibinfo {volume} {5}},\
  \bibinfo {pages} {596--598} (\bibinfo {year} {1991})}\BibitemShut {NoStop}%
\bibitem [{\citenamefont {Schulenborg}\ \emph {et~al.}(2023)\citenamefont
  {Schulenborg}, \citenamefont {Kr\o{}jer}, \citenamefont {Burrello},
  \citenamefont {Leijnse},\ and\ \citenamefont
  {Flensberg}}]{Schulenborg2023Mar}%
  \BibitemOpen
  \bibfield  {author} {\bibinfo {author} {\bibfnamefont {J.}~\bibnamefont
  {Schulenborg}}, \bibinfo {author} {\bibfnamefont {S.}~\bibnamefont
  {Kr\o{}jer}}, \bibinfo {author} {\bibfnamefont {M.}~\bibnamefont {Burrello}},
  \bibinfo {author} {\bibfnamefont {M.}~\bibnamefont {Leijnse}},\ and\ \bibinfo
  {author} {\bibfnamefont {K.}~\bibnamefont {Flensberg}},\ }\bibfield  {title}
  {\bibinfo {title} {Detecting majorana modes by readout of poisoning-induced
  parity flips},\ }\href {https://doi.org/10.1103/PhysRevB.107.L121401}
  {\bibfield  {journal} {\bibinfo  {journal} {Phys. Rev. B}\ }\textbf {\bibinfo
  {volume} {107}},\ \bibinfo {pages} {L121401} (\bibinfo {year}
  {2023})}\BibitemShut {NoStop}%
\end{thebibliography}

%

\end{document}